\documentclass[a4paper,11pt]{article}
\pdfoutput=1

\usepackage{jinstpub}

\usepackage{color}
\usepackage[normalem]{ulem}
\usepackage{lineno}
\usepackage{graphicx}
\usepackage{subcaption}
\usepackage{mwe}
\usepackage{amsmath}
\usepackage[]{units}
\usepackage{amssymb}
\usepackage{tabu}
\usepackage{siunitx}
\usepackage{afterpage}

\usepackage{floatrow}
\usepackage[colorinlistoftodos]{todonotes}

\usepackage{helvet}
\def\Put(#1,#2)#3{\leavevmode\makebox(0,0){\put(#1,#2){#3}}}

\newcommand{\Ar}[1]{$\mathrm{{}^{#1}Ar}$\xspace}

\newlength{\figwidth}
\newlength{\fighalfwidth}
\collaboration{MicroBooNE Collaboration}

\title{Measurement of Space Charge Effects in the MicroBooNE LArTPC Using Cosmic Muons}

\author[ii]{P.~Abratenko}
\author[o]{M.~Alrashed}
\author[n]{R.~An}
\author[d]{J.~Anthony}
\author[hh]{J.~Asaadi}
\author[s]{A.~Ashkenazi}
\author[ll]{S.~Balasubramanian}
\author[k]{B.~Baller}
\author[t]{C.~Barnes}
\author[x]{G.~Barr}
% ...

% ...
\affiliation[d]{University of Cambridge, Cambridge CB3 0HE, United Kingdom}
% ...
\affiliation[k]{Fermi National Accelerator Laboratory (FNAL), Batavia, IL 60510, USA}
% ...
\affiliation[n]{Illinois Institute of Technology (IIT), Chicago, IL 60616, USA}
\affiliation[o]{Kansas State University (KSU), Manhattan, KS, 66506, USA}
% ...
\affiliation[s]{Massachusetts Institute of Technology (MIT), Cambridge, MA, 02139, USA}
\affiliation[t]{University of Michigan, Ann Arbor, MI, 48109, USA}
% ...
\affiliation[x]{University of Oxford, Oxford OX1 3RH, United Kingdom}
% ...
\affiliation[hh]{University of Texas, Arlington, TX, 76019, USA}
\affiliation[ii]{Tufts University, Medford, MA, 02155, USA}
% ...
\affiliation[ll]{Wright Laboratory, Department of Physics, Yale University, New Haven, CT, 06520, USA}

\emailAdd{microboone\_info@fnal.gov}

\abstract{
Large liquid argon time projection chambers (LArTPCs), especially those operating near the surface, are susceptible to space charge effects.  In the context of LArTPCs, the space charge effect is the build-up of slow-moving positive ions in the detector primarily due to ionization from cosmic rays, leading to a distortion of the electric field within the detector.  This effect leads to a displacement in the reconstructed position of signal ionization electrons in LArTPC detectors (``spatial distortions''), as well as to variations in the amount of electron-ion recombination experienced by ionization throughout the volume of the TPC.  We present techniques that can be used to measure and correct for space charge effects in large LArTPCs by making use of cosmic muons, including the use of track pairs to unambiguously pin down spatial distortions in three dimensions.  The performance of these calibration techniques are studied using both Monte Carlo simulation and MicroBooNE data, utilizing a UV laser system as a means to estimate the systematic bias associated with the calibration methodology.
}


\begin{document}
\maketitle
\flushbottom

\section{Introduction} \label{sec:intro}

To accurately reconstruct the trajectories of charged particles that travel through a liquid argon time projection chamber (LArTPC), it is essential to precisely know the magnitude and direction of the electric field throughout the active volume of the TPC.  The electric field is designed to be uniform throughout the TPC, established by the cathode, anode plane wires, and field cage geometry.  However, electric field non-uniformities, such as those due to space charge effects (SCE), result in ``spatial distortions'' or modifications to the trajectories of charged-particle-induced ionization electron clusters as they drift to the TPC sense wire planes~\cite{SCEmooney}.

\begin{figure}[tb]
\begin{center}
\includegraphics[width=0.55\linewidth]{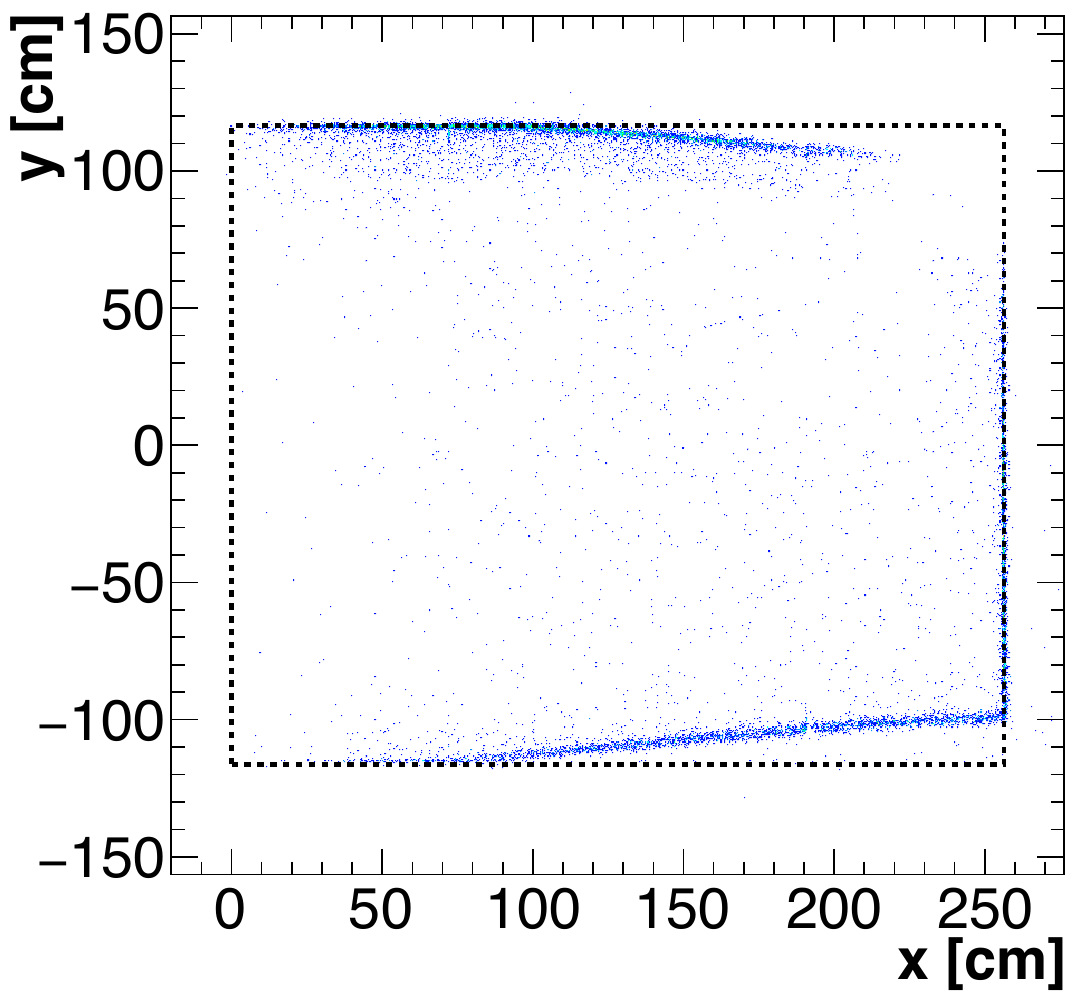}
\Put(-80,405){\fontfamily{phv}\selectfont \textbf{MicroBooNE}}
\end{center}
\caption{Entry/exit points of reconstructed cosmic muon tracks coincident with a signal from a muon counter located outside of the cryostat.  The points are shown in the $x-y$ plane.  In the absence of SCE and the associated non-uniform electric field in the detector volume, the points should be located strictly along the TPC boundaries (dashed lines).  The anode is located at $x = \SI{0}{cm}$ while the cathode is at $x = \SI{256}{cm}$.} \label{fig:distortion}
\end{figure}

A study of such effects at MicroBooNE~\cite{UBDetector}, a LArTPC neutrino detector at Fermilab that is 85 tons in active mass, was previously performed using muon tracks reconstructed in the TPC.  These tracks were required to pass through a small cosmic muon tagger~\cite{ubMuCS} on top of the cryostat in order to reconstruct their position in the drift direction.  Figure~\ref{fig:distortion} shows the entry/exit points of reconstructed tracks associated with muons from data in the rectangular MicroBooNE TPC projected onto the $x-y$ plane, where $x$ is aligned opposite to the ionization drift direction ($x\in[0,256]$~cm), $y$ is the vertical direction ($y\in[-116,116]$~cm), along the zenith axis, and $z$ is the beam direction ($z\in[0,1037]$~cm).  This simple map of the TPC volume, with the track points reconstructed under the assumption of uniform electric field, is indicative of a distorted electric field within the detector.  Instead of being located strictly along the TPC boundaries, the entry/exit points exhibit an offset from the edges of the TPC that increases in magnitude as the track origination point is further from the anode in $x$.  Trajectories of ionization electrons originating from the cathode, which undergo the longest possible drift before being detected by the sense wire planes, tend to be distorted more than those originating closer to the anode.  This significantly complicates the reconstruction of charged particle tracks, both in terms of spatial and calorimetric information.

A continuous, spatially varying distribution of excess electrons or ions within a drift volume, referred to as space charge, can affect the operation of a broad range of detectors involving drifting charge as the signal of interest~\cite{ChildPaper,LangmuirPaper}.  Space charge is produced in LArTPC detectors from the build-up of slow-moving positive ions, which originate from electron-ion pair creation via ionization activity in the detector.  As MicroBooNE is a surface detector with negligible overburden, the cosmic muon flux (20--30 muons in the active volume per \SI{4.8}{ms} readout window at the nominal drift electric field of \SI{273.9}{V/cm}) creates a significant amount of charge build-up from slow-moving positive argon ions.  This charge build-up substantially impacts the drift electric field within the TPC volume, as the argon ion drift velocity (approximately \SI{4}{mm/s}~\cite{ionvel}) is roughly five orders of magnitude smaller than the ionization electron drift velocity (approximately \SI{1.1}{mm/{\micro}s}; see section~\ref{sec:method_driftvel}) in the MicroBooNE TPC.  The resulting distortions in the drift electric field lead to modifications to the paths and total drift time of ionization electrons, as well as a varying amount of electron-ion recombination~\cite{ANrecomb}, which is sensitive to the magnitude of the electric field, in different parts of the TPC~\cite{calibration}.  These effects, if not taken into account, would result in incorrect reconstruction of particle position and deposited energy.  The observation of distorted tracks illustrated in figure~\ref{fig:distortion} is indicative of the presence of SCE in the MicroBooNE TPC active volume.  Because the positive ions tend to draw drifting ionization electrons closer to the center of the TPC, with the effect more pronounced when the drift time is longer, tracks originating closer to the cathode have reconstructed entry/exit points that are more offset from the edges of the TPC active volume.

We present a study comparing spatial distortions from the SCE predicted in simulation with those observed in MicroBooNE data.  This work focuses on estimating SCE in the detector using cosmic muons, and is complementary to efforts aimed at extracting the electric field throughout the detector using a UV laser system at MicroBooNE~\cite{ubLaser,laser_calib}.  Spatial distortions near the faces of the TPC are probed using single cosmic muon tracks and those throughout the bulk of the TPC are estimated using nearly intersecting pairs of cosmic muon tracks; pairs of tracks allow for the spatial distortions in the middle of the TPC to be unambiguously determined in three dimensions.  The spatial distortions estimated with this method are shown to be consistent with the presence of SCE in the MicroBooNE LArTPC.  Application of SCE corrections derived from this method on three-dimensional points along the trajectories of reconstructed laser tracks, produced using the aforementioned UV laser system, are shown to straighten the tracks.  Furthermore, the estimated spatial distortions can be used to determine the underlying electric field distortions, allowing us to account for the varying amount of electron-ion recombination throughout the detector, which is relevant for both particle identification and particle energy measurements.

%This article is structured as follows: first, we briefly discuss the simulation of SCE used to compare to observations of the effect in data (section~\ref{sec:sim}).  Following this, we describe the datasets used in the study, including a discussion of the cosmic muon track sample that the analysis utilizes (section~\ref{sec:datatracks}).  Next, the methodology used to determine the spatial distortions throughout the bulk of the detector using cosmic muon tracks is explained (section~\ref{sec:method}).  Results of the estimation of SCE in the MicroBooNE LArTPC are then shown, including a data-driven validation of the performance of an associated SCE calibration on cosmic muon tracks (section~\ref{sec:results}).  The calibration results are followed by the presentation of a study of systematic bias using MicroBooNE's UV laser system in the next section (section~\ref{sec:systbias}).  We then provide a brief study of the time dependence of SCE at MicroBooNE (section~\ref{sec:timedep}).  Finally, we discuss conclusions from this work and potential applications to future LArTPC neutrino experiments (section~\ref{sec:conclusions}).

\section{Simulation of Space Charge Effects} \label{sec:sim}

We have simulated the impact of space charge on the electric field within the TPC, along with the distortions in reconstructed ionization electron position at different points within the TPC volume.  This simulation uses a Fourier series solution to the boundary value problem to solve for the electric field on a three-dimensional grid within the TPC volume, an interpolation between the grid points using radial basis functions to find the electric field everywhere in the TPC, and ray tracing using the RKF45 (``Runge-Kutta-Fehlberg'') method~\cite{rkf} to simulate the expected distortions in reconstructed position of ionization electron clusters.  This last step requires knowing the dependence of the drift velocity on the electric field, $v(E)$.  The SCE simulation makes use of a drift velocity model that is established using a fifth-order polynomial fit to ICARUS T600 drift velocity measurements~\cite{ICARUSdriftvel} performed at a variety of electric fields and at the same cryogenic temperature as at MicroBooNE (\SI{89}{K}).  The data-driven calibration results shown later use an improved drift velocity model that includes a drift velocity measurement made at MicroBooNE, which is discussed in section~\ref{sec:method_driftvel}.

\begin{figure}
\centering
%made with google docs:
%https://docs.google.com/presentation/d/19mRRFDYG-Z1YSwdxFYrNEaxQO0Z9ZRTl_EJwR_3ds6c/edit?usp=sharing
\includegraphics[width=.5\textwidth]{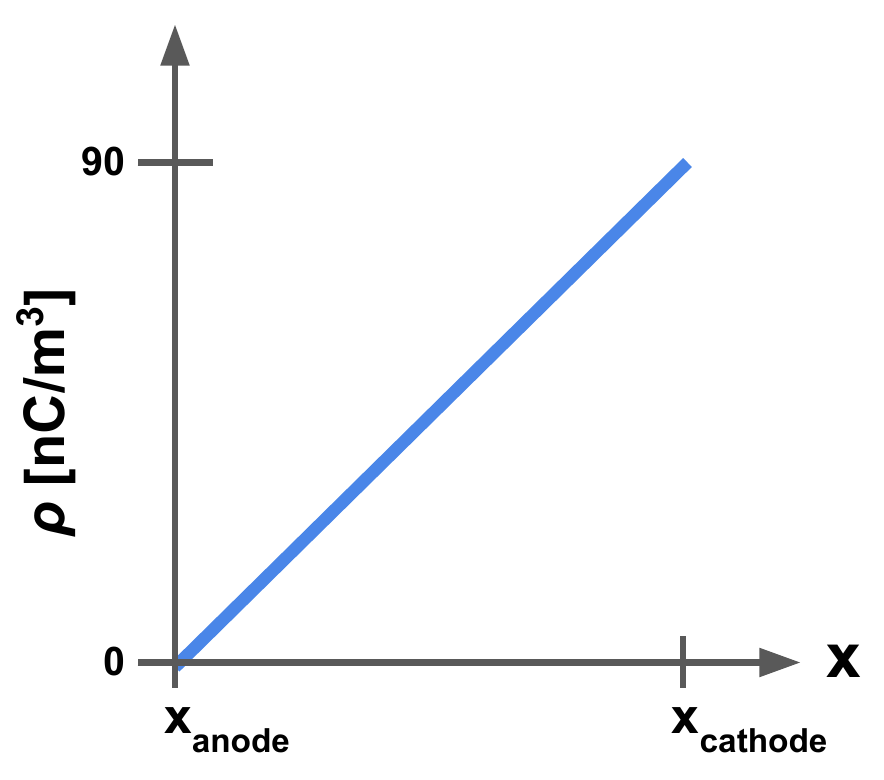}
\caption{Space charge density $\rho$ as a function of the $x$ position assumed in the simulation.  The space charge density at the cathode, $90$ $\mathrm{nC}/\mathrm{m}^3$, reflects both the expected rate of cosmic ray charge deposition and effects of recombination.  The distribution is independent of $y$ and $z$ in the simulation (an approximation).} \label{fig:simSCDist}
\end{figure}

For the SCE simulation, we assume that the charge deposition rate from cosmic muons is uniform across the TPC volume.  Based on the known angular spectrum of cosmic rays~\cite{PDG}, which initially produce roughly 50,000 electron-ion pairs per centimeter, we arrive at an ion generation rate of $1.6{\times}10^{-10}~\mathrm{C}/\mathrm{m}^{3}/\mathrm{s}$ at a drift electric field of \SI{273.9}{V/cm}, including the impact from electron-ion recombination.  Ignoring higher-order effects of the electric field distortions on the space charge configuration itself, we approximate the space charge density as linear with respect to the distance from the anode plane and independent of $y$ and $z$.  The linear space charge density profile used in the simulation is shown in figure~\ref{fig:simSCDist} for a drift electric field of \SI{273.9}{V/cm}.

\begin{figure}[p]
\centering
  \begin{subfigure}{0.41\textwidth}
    \centering
    \includegraphics[width=.99\textwidth]{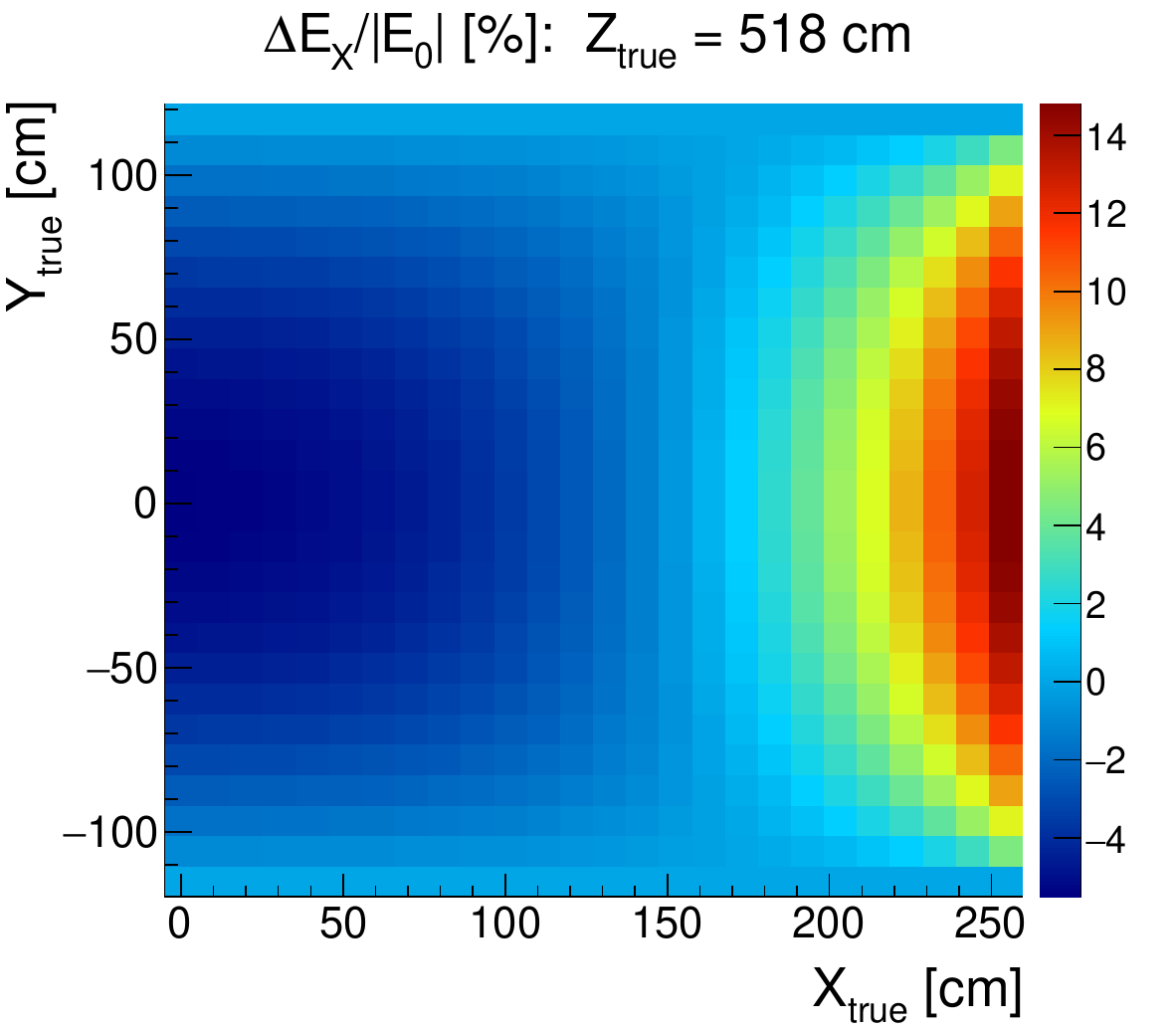}
    \caption{}
  \end{subfigure}
  \begin{subfigure}{0.41\textwidth}
    \centering
    \includegraphics[width=.99\textwidth]{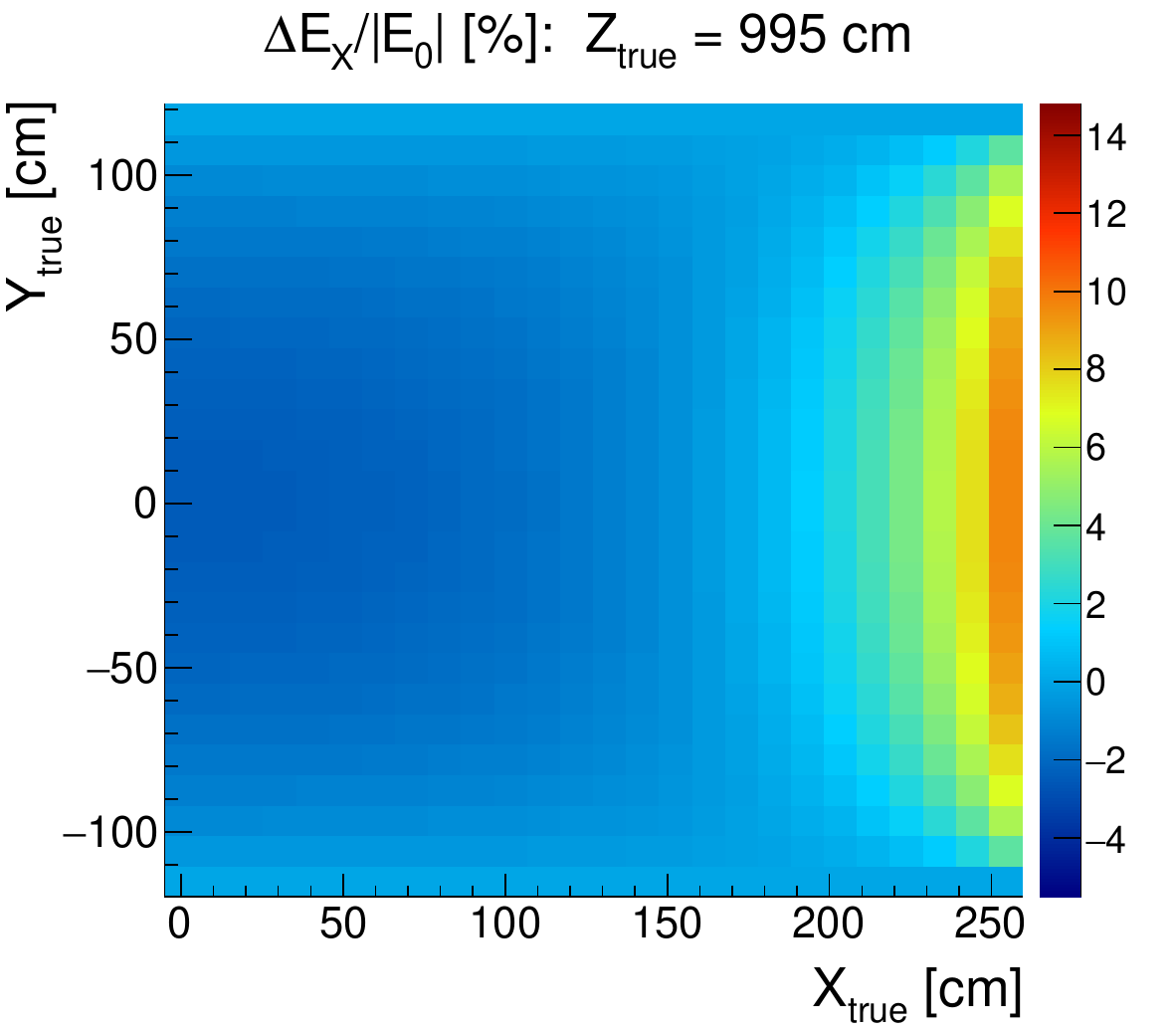}
    \caption{}
  \end{subfigure}
  \\
  \vspace{3mm}
  \begin{subfigure}{0.41\textwidth}
    \centering
    \includegraphics[width=.99\textwidth]{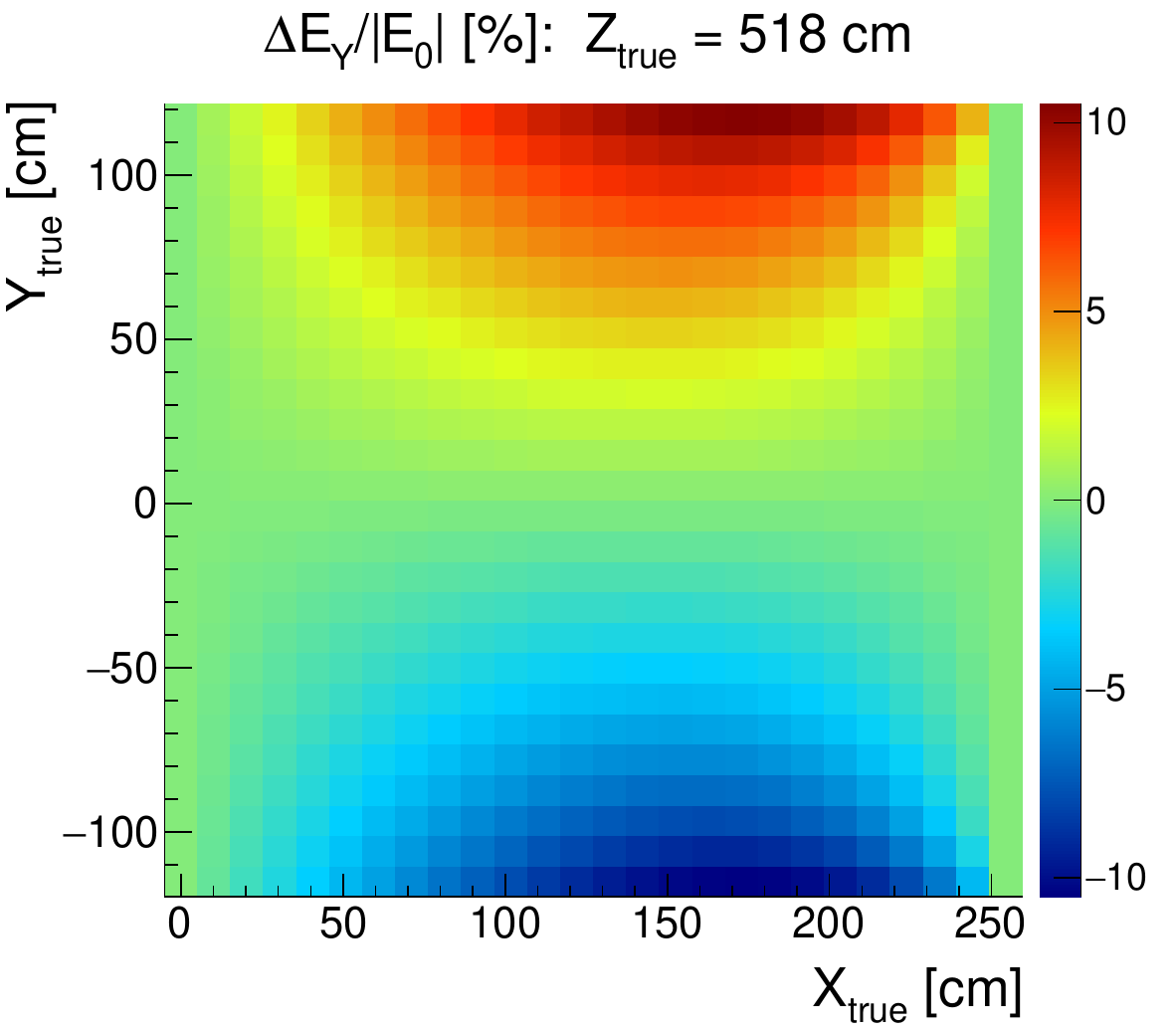}
    \caption{}
  \end{subfigure}
  \begin{subfigure}{0.41\textwidth}
    \centering
    \includegraphics[width=.99\textwidth]{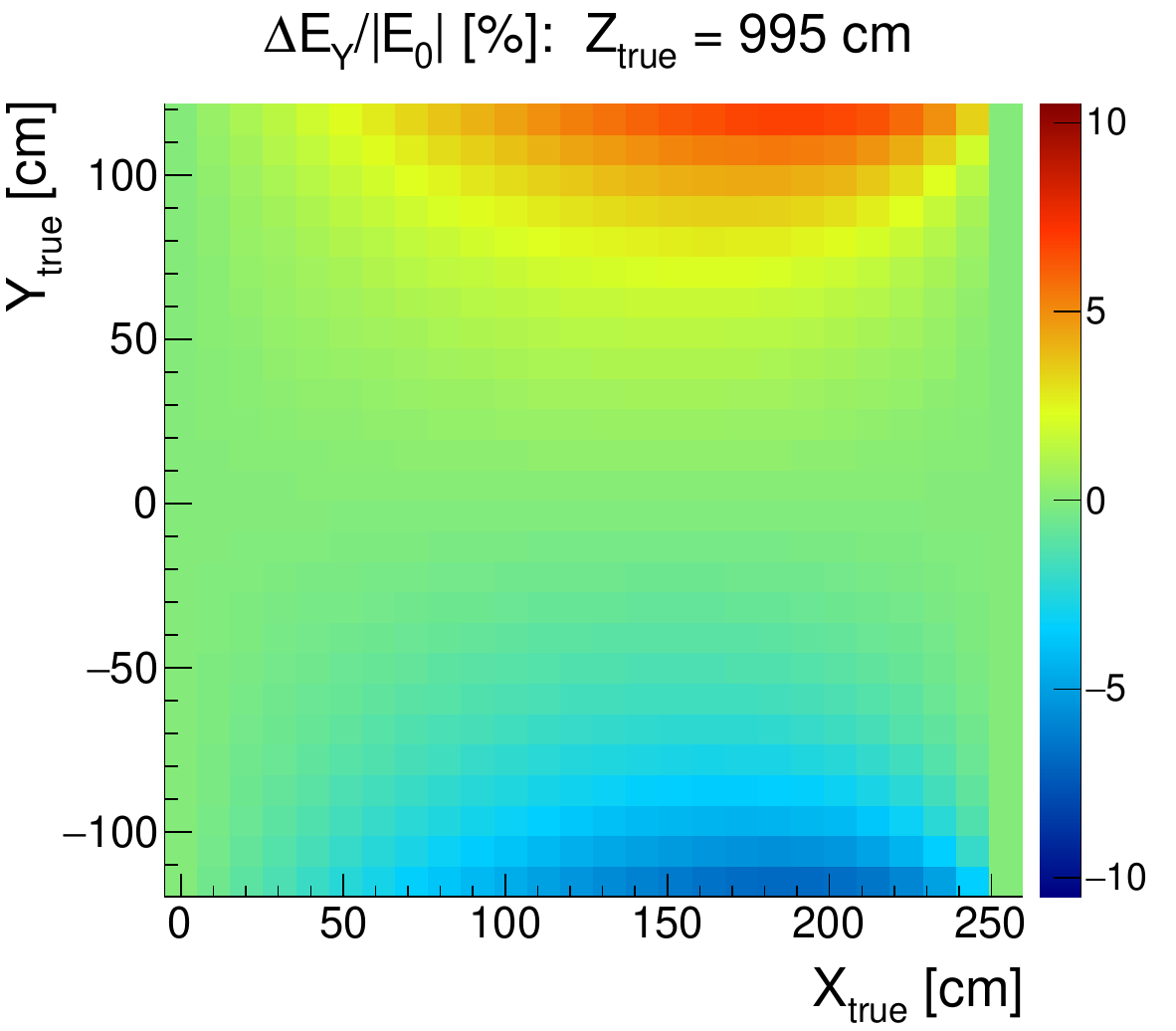}
    \caption{}
  \end{subfigure}
  \\
  \vspace{3mm}
  \begin{subfigure}{0.41\textwidth}
    \centering
    \includegraphics[width=.99\textwidth]{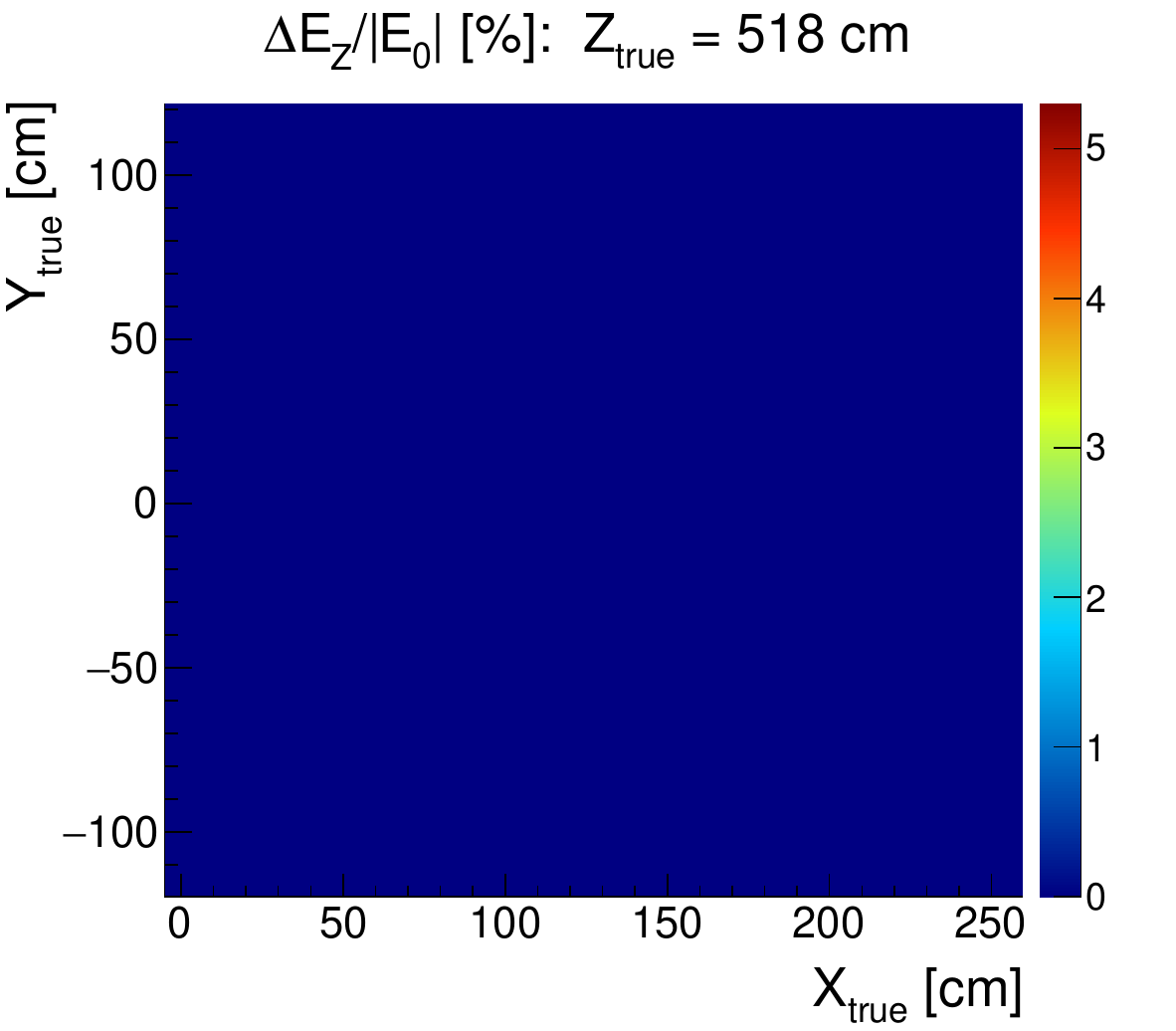}
    \caption{}
  \end{subfigure}
  \begin{subfigure}{0.41\textwidth}
    \centering
    \includegraphics[width=.99\textwidth]{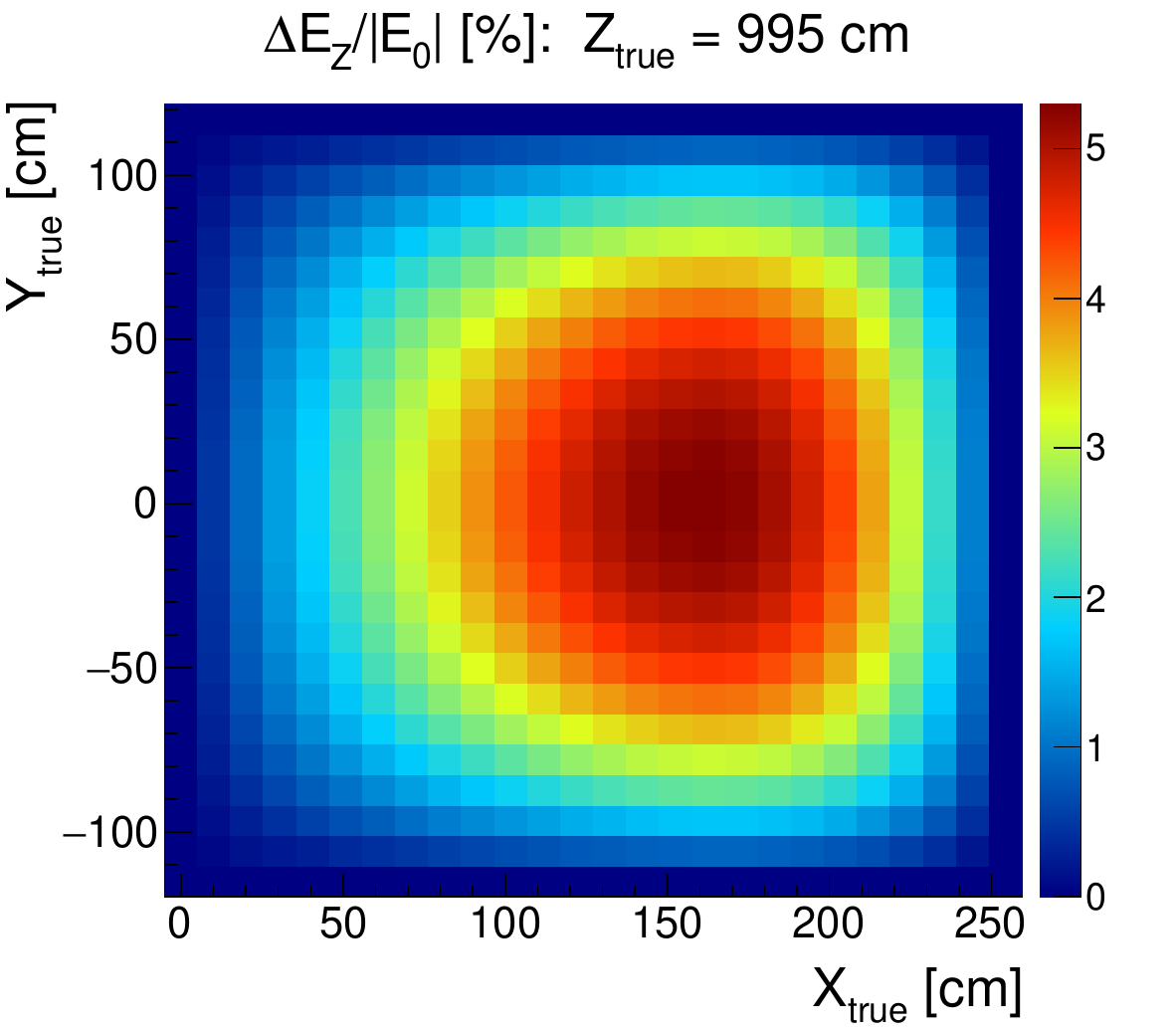}
    \caption{}
  \end{subfigure}
  \\
\Put(36,570){\fontfamily{phv}\selectfont \textbf{MicroBooNE}}
\Put(36,545){\fontfamily{phv}\selectfont \textbf{Simulation}}
  \caption{Illustration of the simulated effects of space charge on the drift electric field in the MicroBooNE TPC.  Results are shown for the effect on the (a, b) $x$-component, (c, d) $y$-component, and (e, f) $z$-component of the electric field.  The electric field distortions are normalized to the nominal drift electric field magnitude ($E_{0}$) of \SI{273.9}{V/cm} and are plotted as a function of the true position in the TPC.  Simulation results are shown both for (a, c, e) a central slice in $z$ and (b, d, f) a slice in $z$ close to the downstream end of the TPC.} \label{fig:simexample_Evals}
\end{figure}

\begin{figure}[p]
\centering
  \begin{subfigure}{0.41\textwidth}
    \centering
    \includegraphics[width=.99\textwidth]{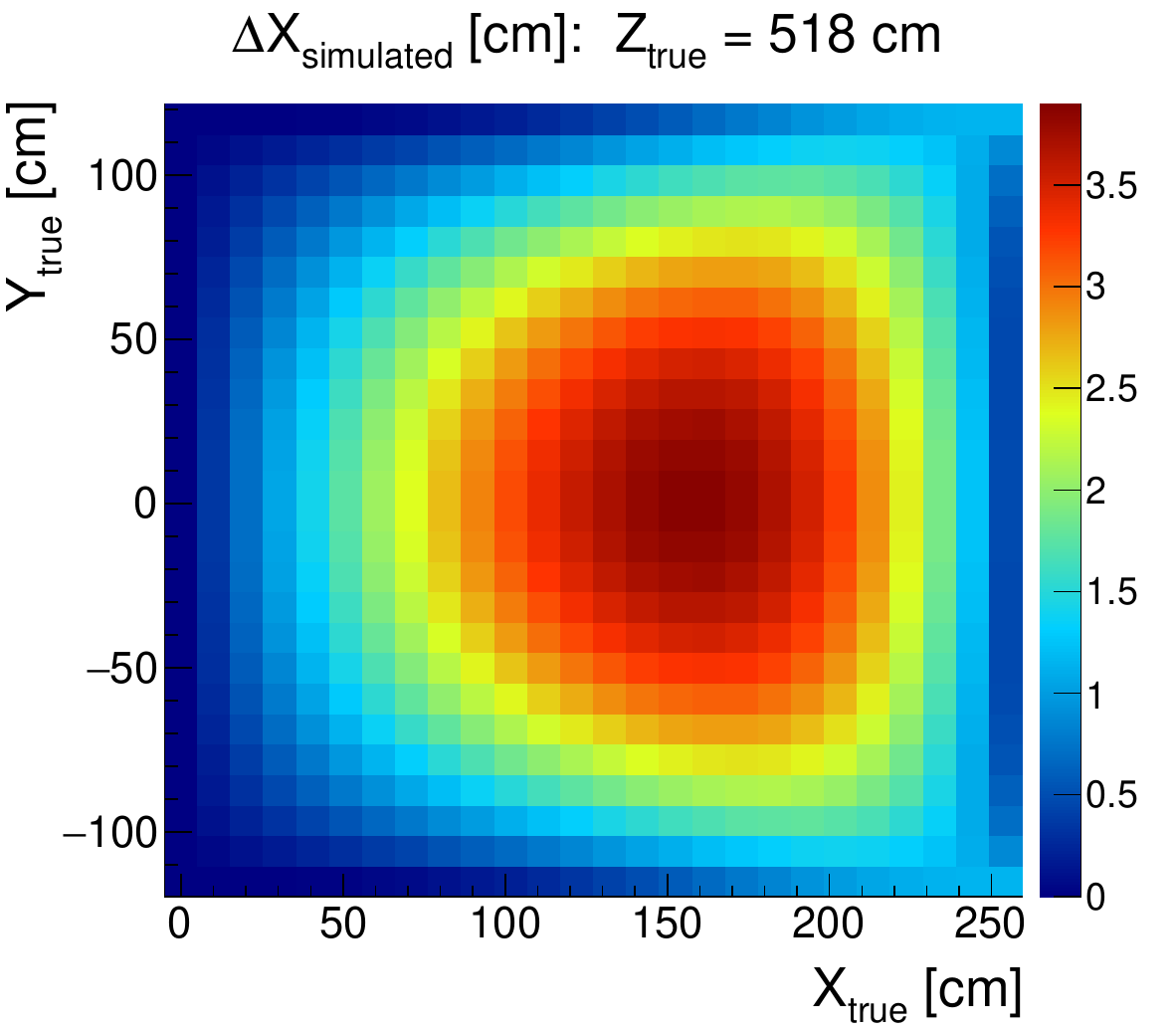}
    \caption{}
  \end{subfigure}
  \begin{subfigure}{0.41\textwidth}
    \centering
    \includegraphics[width=.99\textwidth]{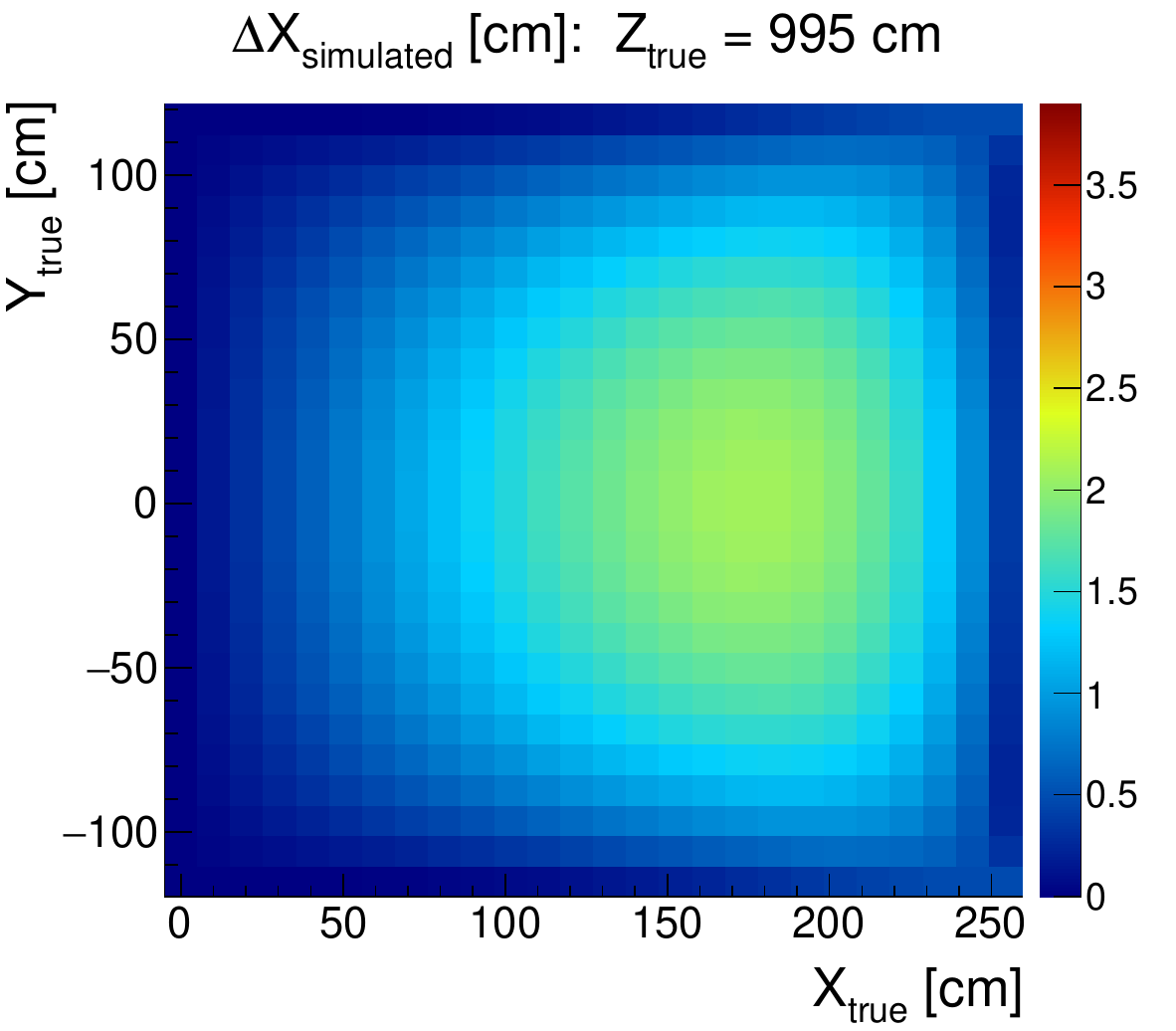}
    \caption{}
  \end{subfigure}
  \\
  \vspace{3mm}
  \begin{subfigure}{0.41\textwidth}
    \centering
    \includegraphics[width=.99\textwidth]{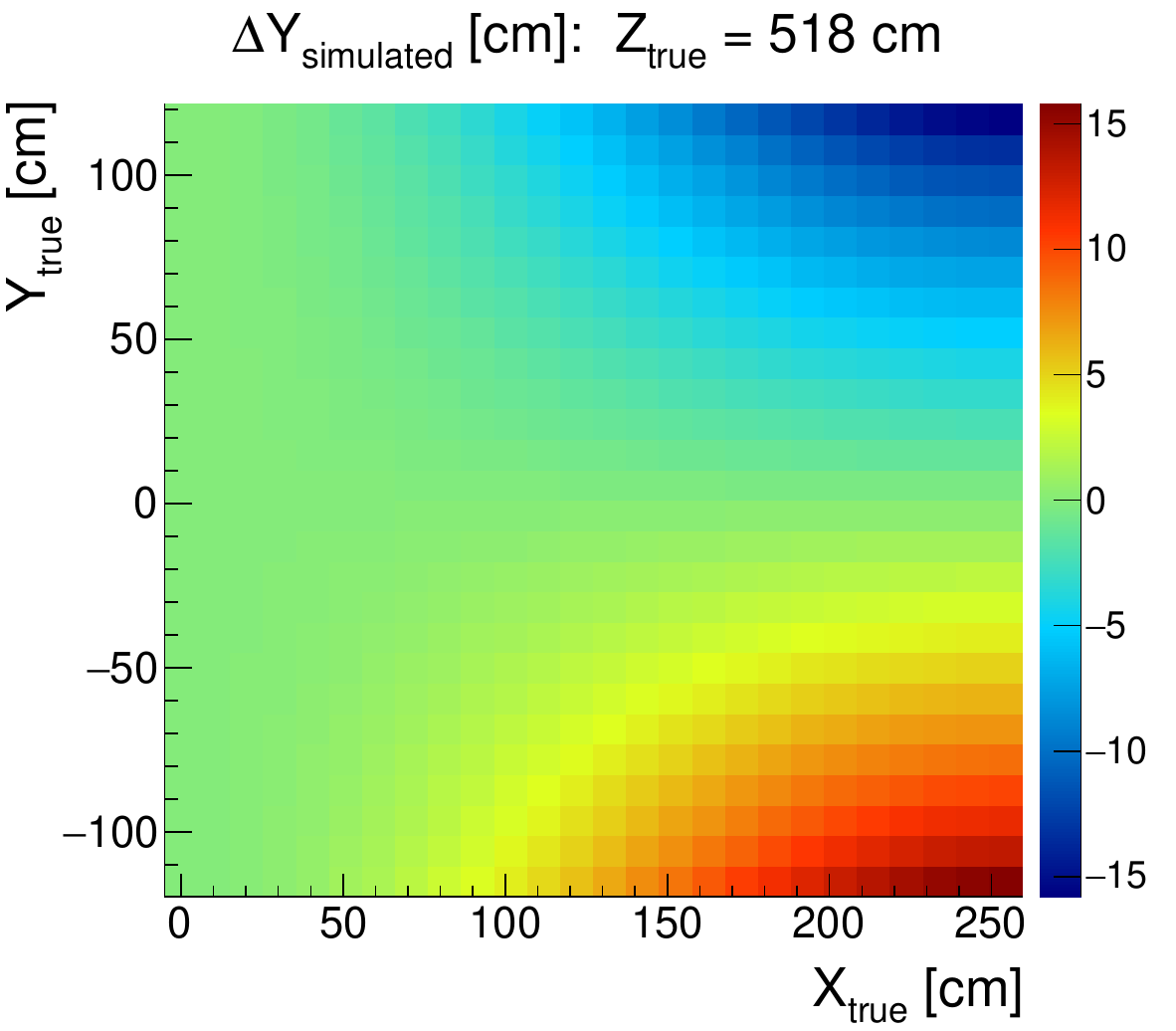}
    \caption{}
  \end{subfigure}
  \begin{subfigure}{0.41\textwidth}
    \centering
    \includegraphics[width=.99\textwidth]{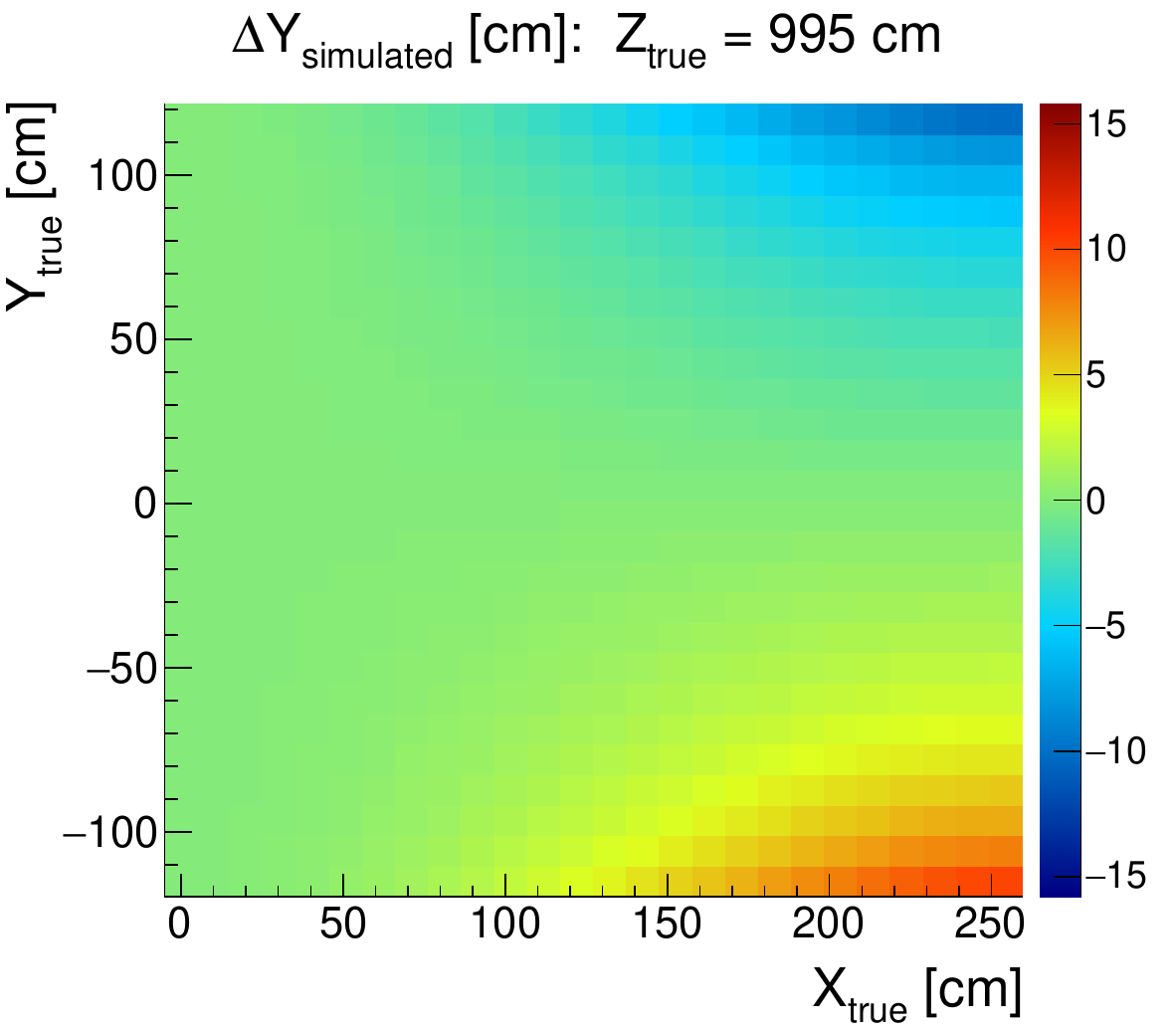}
    \caption{}
  \end{subfigure}
  \\
  \vspace{3mm}
  \begin{subfigure}{0.41\textwidth}
    \centering
    \includegraphics[width=.99\textwidth]{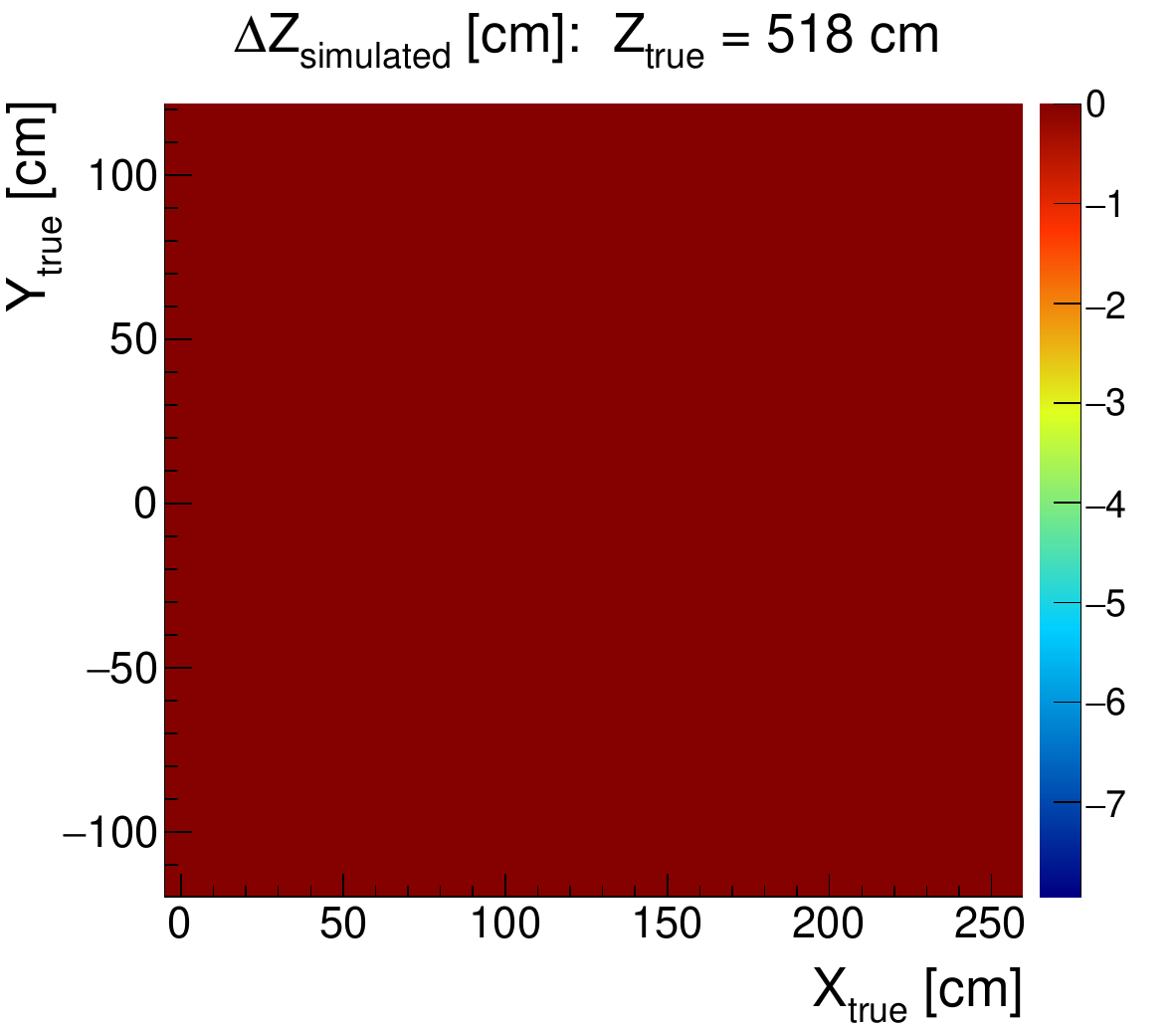}
    \caption{}
  \end{subfigure}
  \begin{subfigure}{0.41\textwidth}
    \centering
    \includegraphics[width=.99\textwidth]{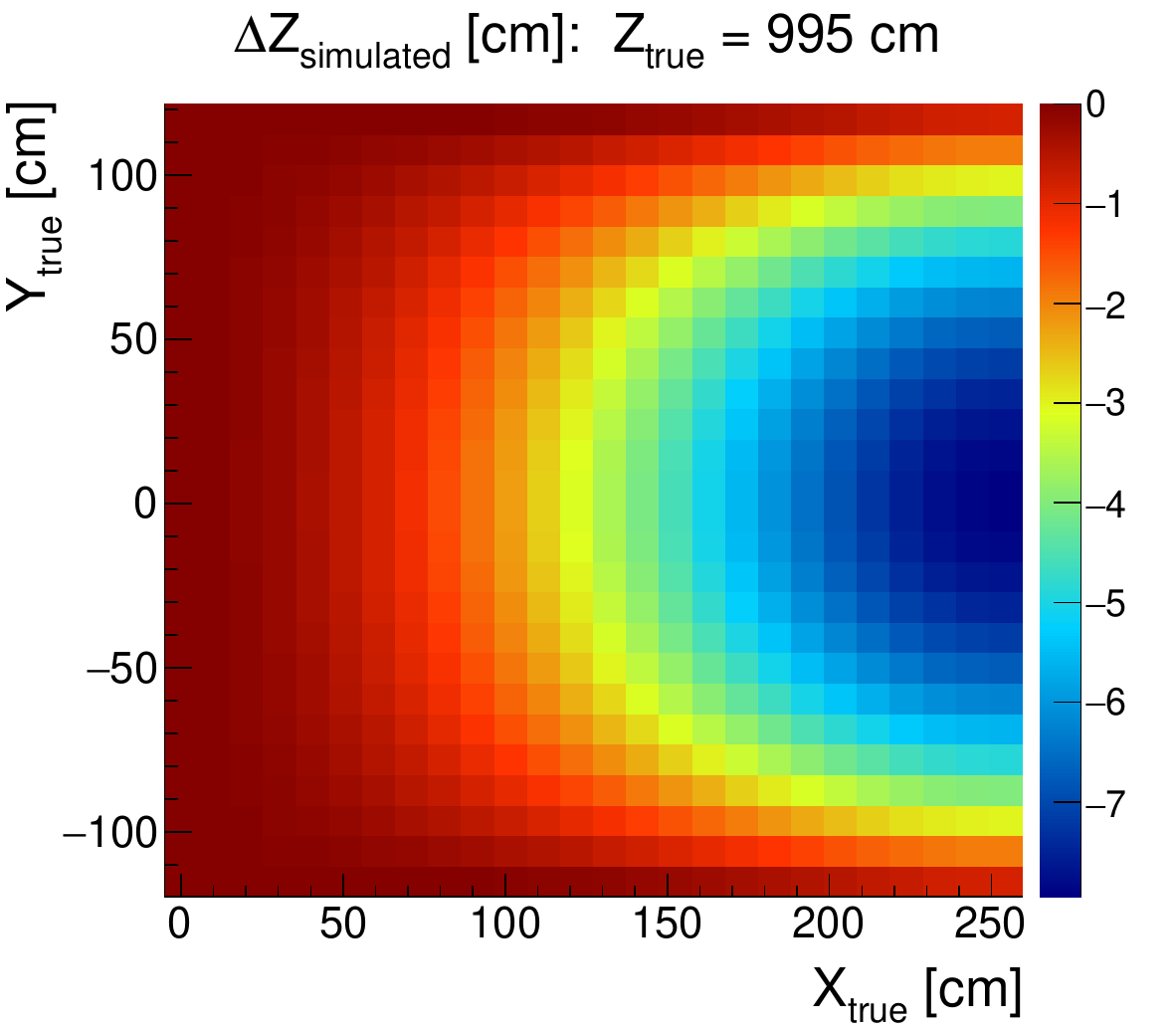}
    \caption{}
  \end{subfigure}
  \\
\Put(36,570){\fontfamily{phv}\selectfont \textbf{MicroBooNE}}
\Put(36,545){\fontfamily{phv}\selectfont \textbf{Simulation}}
\caption{Illustration of the simulated effects of space charge on the distortions in reconstructed ionization electron cluster position in the MicroBooNE TPC.  Results are shown for the spatial distortions in (a, b) $x$, (c, d) $y$, and (e, f) $z$.  The distortions in reconstructed ionization electron cluster position are shown in units of cm and are plotted as a function of the true position in the TPC.  Simulation results are shown both for (a, c, e) a central slice in $z$ and (b, d, f) a slice in $z$ close to the downstream end of the TPC.} \label{fig:simexample_Dvals}
\end{figure}

Representative samples of simulation results are shown in figure~\ref{fig:simexample_Evals} and figure~\ref{fig:simexample_Dvals}, which illustrate the impact of space charge on both the drift electric field (figure~\ref{fig:simexample_Evals}) as well as the distortions in reconstructed ionization electron cluster position (figure~\ref{fig:simexample_Dvals}).  At MicroBooNE's drift electric field of \SI{273.9}{V/cm}, the expected maximal impact on the electric field according to the simulation is 10--15\% in both the drift and transverse directions.  The impact of space charge near the center of the detector is primarily in the $x$ direction due to cancellations from similar amounts of space charge at smaller/larger values of $y$ and $z$.  On the other hand, the impact of space charge near the edges of the detector is mainly in the $y$ and/or $z$ directions due to the boundary condition imposed by the field cage (nominal electric field in the $x$ direction) and the much larger asymmetry of space charge in the $y$/$z$ directions near the TPC edges.

The simulation provides a useful estimation of the distortions in reconstructed ionization electron cluster position within the MicroBooNE TPC and elucidates basic features that we might expect in the data.  However, there are several limitations of the simulation that motivate the need for a data-driven technique for fully characterizing the effect.  The flow of liquid argon, which moves positive argon ions into or out of the active TPC volume, is not simulated.  Also, the charge deposition from cosmic rays throughout the TPC may not be uniform as a result of enhanced cosmogenic activity near the top of the detector due to interactions in the detector overburden, which would lead to greater ion production rates closer to the top of the TPC active volume.  Finally, the linear space charge density assumed in the simulation (see figure~\ref{fig:simSCDist}) approximates the ion drift speed, roughly \SI{4}{mm/s}~\cite{ionvel} at a drift electric field of \SI{273.9}{V/cm}, as constant throughout the TPC, while the electric field distortions arising from the SCE itself break this assumption.

\section{Track Samples and Datasets} \label{sec:datatracks}

Since the SCE impacts the electric field in the TPC and track reconstruction in three dimensions, knowledge of the three-dimensional information associated with tracks is necessary to constrain the effect.  When cosmic ray muons are reconstructed in MicroBooNE, their positions along the drift direction are reconstructed with an offset proportional to the difference between the time when the track passes through the TPC and the TPC readout trigger time, as the charge takes a finite time to drift to the anode before any signals are observed.  This quantity, referred to as $t_{0}$, enables determination of the true position of the track in the drift direction.  While a variety of methods exist to determine $t_{0}$ for a given cosmic muon track, the $t_{0}$-tagging method used in the present work is briefly summarized below.

\begin{figure}[tb]
  \centering
  \begin{subfigure}{0.46\textwidth}
    \centering
    \includegraphics[width=.99\textwidth]{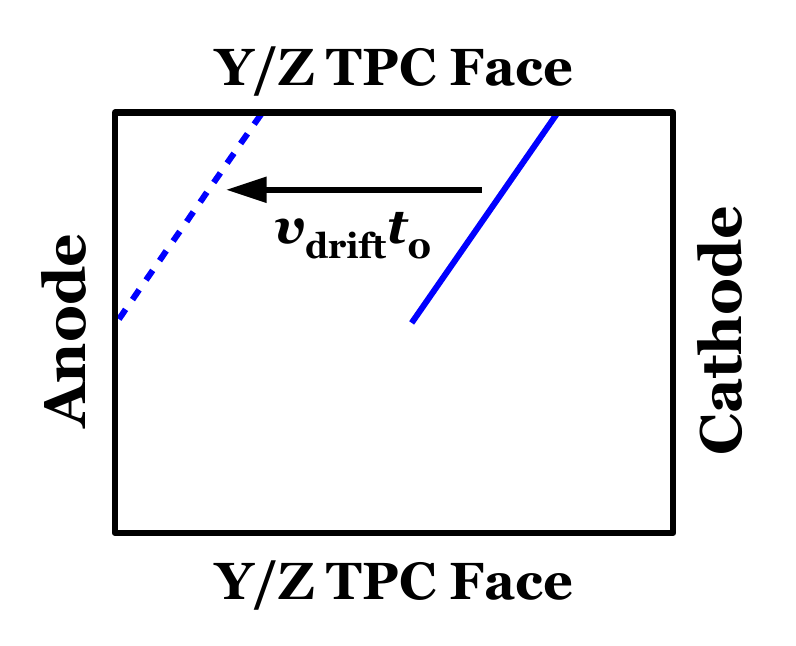}
    \caption{}
  \end{subfigure}
  \begin{subfigure}{0.46\textwidth}
    \centering
    \includegraphics[width=.99\textwidth]{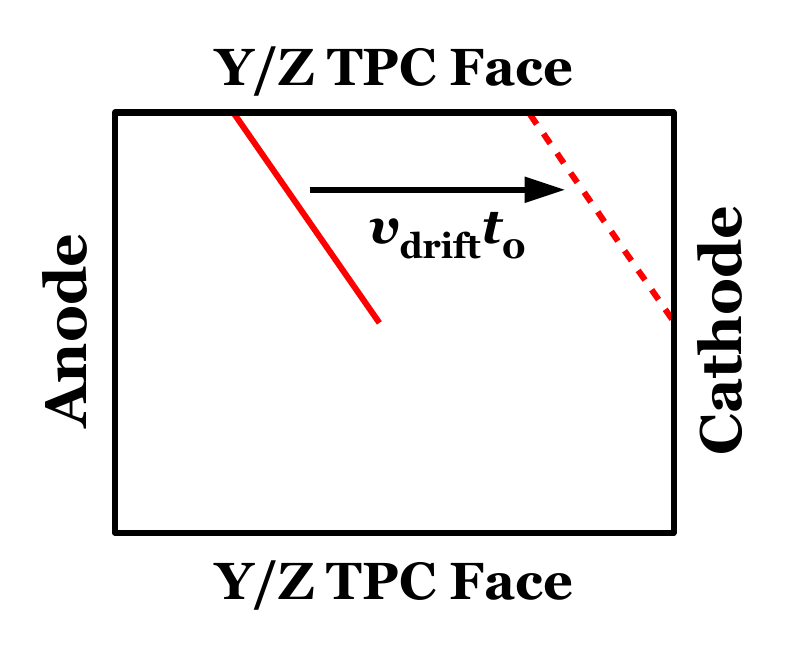}
    \caption{}
  \end{subfigure}
  \caption{Illustration of the $t_{0}$-tagging method used in these studies.  The drift coordinates of cosmic ray tracks are corrected using the assumption that the cosmic ray is through-going, requiring the track to pass through only one TPC face in $y$ or $z$, and using the angle of the track in the $y-x$ plane (or $z-x$ plane) to determine if the track is (a) anode piercing or (b) cathode piercing.  The known ionization electron drift velocity, $v_{\mathrm{drift}}$, is then used to correct the drift coordinate of the cosmic ray track.  The track before and after correction is shown by the solid and dashed lines, respectively.  Before the correction (when the $t_{0}$ of the track is assumed to be the trigger time), the track incorrectly appears to stop in the middle of the TPC.}
  \label{fig:t0tagging}
\end{figure}

The $t_{0}$-tagging method is visualized in figure~\ref{fig:t0tagging}.  The track associated with each cosmic ray is reconstructed in three dimensions using the Pandora multi-algorithm pattern recognition software~\cite{Pandora} employed on TPC signals, either simulated or collected with the MicroBooNE detector.  Pandora builds tracks by constructing two-dimensional clusters from proximate ``hits,'' which are distinct depositions of charge on wires.  These two-dimensional clusters are combined across the three wire planes to form three-dimensional clusters of deposited charge, which are reconstructed into particle tracks if they are line-like in nature.  Cosmic rays are assumed to be through-going (passing through two faces of the TPC).  This is a good assumption as the vast majority of cosmic rays have high enough momentum to pass through the TPC completely without stopping.  Next, the track is required to either enter or exit through exactly one of four faces of the TPC: top (high $y$), bottom (low $y$), upstream (low $z$), or downstream (high $z$).  If this condition is met, and the cosmic ray is through-going, the cosmic ray must either enter or exit through the anode or cathode.  Depending on the angle that the reconstructed track makes in the $y-x$ plane (or $z-x$ plane), the cosmic ray is determined to be either ``anode piercing'' or ``cathode piercing'' (see figure~\ref{fig:t0tagging}), and the value of $t_{0}$ is assigned to the time associated with the signal peak on the waveform at the part of the track closest to either the anode or cathode, respectively.  A ``flash'' of light seen in one or more of MicroBooNE's 32 eight-inch photomultiplier tubes (PMTs)~\cite{ubPMT} is required to be found at the same time as the determined $t_{0}$ value (within \SI{1}{{\micro}s}) in order to increase the purity of the $t_{0}$-tagging technique.  In principle, it may be possible to obtain the $t_{0}$ information for cosmic muon tracks using strictly the PMT flash information, though this is complicated for near-surface LArTPC detectors by the presence of many cosmic muon tracks in a single event and associated ambiguities with matching TPC tracks to PMT flashes (an active topic of research).

The drift coordinate of each $t_{0}$-tagged track is obtained using the known drift velocity ($v_{\mathrm{drift}}$) of ionization electrons in liquid argon at an electric field of \SI{273.9}{V/cm}, \SI{1.098}{mm/{\micro}s}~(see section~\ref{sec:method_driftvel}).  Validations using Monte Carlo simulation and data-driven methods utilizing an external cosmic ray tagger~\cite{ubMuCS} have shown this $t_{0}$-tagging method reconstructs the track $t_{0}$ correctly $>$98\% of the time for anode-piercing tracks and $>$97\% of the time for cathode-piercing tracks.

For the study of MicroBooNE data, multiple datasets are used.  The study uses TPC readout windows where the incoming neutrino beam is off, giving an event which only includes cosmic ray activity.  Approximately 5M events of this type were collected; however, to achieve sufficient statistics for the calibration, only 200k events are needed.  For SCE time dependence studies, which require greater statistics, all 5M events are used (see section~\ref{sec:timedep}).  For Monte Carlo (MC) simulation studies of cosmic muons, approximately 1M events are generated using the CORSIKA generator~\cite{CORSIKA}.  This uses the nominal MicroBooNE detector simulation~\cite{UBnoise,SP1,SP2,Pandora} as well as the simulation of SCE described in section~\ref{sec:sim}.  Both data and simulation use the same event reconstruction chain~\cite{SP1,SP2,Pandora}.  Each event in data contains 1.5~$t_{0}$-tagged tracks on average, while each CORSIKA MC event contains 3.0~$t_{0}$-tagged tracks on average.  The difference between these two $t_{0}$-tagging rates is largely due to CORSIKA overestimating the overall cosmic ray rate, though the difference in SCE in data and MC contributes to this rate difference as well.

\section{Calibration Methodology} \label{sec:method}

To estimate SCE throughout the detector, we first determine the offsets in reconstructed charge position (``spatial distortions'') that arise due to the underlying electric field variations.  The underlying electric field distortions can be extracted from the spatial distortion map with the application of Maxwell's equations and knowledge of the nonlinear relationship of drift velocity to electric field (see section~\ref{sec:method_driftvel}).  Furthermore, it is advantageous to use reconstructed charge position as opposed to the amount of reconstructed charge to isolate the impact of SCE within the TPC, as a variety of other detector effects, such as attachment of ionization to electronegative impurities in the liquid argon, can impact the amount of charge collected at the TPC wires.  Finally, the position of reconstructed charge deposits in the detector must be corrected using the spatial offsets derived from a data-driven calibration.  In addition to spatial distortions compromising particle reconstruction, the failure to correct for SCE may result in unwanted variations in measured ionization charge per unit length throughout the detector due to localized ``squeezing'' or ``stretching'' of charge.

To determine the spatial distortions from SCE throughout the MicroBooNE detector, there are two options: make use of the UV laser system installed at either end of the TPC~\cite{ubLaser}, or utilize through-going cosmic muons.  While the UV laser system has the advantage of knowledge of the track's true position, it illuminates a restricted region of the detector volume due to the limited motion of the reflecting mirrors.  The set of locations in the detector where two laser beams can nearly cross, allowing for an umabiguous three-dimensional spatial correction, is limited.  In contrast, the cosmic muon calibration is able to measure SCE distortions in nearly all regions of the TPC.  The study presented in this work explores the second option to constrain the spatial distortions associated with SCE.  A separate effort explores the extraction of the spatial distortion map using the UV laser system~\cite{laser_calib}; these two maps can be merged to obtain greater precision and coverage, and the underlying electric field distortions can be extracted from the combined spatial distortion map.  The electric field distortion map can then be used to correct for variations in the amount of electron-ion recombination experienced by deposited charge throughout the TPC volume.

\begin{figure}[tb]
\centering
\includegraphics[width=.99\textwidth]{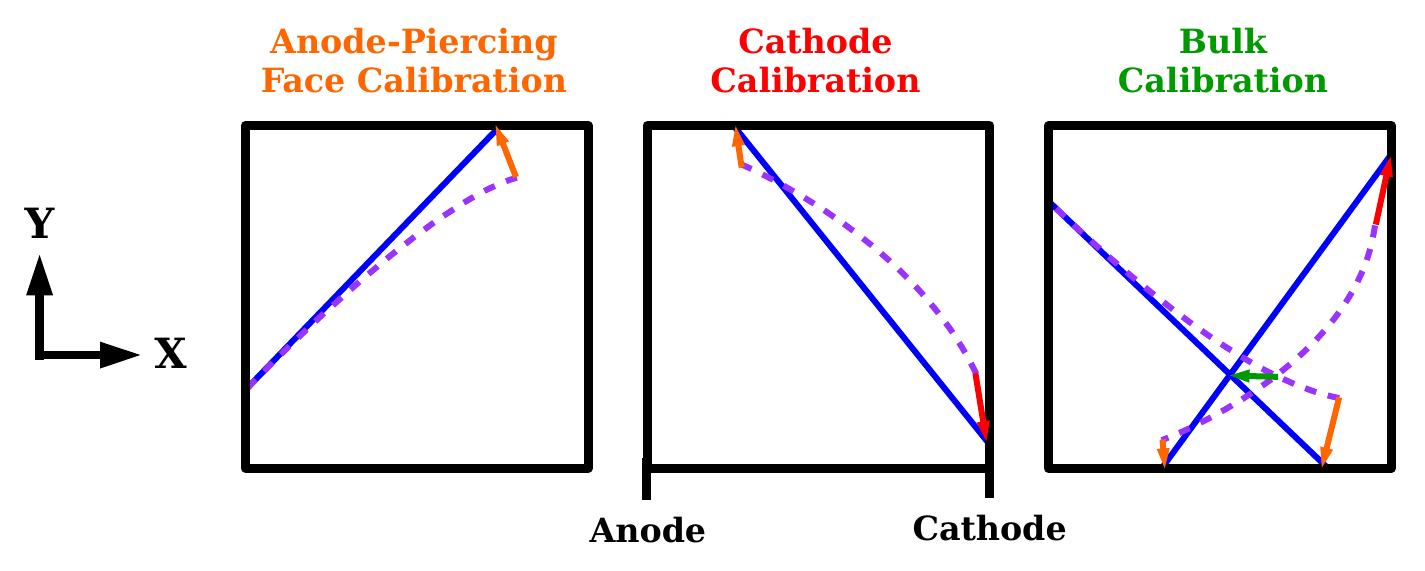}
\caption{Illustration of the three steps of the calibration method employed in order to estimate spatial offsets associated with SCE at MicroBooNE.  Included is the anode-piercing face calibration (left), cathode calibration (center), and bulk calibration (right).  The first two steps (orange and red arrows) correct the end points of the ``true'' track trajectories, while the third step (green arrow) yields a spatial offset map throughout the majority of the TPC volume to be used for correcting the reconstructed position of ionization charge.} \label{fig:calibmethod}
\end{figure}

One method for obtaining the spatial distortion map using through-going cosmic muons is presented in figure~\ref{fig:calibmethod}.  The main principle of this technique is to find the ``true'' end points of $t_{0}$-tagged cosmic muon tracks that pierce the TPC faces to estimate the actual trajectory of the particles through the TPC, ignoring effects of multiple Coulomb scattering (which are averaged out when using a large sample of tracks); the ``true'' trajectory of the muon through the TPC can then be compared to the reconstructed track to estimate the impact of SCE.  The individual steps of this calibration technique are discussed in detail below.

\subsection{Measurement of Nominal Drift Velocity} \label{sec:method_driftvel}

Before carrying out the SCE calibration procedure, it is necessary to determine the nominal drift velocity of ionization electrons in the TPC at MicroBooNE's nominal drift electric field of \SI{273.9}{V/cm}.  The nominal drift velocity is used in the first pass of particle reconstruction, which assumes a uniform electric field and a constant drift velocity, to convert times associated with ionization charge arriving at the TPC wire planes into physical distance in the drift direction.

A set of tracks crossing both the anode and cathode that is a subset of the sample of 5M events described in section~\ref{sec:datatracks} is used for this measurement and amounts to 27k tracks in total.  The drift times associated with ionization from the cathode side of the tracks forms a distribution that peaks at \SI{2.321}{ms}, corresponding to the maximum possible drift time associated with ionization in the TPC.  The measured cathode-anode drift distance is \SI{254.8}{cm} at room temperature, which is calculated to shrink to \SI{254.4}{cm} at \SI{89}{K}, the operating temperature.  An additional \SI{0.5}{cm} is added to account for cumulative space charge effects in the drift direction for charge originating near the cathode, as predicted by the SCE simulation discussed in section~\ref{sec:sim}.  This leads to a calculated nominal drift velocity of \SI[separate-uncertainty=true,multi-part-units=single]{1.098{\pm}0.004}{mm/{\micro}s}.  The uncertainty is dominated by a 0.3\% uncertainty that is associated with variations in the cathode-anode drift distance across the detector, estimated using TPC survey data taken prior to installation of the MicroBooNE TPC into the cryostat.  Another contribution is a 0.2\% uncertainty associated with the SCE correction, which may be different in data; a relative uncertainty of 100\% is associated with this a priori correction, from which the contribution of 0.2\% to the uncertainty on the drift velocity is derived.  Additional 0.1\% uncertainties on this measurement arise from the statistical uncertainty on the timing measurement as well as potential bias associated with reconstructing the length of the muon tracks.  These four sources of uncertainty added in quadrature yield a 0.4\% uncertainty on the drift velocity measurement.

\begin{figure}[tb]
\centering
  \begin{subfigure}{0.49\textwidth}
    \centering
    \includegraphics[width=.99\textwidth]{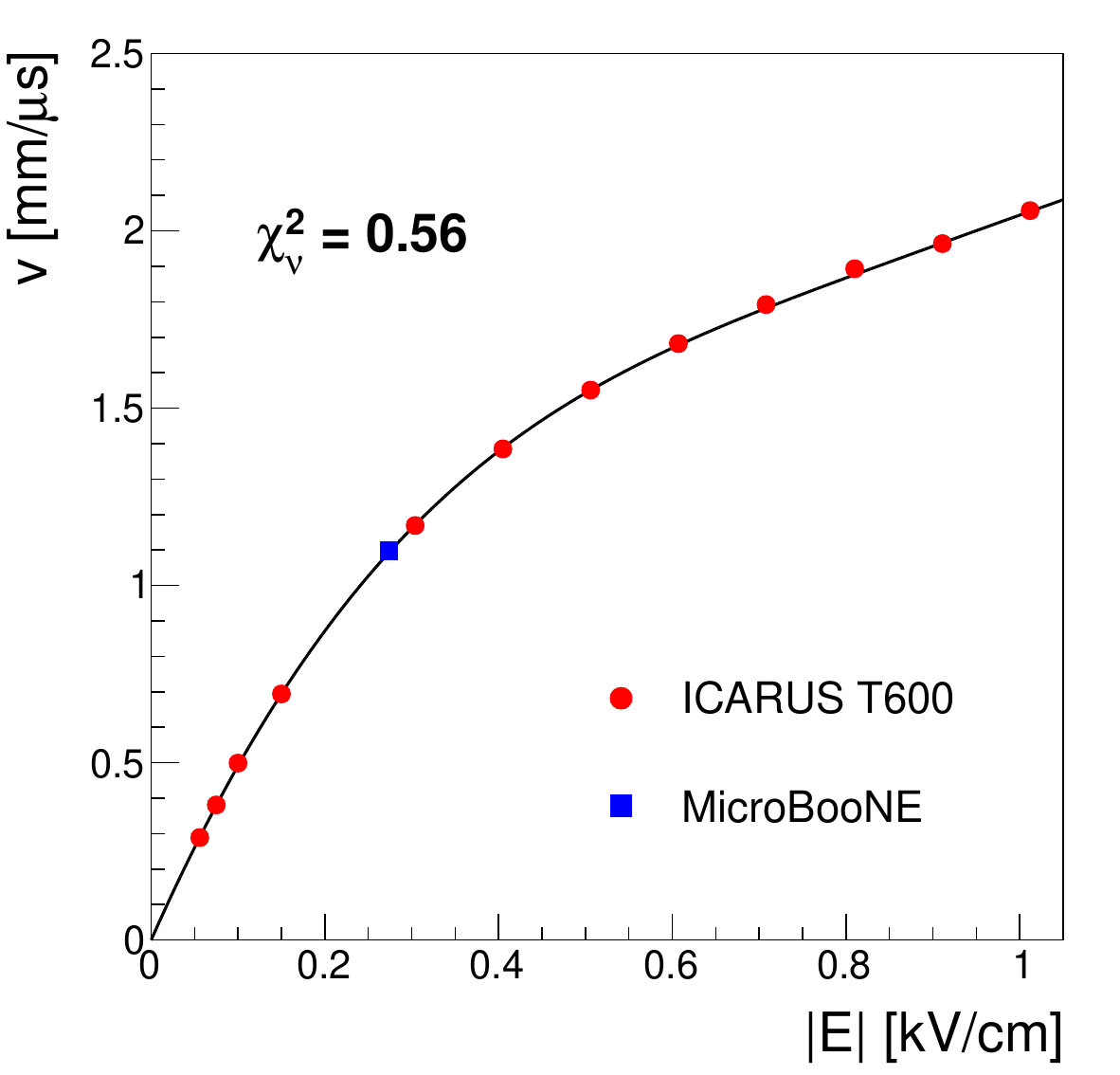}
    \caption{}
  \end{subfigure}
  \begin{subfigure}{0.49\textwidth}
    \centering
    \includegraphics[width=.99\textwidth]{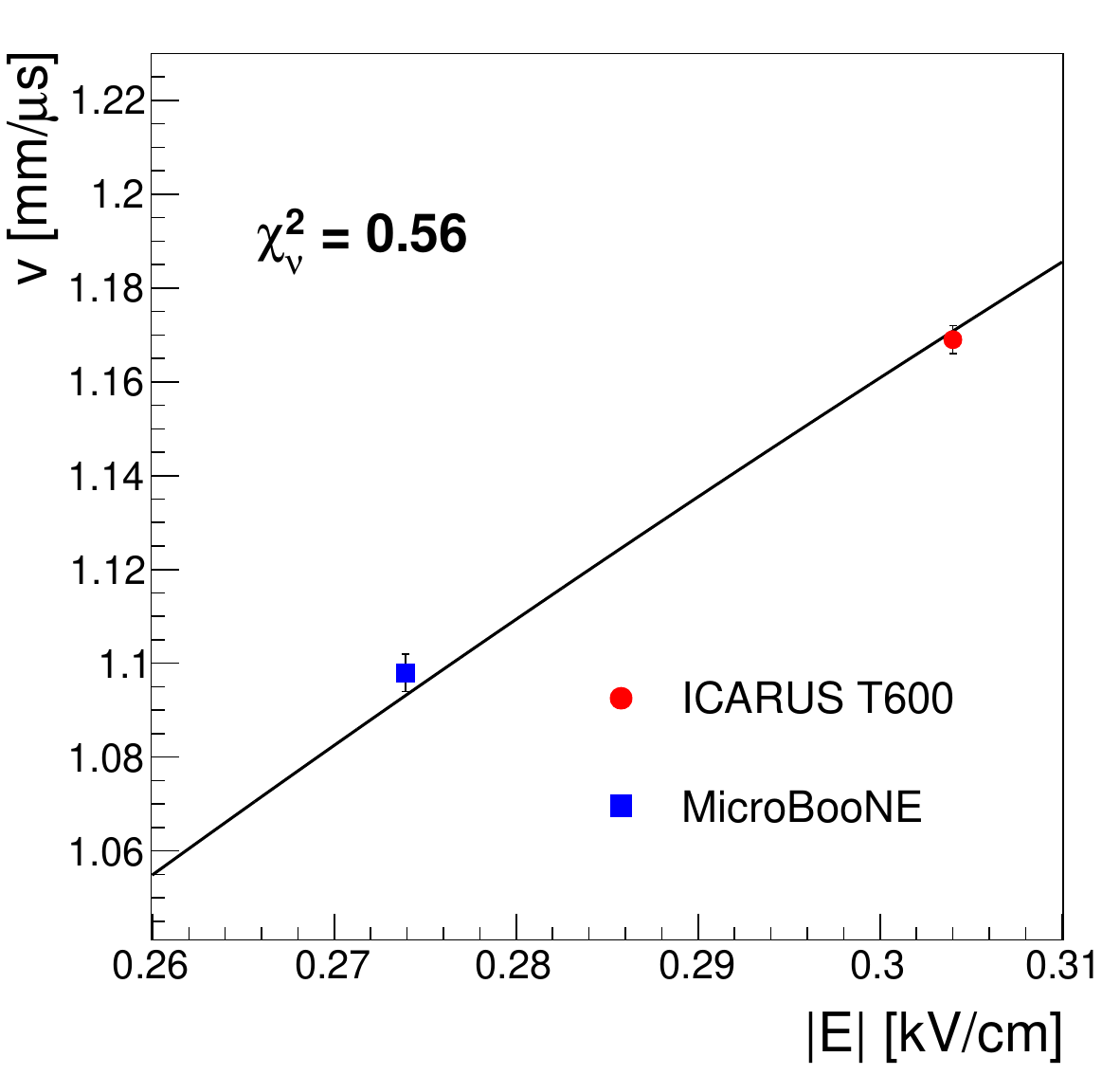}
    \caption{}
  \end{subfigure}
\Put(-300,180){\fontfamily{phv}\selectfont \textbf{MicroBooNE}}
\caption{MicroBooNE drift velocity measurement ($v_{0} = \SI[separate-uncertainty=true,multi-part-units=single]{1.098{\pm}0.004}{mm/{\micro}s}$ measured at $|E_{0}| = \SI{0.2739}{kV/cm}$) compared to drift velocity measurements made at various electric field values at the ICARUS T600 experiment~\cite{ICARUSdriftvel}; shown are both (a) the full electric field range and (b) the region of parameter space near the MicroBooNE electric field and closest ICARUS T600 measurement.  The combined dataset is fitted to a fifth-order polynomial function (solid curve), resulting in an excellent fit ($\chi^{2}_{\nu}$ = 0.56, where $\chi^{2}_{\nu}$ is the reduced chi-square statistic associated with the fit).  All measurements shown were made at an operating temperature of \SI{89}{K}.}
\label{fig:driftvel}
\end{figure}

The result of the nominal drift velocity measurement is shown in figure~\ref{fig:driftvel}, along with previous measurements made in the ICARUS T600 detector at a variety of different electric fields~\cite{ICARUSdriftvel}.  These measurements are directly comparable as the operating temperature was \SI{89}{K} for both detectors.  Very good agreement is observed between the MicroBooNE measurement and the ICARUS T600 measurements, the latter minimally impacted by space charge effects due to the detector being underground.  For this reason, the results shown in this paper make use of a drift velocity model that is formed by fitting a fifth-order polynomial to the ICARUS T600 data combined with the MicroBooNE nominal drift velocity measurement.  The fit parameters associated with this model are given in table~\ref{tab:driftvel}.

\begin{table}[tbh]
  \centering
  \begin{tabu}{c|[2pt]c}
    Fit Parameter    & Value \\ \tabucline[2pt]{-}
    $p_{0}$          & 0.0          \\ \hline
    $p_{1}$          & 5.534          \\ \hline
    $p_{2}$          & -6.531          \\ \hline
    $p_{3}$          & 3.208          \\ \hline
    $p_{4}$          & 0.3897          \\ \hline
    $p_{5}$          & -0.5562          
  \end{tabu}
  \caption{Fit parameters associated with the drift velocity model used in the SCE calibration procedure; the fit function is a fifth-order polynomial, $v(E)=p_{0}+p_{1}E+p_{2}E^{2}+p_{3}E^{3}+p_{4}E^{4}+p_{5}E^{5}$.  In this parametrization, $E$ is in units of \SI{}{kV/cm} and $v$ is in units of $\SI{}{mm/{\micro}s}$.  $p_{0}$ is forced to zero in the fit.}
\label{tab:driftvel}
\end{table}

\subsection{Anode-Piercing Face Calibration} \label{sec:method_step1}

The set of anode-piercing tracks from the sample described in section~\ref{sec:datatracks} is used in the first step of the calibration, which is illustrated in figure~\ref{fig:calibmethod}.  Tracks that pierce (enter or exit) through the anode plane experience no spatial distortions from SCE at the anode due to the negligible distance the ionization electrons drift before reaching the collection wires.  As a result, no calibration is needed for the face of the TPC coinciding with the anode plane.  However, the other five TPC faces require a calibration to constrain the true point of entry or exit of the cosmic muon.

\begin{figure}[tb]
\centering
\includegraphics[width=.99\textwidth]{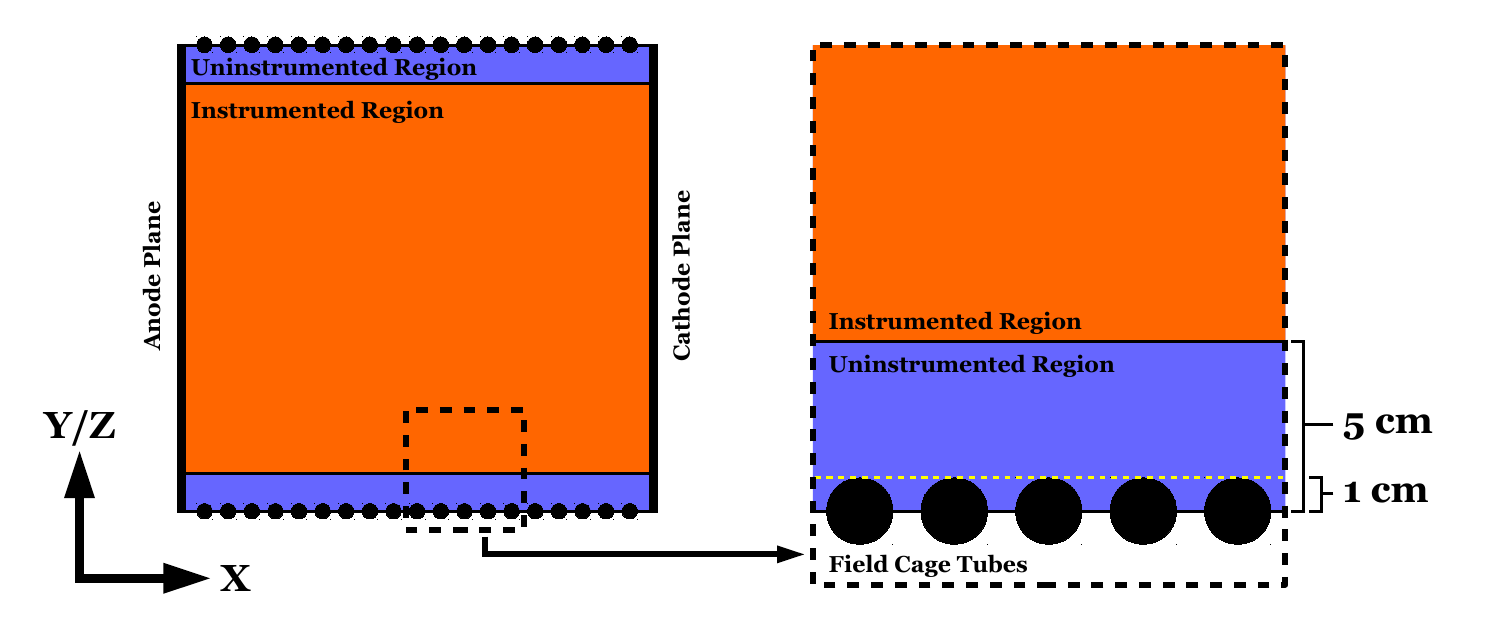}
\caption{Illustration highlighting the \SI{5}{cm} gap between the field cage and instrumented TPC volume in the MicroBooNE TPC, located at the edges of the TPC in $y$ and $z$.  Given the \SI{1}{cm} radius of the field cage tubes, this leads to a \SI{4.5}{cm} offset on average in the assumed true charge deposition location for ionization deposited near the edges of the TPC in the $y$ and $z$ directions.  The overall detector dimensions are not drawn to scale in this illustration.}
\label{fig:gapcartoon}
\end{figure}

A \SI{5}{cm} gap between the field cage and instrumented TPC volume impacts the assumed ``true'' position of ionization charge deposition prior to drift.  This gap is illustrated in figure~\ref{fig:gapcartoon}.  The radius of the field cage tubes is \SI{1}{cm}, leading to an average offset of \SI{4.5}{cm} in the true location of charge deposition for ionization near the edges of the TPC in the $y$ and $z$ directions.  This additional offset impacts the first step of the calibration methodology presented below and must be accounted for in order to obtain a more accurate calibration in data events.  This offset is not necessary to apply to simulated events, as the instrumented TPC volume is assumed to end at the field cage in MicroBooNE simulation.

The first step of the track-based calibration procedure uses the end point of the anode-piercing track that does not intersect the anode plane to determine this correction in three dimensions, calibrating every other TPC face aside from the anode and the cathode.  The calibration maps are constructed using the median offset values across all tracks passing through the same two-dimensional voxel at the TPC face, with a voxel size that is approximately 10~cm in both dimensions.  The component of the correction vector orthogonal to the TPC face is found from the transverse distance of the track end to the TPC face, adding an additional \SI{4.5}{cm} to account for the average gap between the field cage and instrumented TPC volume that is discussed above (data events only).  The other two components are identified using predictions from the simulation described in section~\ref{sec:sim}.  For instance, at the top of the TPC, the transverse distance of the track end to the top of the TPC yields ${\Delta}y_{\mathrm{reco}}$, and the other two components are determined by
\begin{equation}
{\Delta}x_{\mathrm{reco}} = {\Delta}y_{\mathrm{reco}} \times {\Delta}x_{\mathrm{sim}}(x,z)/{\Delta}y_{\mathrm{sim}}(x,z)
\end{equation}
and
\begin{equation}
{\Delta}z_{\mathrm{reco}} = {\Delta}y_{\mathrm{reco}} \times {\Delta}z_{\mathrm{sim}}(x,z)/{\Delta}y_{\mathrm{sim}}(x,z),
\end{equation}
where ${\Delta}x_{\mathrm{reco}}$, ${\Delta}y_{\mathrm{reco}}$, and ${\Delta}z_{\mathrm{reco}}$ are the reconstructed spatial offsets in the $x$, $y$, and $z$ dimensions, respectively, and e.g.~${\Delta}x_{\mathrm{sim}}(x,z)/{\Delta}y_{\mathrm{sim}}(x,z)$ is the ratio of spatial offsets in the $x$ and $y$ dimensions as predicted by the SCE simulation.  While this step of the calibration relies on the predictions of the simulation, the dependence is minimal because the offset in the direction orthogonal to the TPC face (by far the largest component of the correction vector) is derived in a data-driven fashion.

This measurement is only made for regions far enough away from the anode such that the transverse spatial distortions associated with space charge can be measured.  Due to the average gap between the field cage and instrumented TPC volume of \SI{4.5}{cm}, the inward spatial migration of track end points can not be detected until the magnitude of the spatial distortion is at least \SI{4.5}{cm}.  This leads to a lack of data-driven measurements in the regions closest to the anode wire planes ($x<\SI{100}{cm}$) at each TPC face.  In order to fill in the transverse spatial distortion map (e.g.~${\Delta}y$ at the top face of the TPC) in the regions defined by $x<\SI{100}{cm}$, the simulated transverse spatial distortion map is used instead in these regions, scaled according to the measurements in data made at larger $x$ values where data-driven measurements are possible.  The other components of the distortion map at the TPC face in question are computed as described above, in a similar fashion to parts of the map at larger $x$ values.

\subsection{Cathode Calibration} \label{sec:method_step2}

While the first part of the calibration described above targets the upstream, downstream, top, and bottom faces of the detector, the cathode remains uncalibrated.  The second step of the calibration procedure, illustrated in the middle part of figure~\ref{fig:calibmethod}, aims to specifically calibrate the ``true'' entry or exit point of the cosmic muon at the cathode plane.  The dedicated SCE simulation is leveraged to perform this part of the calibration.  The results of the first step of the calibration are used to weight the SCE simulation such that the spatial offsets at the cathode match the data-driven calibration at the top and bottom of the TPC (i.e.~${\Delta}y_{\mathrm{reco}}(y_{\mathrm{top}},z)$ and ${\Delta}y_{\mathrm{reco}}(y_{\mathrm{bot}},z)$, defined below, are derived from the first step of the calibration procedure).  A linear interpolation between the top and bottom of the TPC is used to obtain the offsets at the cathode as a function of $y$.  The cathode offsets are obtained from
\begin{equation}
F(y)=(y-y_{\mathrm{bot}})/(y_{\mathrm{top}}-y_{\mathrm{bot}}),
\end{equation}
\begin{equation}
\begin{split}
  S(y,z) = \, & F(y){\Delta}y_{\mathrm{reco}}(y_{\mathrm{top}},z)/{\Delta}y_{\mathrm{sim}}(y_{\mathrm{top}},z) \\
           & + (1-F(y)){\Delta}y_{\mathrm{reco}}(y_{\mathrm{bot}},z)/{\Delta}y_{\mathrm{sim}}(y_{\mathrm{bot}},z),
\end{split}
\end{equation}
\begin{equation}
{\Delta}x_{\mathrm{reco}}(y,z) = S(y,z) \times {\Delta}x_{\mathrm{sim}}(y,z),
\end{equation}
\begin{equation}
{\Delta}y_{\mathrm{reco}}(y,z) = S(y,z) \times {\Delta}y_{\mathrm{sim}}(y,z),
\end{equation}
and
\begin{equation}
{\Delta}z_{\mathrm{reco}}(y,z) = S(y,z) \times {\Delta}z_{\mathrm{sim}}(y,z),
\end{equation}
where $y_{\mathrm{top}}$ ($y_{\mathrm{bot}}$) is the vertical position of the top (bottom) of the TPC, ${\Delta}x_{\mathrm{reco}}$, ${\Delta}y_{\mathrm{reco}}$, and ${\Delta}z_{\mathrm{reco}}$ are the reconstructed spatial offsets at the cathode plane in the $x$, $y$, and $z$ dimensions, respectively, ${\Delta}x_{\mathrm{sim}}$, ${\Delta}y_{\mathrm{sim}}$, and ${\Delta}z_{\mathrm{sim}}$ are the simulated spatial offsets at the cathode plane in the $x$, $y$, and $z$ dimensions, respectively, $F(y)$ is the linear interpolation factor described above, and $S(y,z)$ is the data-driven scale factor used to adjust the simulated spatial offsets at the cathode for use in the calibration.  As with the previous step of the calibration, this calibration step uses two-dimensional voxels with a size that is approximately 10~cm in both dimensions.

\subsection{TPC Bulk Calibration} \label{sec:method_step3}

The final step of the calibration procedure, visualized on the right in figure~\ref{fig:calibmethod}, seeks to find the spatial distortions in the bulk of the TPC volume.  After the first two steps of the data-driven calibration, described above, the ``true'' end points of the through-going cosmic muon tracks are known at all six faces of the TPC.  By drawing a straight line between these two points, the true trajectory of the cosmic muon can be approximated.  By comparing to the track reconstructed using the TPC readout, the spatial distortions associated with SCE in the middle of the TPC can be inferred.  However, to obtain point-to-point corrections in three dimensions, single tracks are not sufficient as there are ambiguities in associating points on the reconstructed track with points on the true trajectory.  Instead, pairs of cosmic muon tracks are utilized, comparing the near-crossing point of the ``truth tracks'' (estimated true trajectories of the cosmic muons) to the near-crossing point of the reconstructed tracks: a vector starting at the latter point and ending at the former is the three-dimensional correction vector at the latter point in reconstructed coordinate space.  A requirement that both truth tracks and both reconstructed tracks come within 1~cm of each other at the near-crossing point is enforced to improve the precision of the calibration.  It is not required for the tracks to come from the same TPC readout event, given that the time-evolution of SCE at MicroBooNE is small (see section~\ref{sec:timedep}).  Using voxels with a size that is approximately \SI{10}{cm} in all dimensions, on a voxel-by-voxel basis the spatial offsets are determined (independently for ${\Delta}x$, ${\Delta}y$, and ${\Delta}z$) by taking the median offset associated with the distribution of offsets obtained from all reconstructed track pairs that have a near-crossing point within the voxel in question.

Sufficient statistics for the bulk calibration can be reached with 200k $t_{0}$-tagged cosmic muon tracks. Additionally, multiple Coulomb scattering leads to deviations between the estimated straight-line trajectory of each cosmic muon and its true trajectory through the TPC.  This effect is minimized by taking the median spatial offset in each voxel.

Gaps in the spatial offset map are expected after the full calibration chain is carried out, as there are some parts of the TPC that lack significant coverage by the set of $t_{0}$-tagged tracks used in the calibration; this is especially true in the middle of the TPC.  By combining the results of the cosmic muon calibration with that of the MicroBooNE UV laser system, complete coverage throughout the TPC is attainable.  Also, given that the spatial offsets are continuous and gradually vary as a function of position within the TPC, it is possible to interpolate across the gaps in the map: in the results shown in this work, a cubic spline followed by a median filter is applied (using a sliding $3\times3\times3$ voxel window for the latter) to fill in these gaps.

\subsection{Calculation of Electric Field Distortions} \label{sec:method_efield}

Once the spatial distortion map is determined throughout the TPC volume, the electric field distortions associated with space charge effects can be computed.  This is done by using the drift velocity model, $v(E)$, discussed in section~\ref{sec:method_driftvel}.  First, the local drift velocity must be calculated from the spatial distortion map~\cite{laser_calib}.  Once the local drift velocity $v(x,y,z)$ is determined throughout the TPC volume, the drift velocity model is used to find the corresponding electric field magnitude throughout the TPC by solving $v(E) = v(x,y,z)$ numerically for each voxel in the TPC.  Results shown later are presented as percentage change from the nominal electric field magnitude.

\section{Results of Calibration} \label{sec:results}

In this section, we present the results of the SCE calibration discussed in section~\ref{sec:method}.  The results presented in section~\ref{sec:results_faces} can be considered intermediate results that primarily serve to demonstrate different features of the SCE in data in comparison with predictions from simulation, particularly at the TPC faces.  The results presented in section~\ref{sec:results_bulk} are the ``final'' results that will ultimately serve as a set of corrections for data events and physics measurements in MicroBooNE, including electric field corrections derived from the measurements of spatial offsets.

\subsection{Measurements at TPC Faces} \label{sec:results_faces}

\begin{figure}[p]
\centering
  \begin{subfigure}{0.99\textwidth}
    \centering
    \includegraphics[width=.99\textwidth]{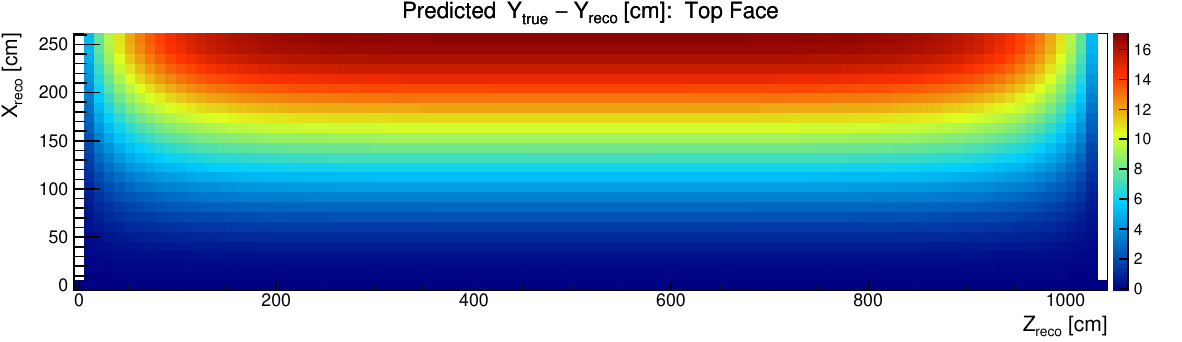}
    \caption{}
  \end{subfigure}
  \\
  \vspace{3mm}
  \begin{subfigure}{0.99\textwidth}
    \centering
    \includegraphics[width=.99\textwidth]{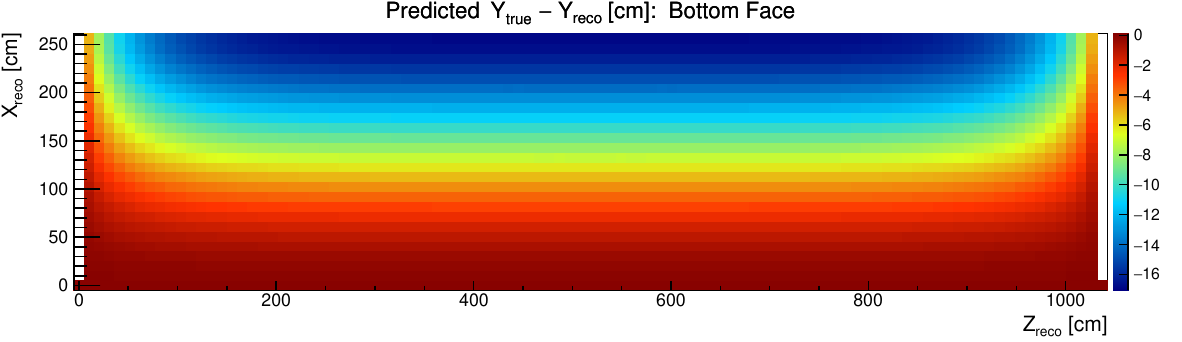}
    \caption{}
  \end{subfigure}
  \\
  \vspace{3mm}
  \begin{subfigure}{0.46\textwidth}
    \centering
    \includegraphics[width=.99\textwidth]{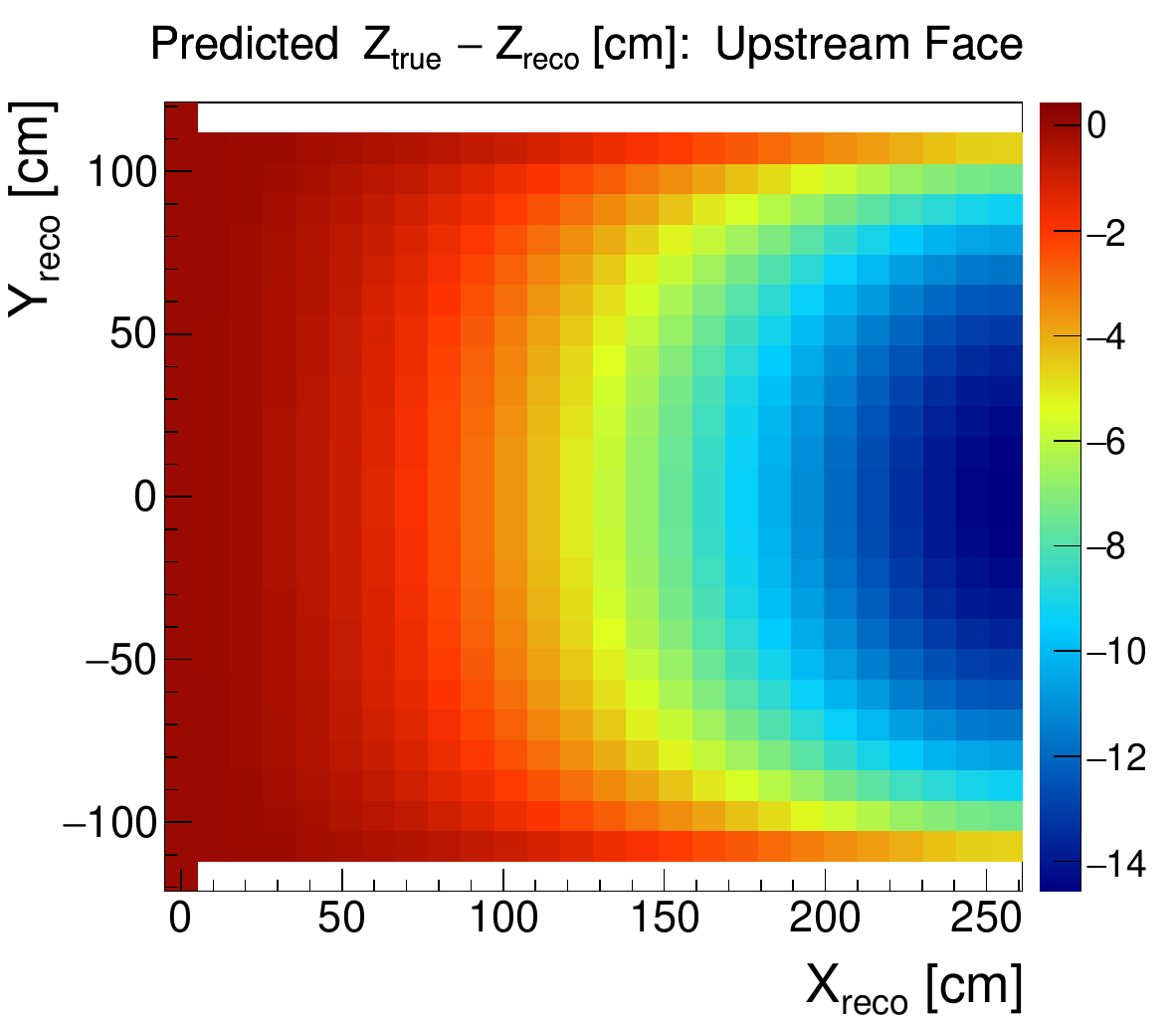}
    \caption{}
  \end{subfigure}
  \begin{subfigure}{0.46\textwidth}
    \centering
    \includegraphics[width=.99\textwidth]{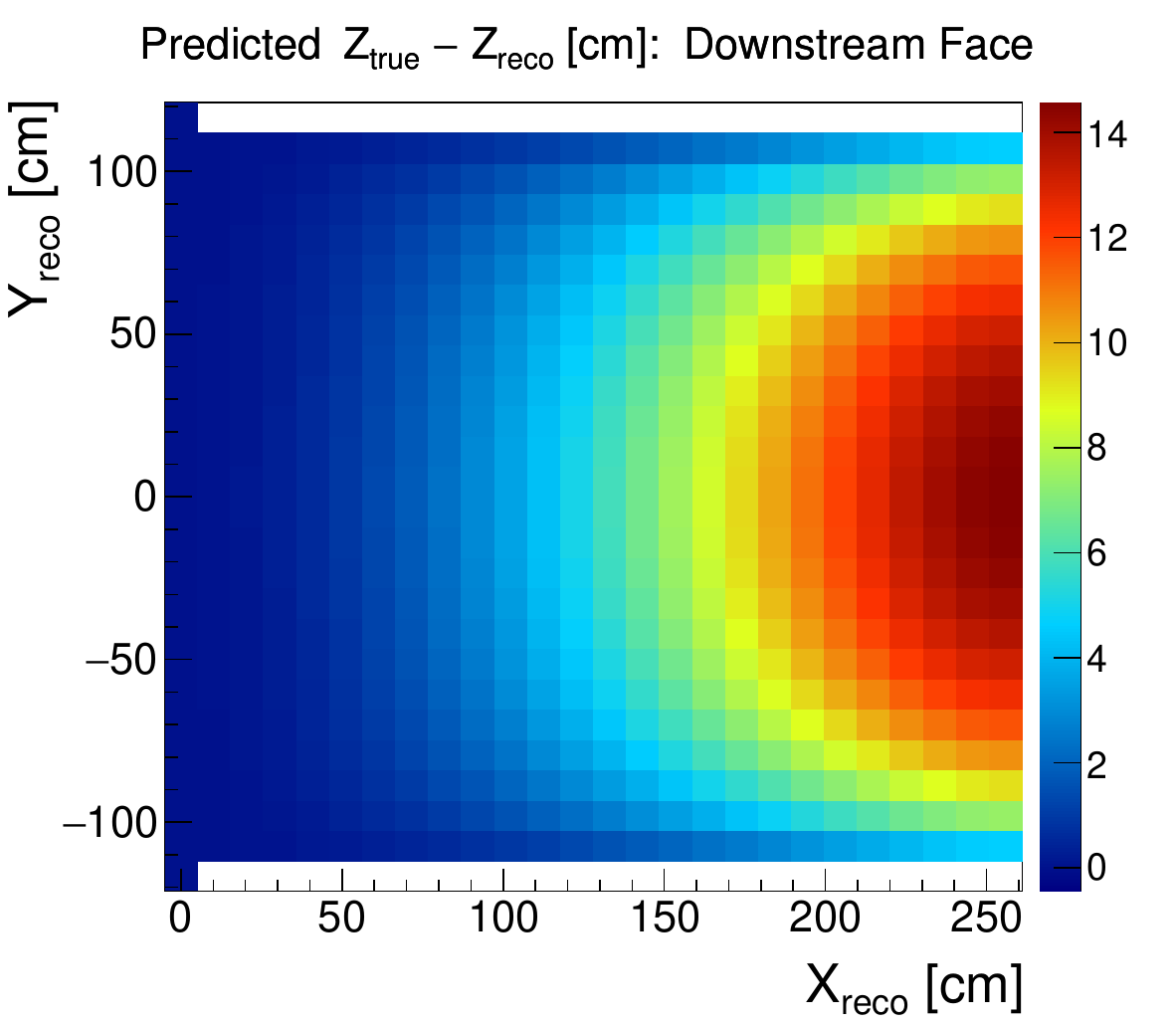}
    \caption{}
  \end{subfigure}
\Put(-360,400){\fontfamily{phv}\selectfont \textbf{MicroBooNE Simulation}}
\caption{Spatial offsets predicted from the SCE simulation at the (a) top of the TPC, (b) bottom of the TPC, (c) upstream TPC face, and (d) downstream TPC face.  Shown are the spatial offsets in the direction orthogonal to each TPC face.  Gaps at the edges of the maps correspond to places in the TPC where charge should not be reconstructed due to the impact of SCE in the detector.} \label{fig:Face_Results_TrueMC}
\end{figure}

Before presenting the raw spatial offset measurements at each of the TPC faces (aside from the cathode), it is instructive to study the predictions from the simulation.  These predictions are shown in figure~\ref{fig:Face_Results_TrueMC}.  Only the spatial offsets perpendicular to each face are shown in this figure, which are most relevant as they are the largest components of the spatial distortions near each TPC face.

\begin{figure}[p]
\centering
  \begin{subfigure}{0.99\textwidth}
    \centering
    \includegraphics[width=.99\textwidth]{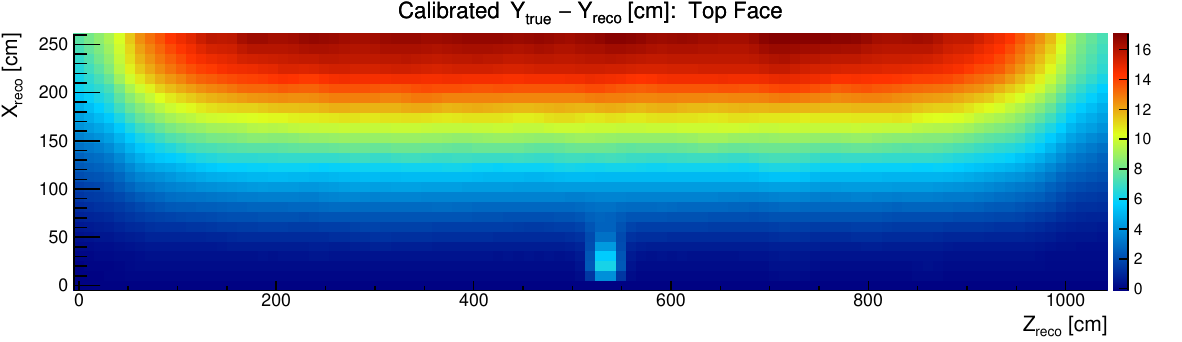}
    \caption{}
  \end{subfigure}
  \\
  \vspace{3mm}
  \begin{subfigure}{0.99\textwidth}
    \centering
    \includegraphics[width=.99\textwidth]{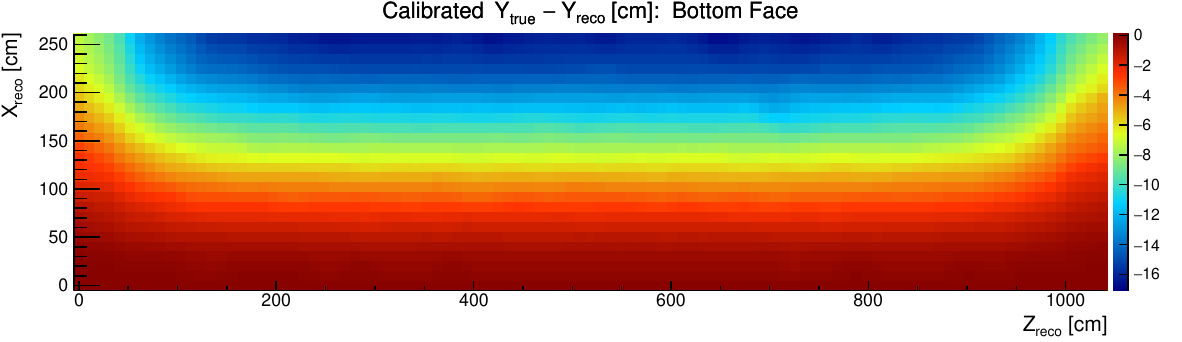}
    \caption{}
  \end{subfigure}
  \\
  \vspace{3mm}
  \begin{subfigure}{0.46\textwidth}
    \centering
    \includegraphics[width=.99\textwidth]{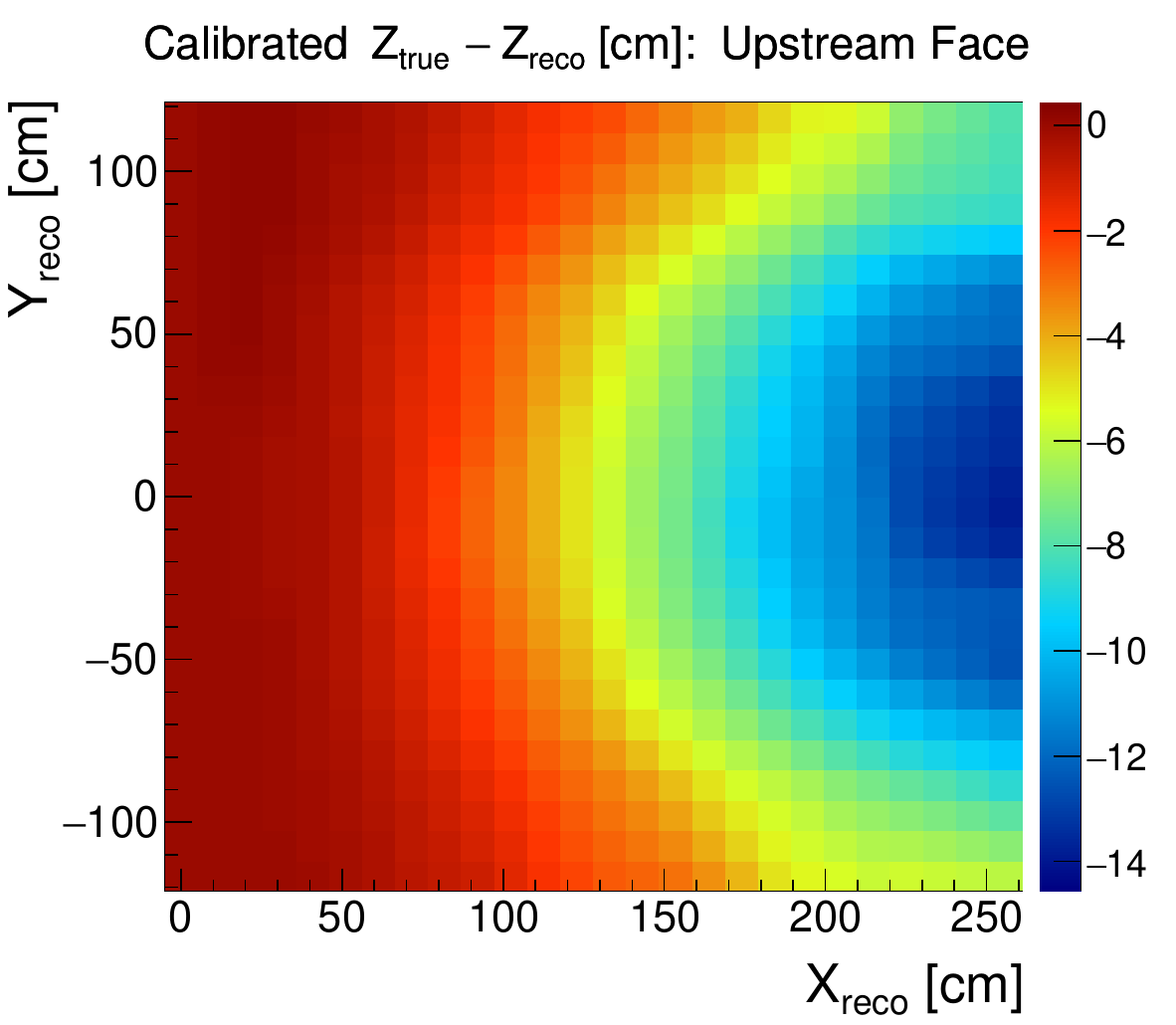}
    \caption{}
  \end{subfigure}
  \begin{subfigure}{0.46\textwidth}
    \centering
    \includegraphics[width=.99\textwidth]{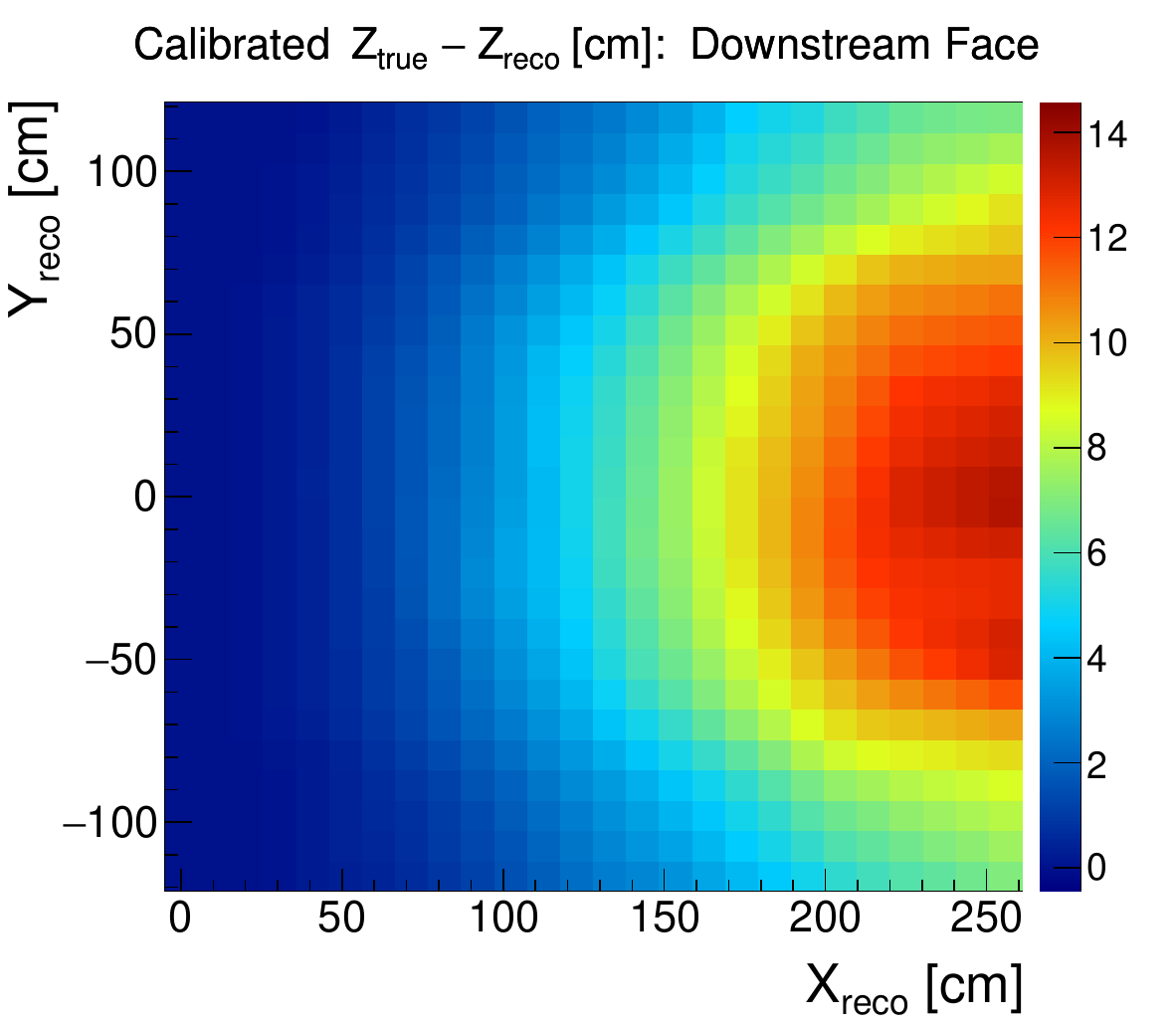}
    \caption{}
  \end{subfigure}
\Put(-360,400){\fontfamily{phv}\selectfont \textbf{MicroBooNE Simulation}}
\caption{Reconstructed spatial offsets in Monte Carlo simulation events at the (a) top of the TPC, (b) bottom of the TPC, (c) upstream TPC face, and (d) downstream TPC face.  Shown are the spatial offsets in the direction orthogonal to each TPC face.  A cubic spline is used to fill in the gaps in the spatial offset maps at the edges of the TPC.} \label{fig:Face_Results_RecoMC}
\end{figure}

\begin{figure}[p]
\centering
  \begin{subfigure}{0.99\textwidth}
    \centering
    \includegraphics[width=.99\textwidth]{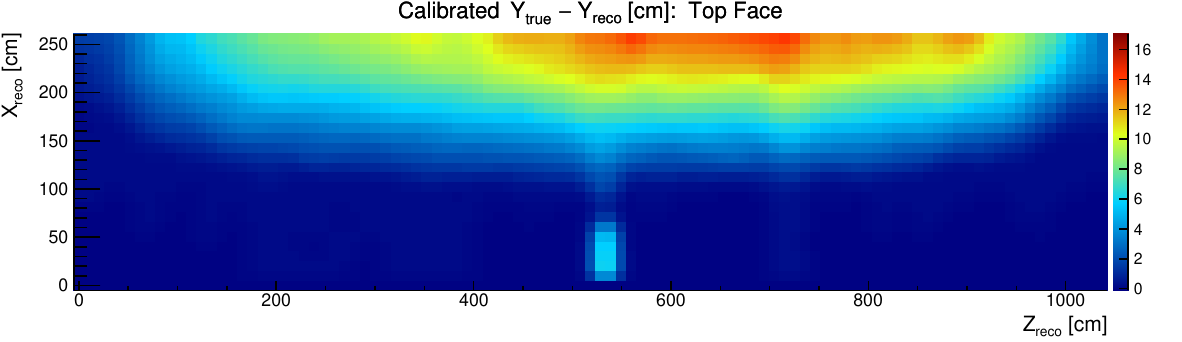}
    \caption{}
  \end{subfigure}
  \\
  \vspace{3mm}
  \begin{subfigure}{0.99\textwidth}
    \centering
    \includegraphics[width=.99\textwidth]{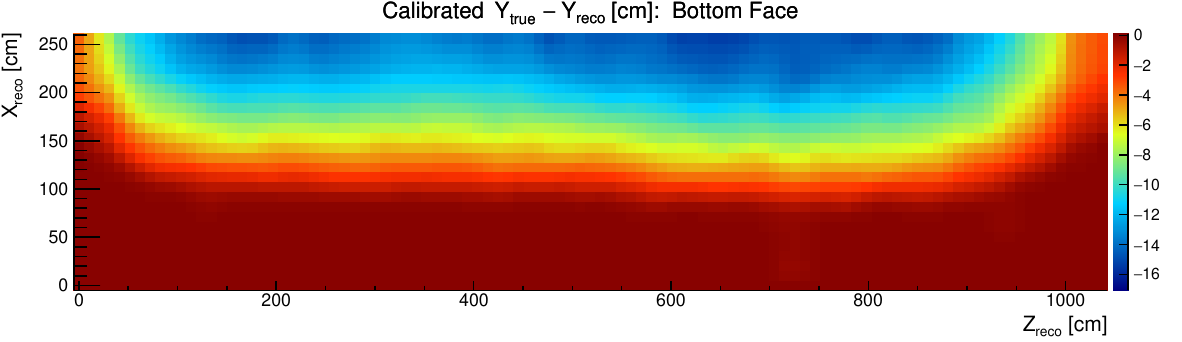}
    \caption{}
  \end{subfigure}
  \\
  \vspace{3mm}
  \begin{subfigure}{0.46\textwidth}
    \centering
    \includegraphics[width=.99\textwidth]{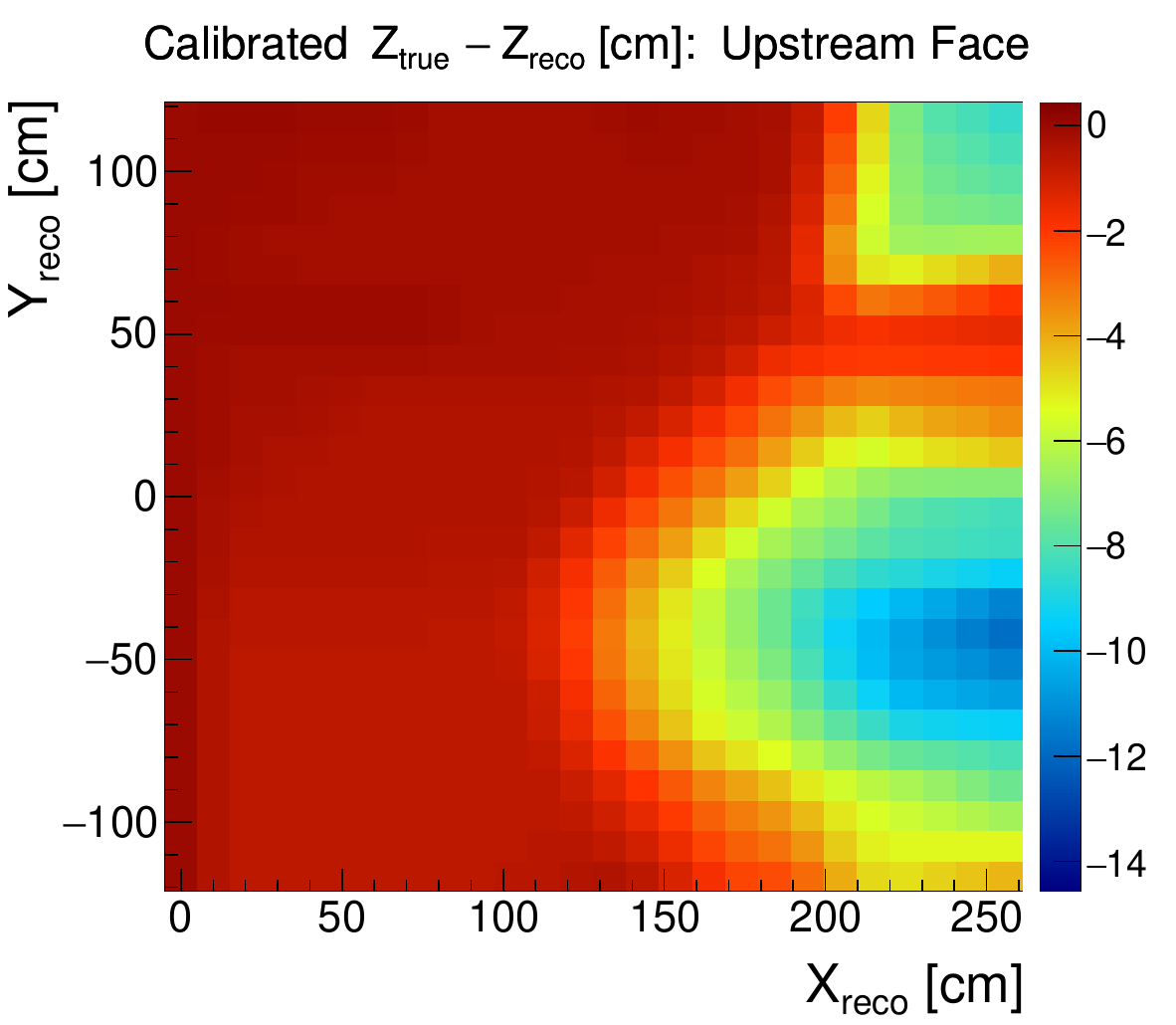}
    \caption{}
  \end{subfigure}
  \begin{subfigure}{0.46\textwidth}
    \centering
    \includegraphics[width=.99\textwidth]{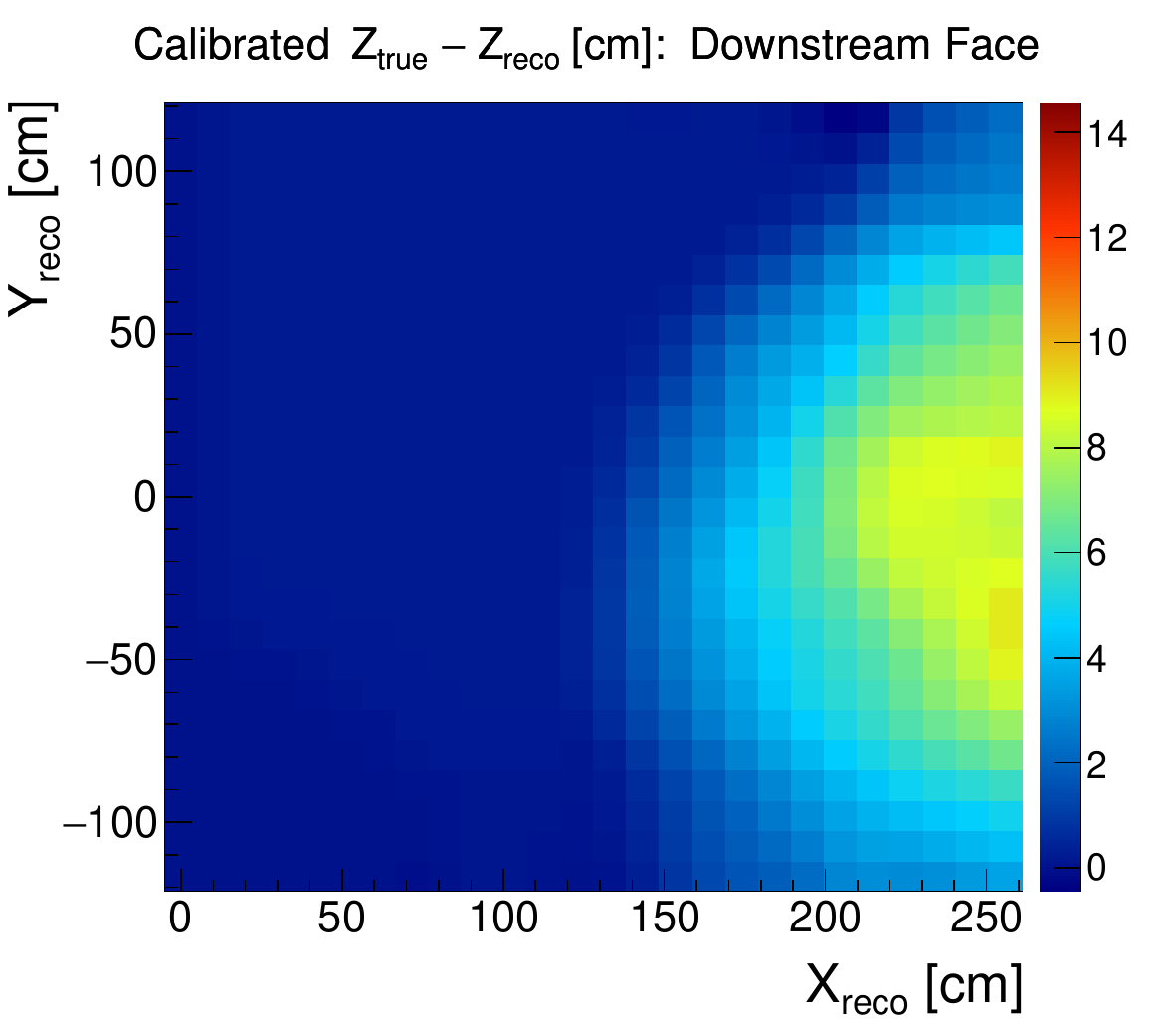}
    \caption{}
  \end{subfigure}
\Put(-360,400){\fontfamily{phv}\selectfont \textbf{MicroBooNE}}
\caption{Reconstructed spatial offsets in MicroBooNE data events at the (a) top of the TPC, (b) bottom of the TPC, (c) upstream TPC face, and (d) downstream TPC face.  Shown are the spatial offsets in the direction orthogonal to each TPC face.  A cubic spline is used to fill in the gaps in the spatial offset maps at the edges of the TPC.} \label{fig:Face_Results_Data}
\end{figure}

Shown in figure~\ref{fig:Face_Results_RecoMC} and figure~\ref{fig:Face_Results_Data} are the measurements of the offsets from the TPC faces discussed in section~\ref{sec:method_step1} for Monte Carlo simulation events and data events, respectively; the offset of \SI{4.5}{cm} associated with the gap between the field cage and instrumented TPC volume is not included in the raw data measurement shown in figure~\ref{fig:Face_Results_Data}, as it is applied at a later stage of the calibration.  Comparing figure~\ref{fig:Face_Results_RecoMC} with figure~\ref{fig:Face_Results_TrueMC}, the spatial offsets predicted by the simulation are largely reproduced after applying the data-driven calibration to Monte Carlo simulation events.  A region of relatively large ${\Delta}y$ is observed at the top of the detector near $z = 5$~m, which the simulation is able to reproduce.  This feature is due to overlapping dead channels from two wire planes near the top of the TPC, leading to charge appearing to be displaced from the top of the detector.  Figure~\ref{fig:Face_Results_Data} illustrates the different impact of SCE within MicroBooNE data in comparison to the simulation results shown in figure~\ref{fig:Face_Results_RecoMC}.  A reduction in the magnitude of the spatial offsets is observed at the top of the detector ($y$), as well as the upstream and downstream ends of the TPC; the difference is reduced once the \SI{4.5}{cm} gap is taken into account, as this leads to an increase in the magnitude of the spatial offsets in data.  Additionally, at the top of the detector, the magnitude of the spatial distortions is less severe at the upstream end of the TPC in comparison to the downstream end.  Finally, a localized reduction of the magnitude of spatial distortions at the upstream part of the TPC is observed in data near the cathode at $y = 0.5$~m, a feature that is not present in the simulation results.  While the origin of this feature has not been identified, it is likely a result of the liquid argon flow pattern in the detector, which could remove space charge from the TPC active volume.

\subsection{Calibration Results in TPC Bulk} \label{sec:results_bulk}

The last step of the SCE calibration, described in section~\ref{sec:method_step3}, leads to a determination of spatial offsets throughout the vast majority of the TPC volume.  It is difficult to visualize the measured distortion map in three dimensions throughout the entire TPC.  We focus on two representative cross-sectional slices of the detector:  one slice in $z$ near the center of the detector ($z = 5.18$~m) and one slice in $z$ near the downstream end of the detector ($z = 9.95$~m).

\begin{figure}[p]
\centering
  \begin{subfigure}{0.41\textwidth}
    \centering
    \includegraphics[width=.99\textwidth]{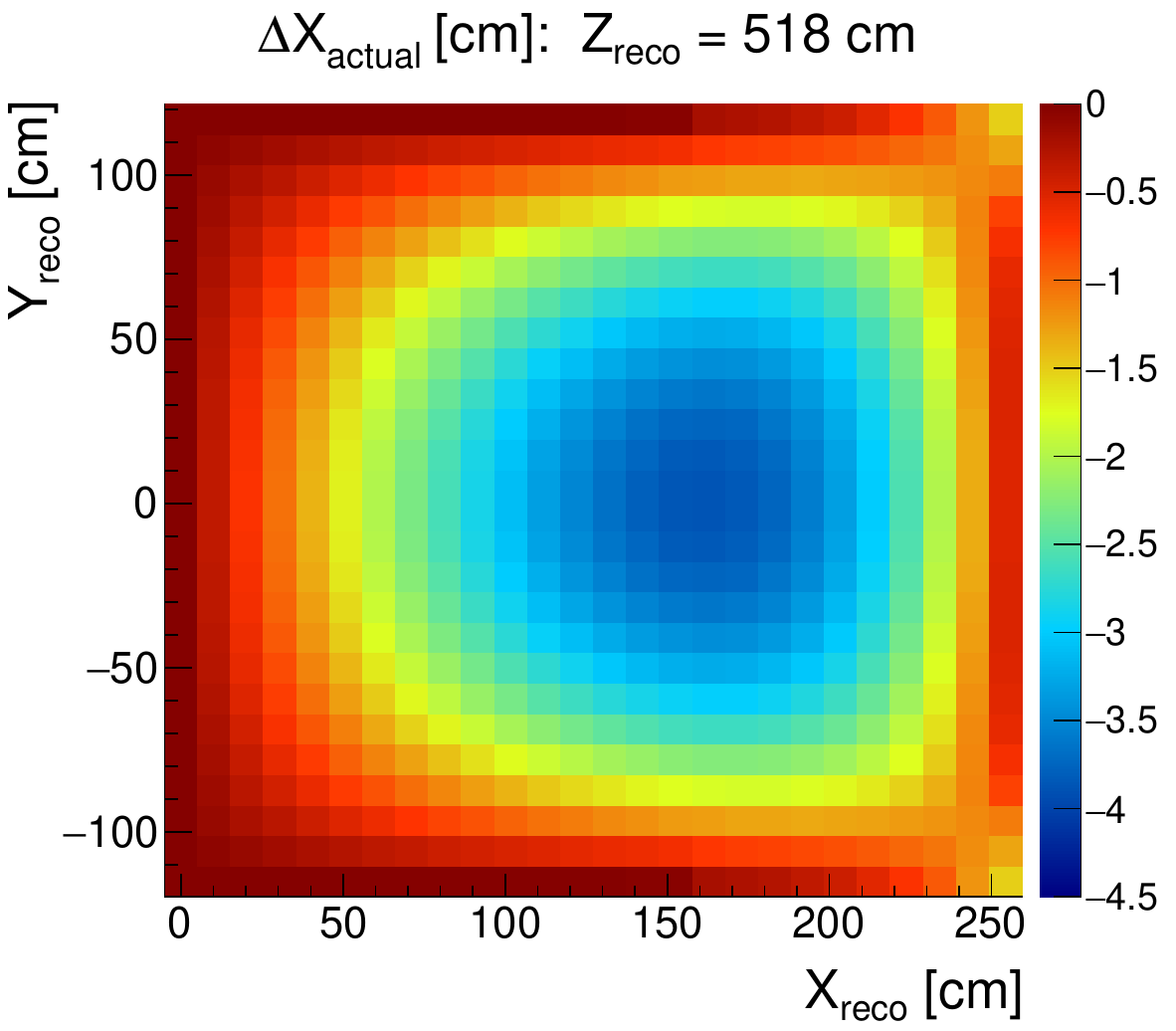}
    \caption{}
  \end{subfigure}
  \begin{subfigure}{0.41\textwidth}
    \centering
    \includegraphics[width=.99\textwidth]{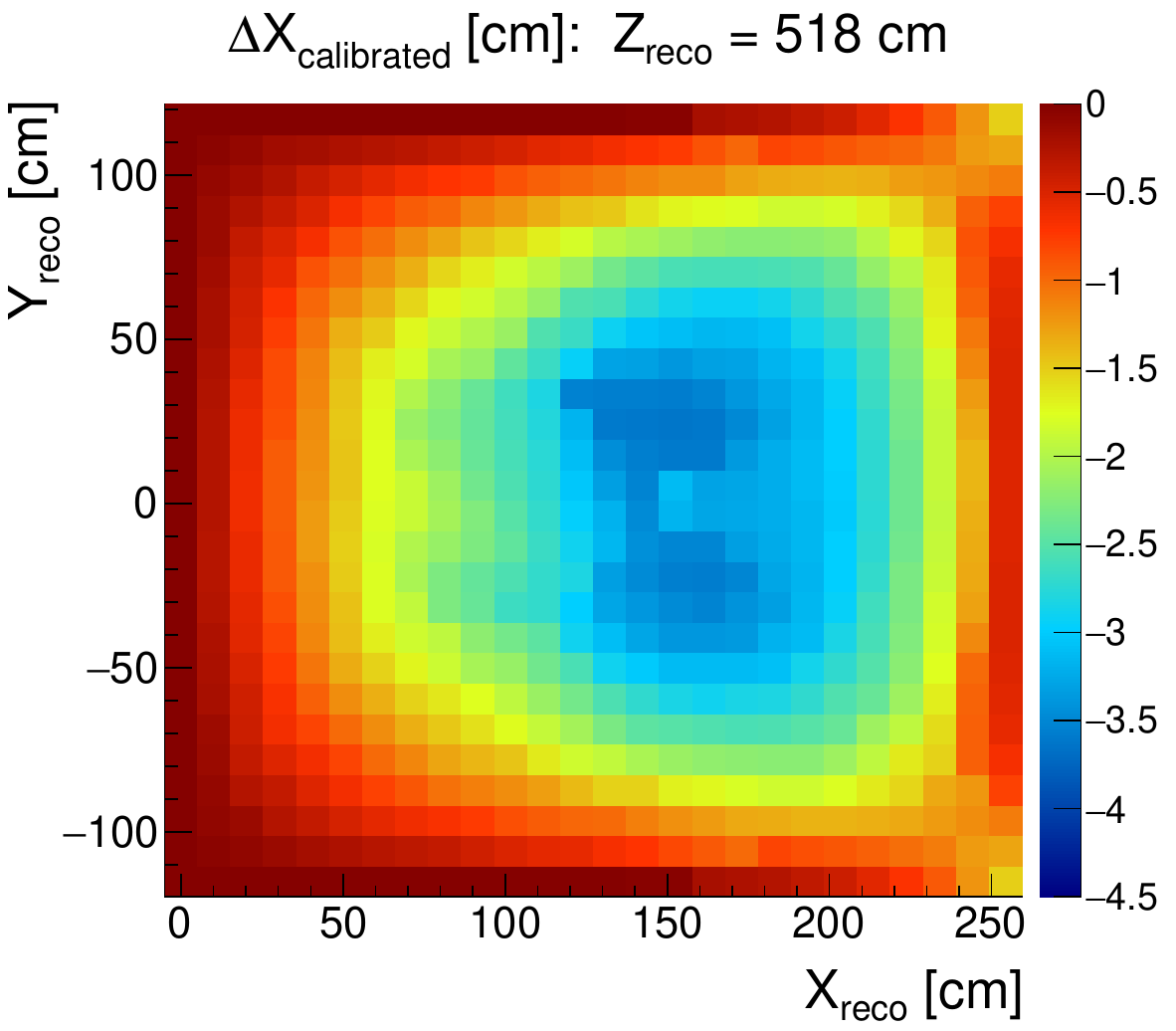}
    \caption{}
  \end{subfigure}
  \\
  \vspace{3mm}
  \begin{subfigure}{0.41\textwidth}
    \centering
    \includegraphics[width=.99\textwidth]{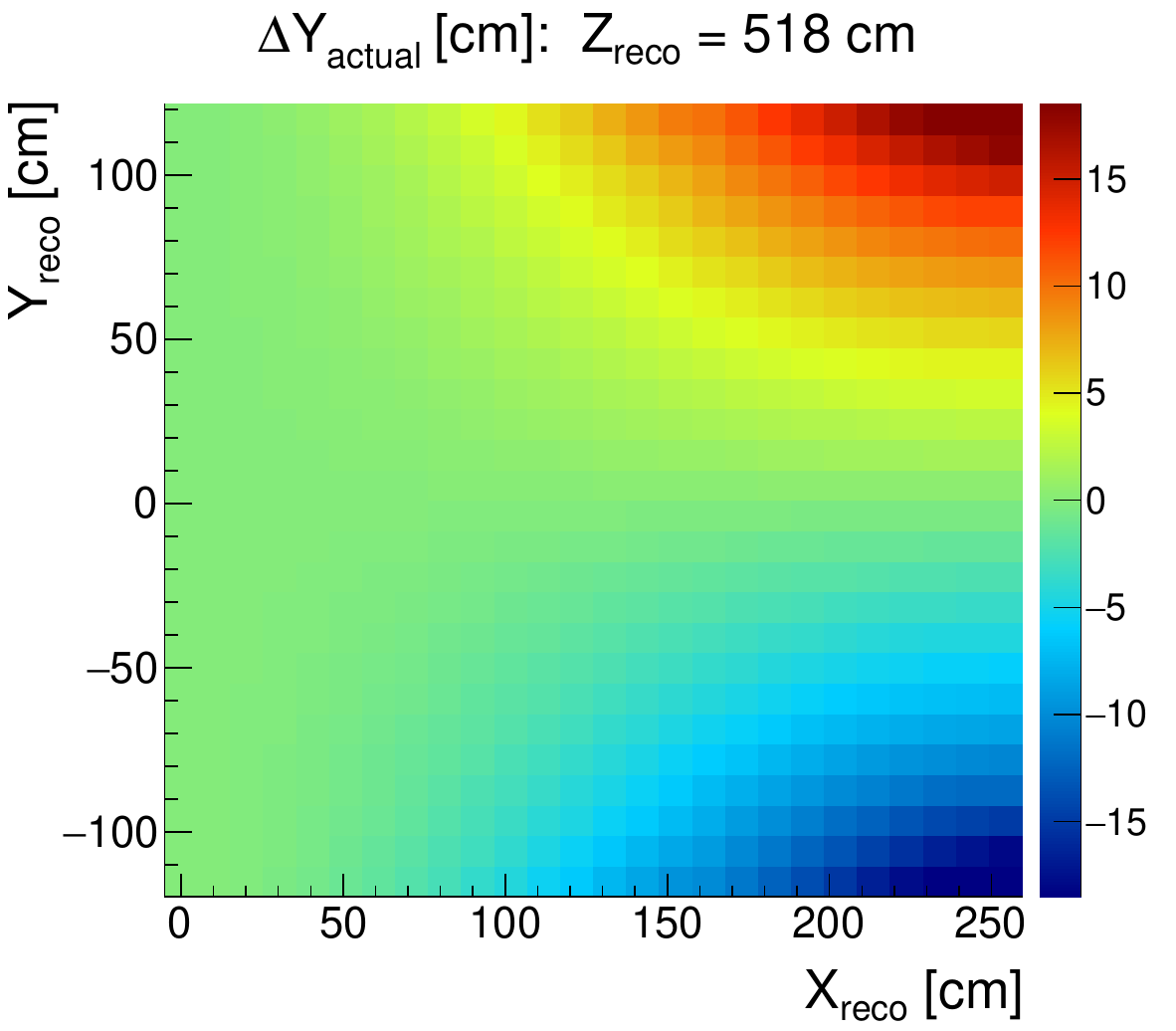}
    \caption{}
  \end{subfigure}
  \begin{subfigure}{0.41\textwidth}
    \centering
    \includegraphics[width=.99\textwidth]{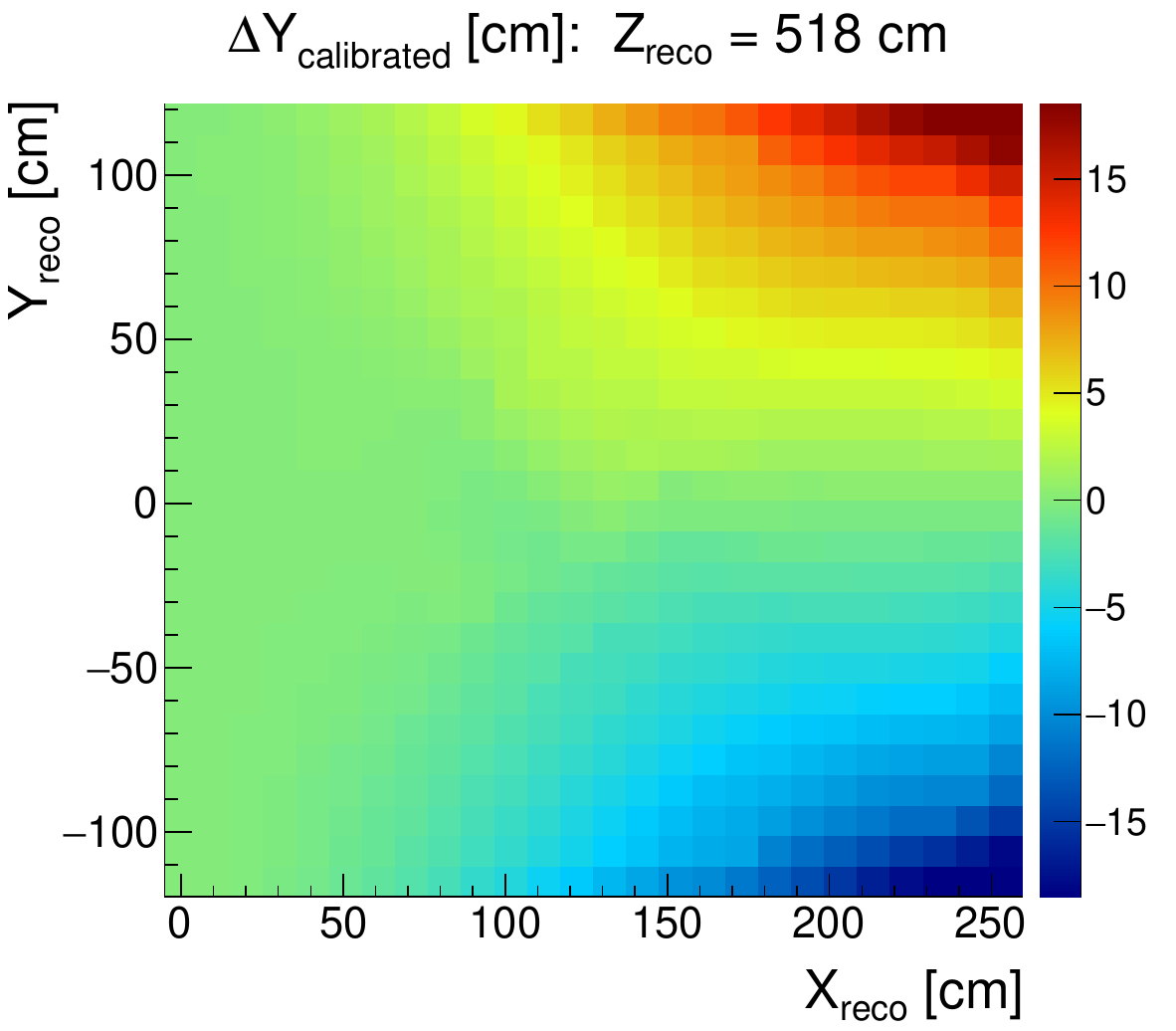}
    \caption{}
  \end{subfigure}
  \\
  \vspace{3mm}
  \begin{subfigure}{0.41\textwidth}
    \centering
    \includegraphics[width=.99\textwidth]{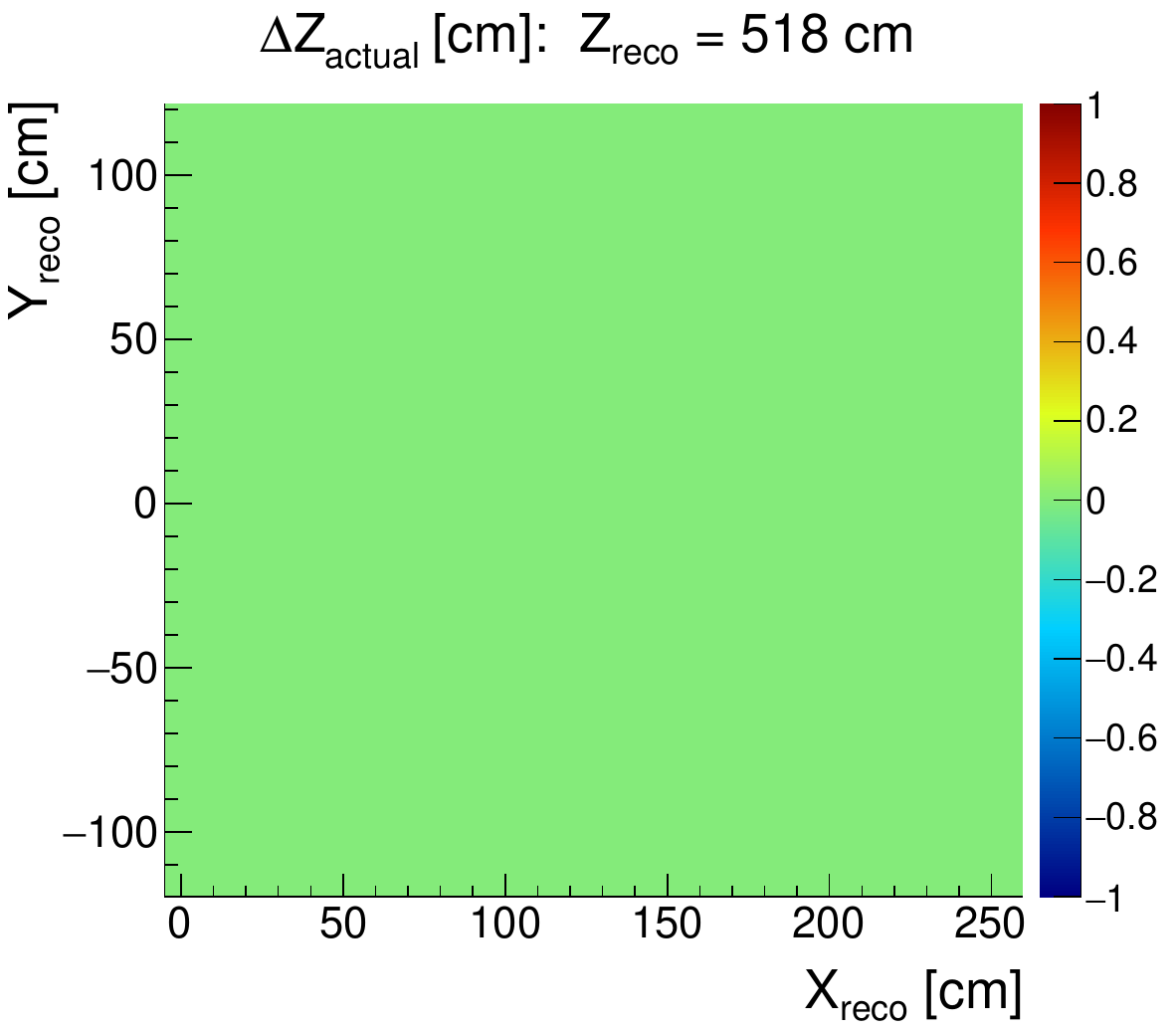}
    \caption{}
  \end{subfigure}
  \begin{subfigure}{0.41\textwidth}
    \centering
    \includegraphics[width=.99\textwidth]{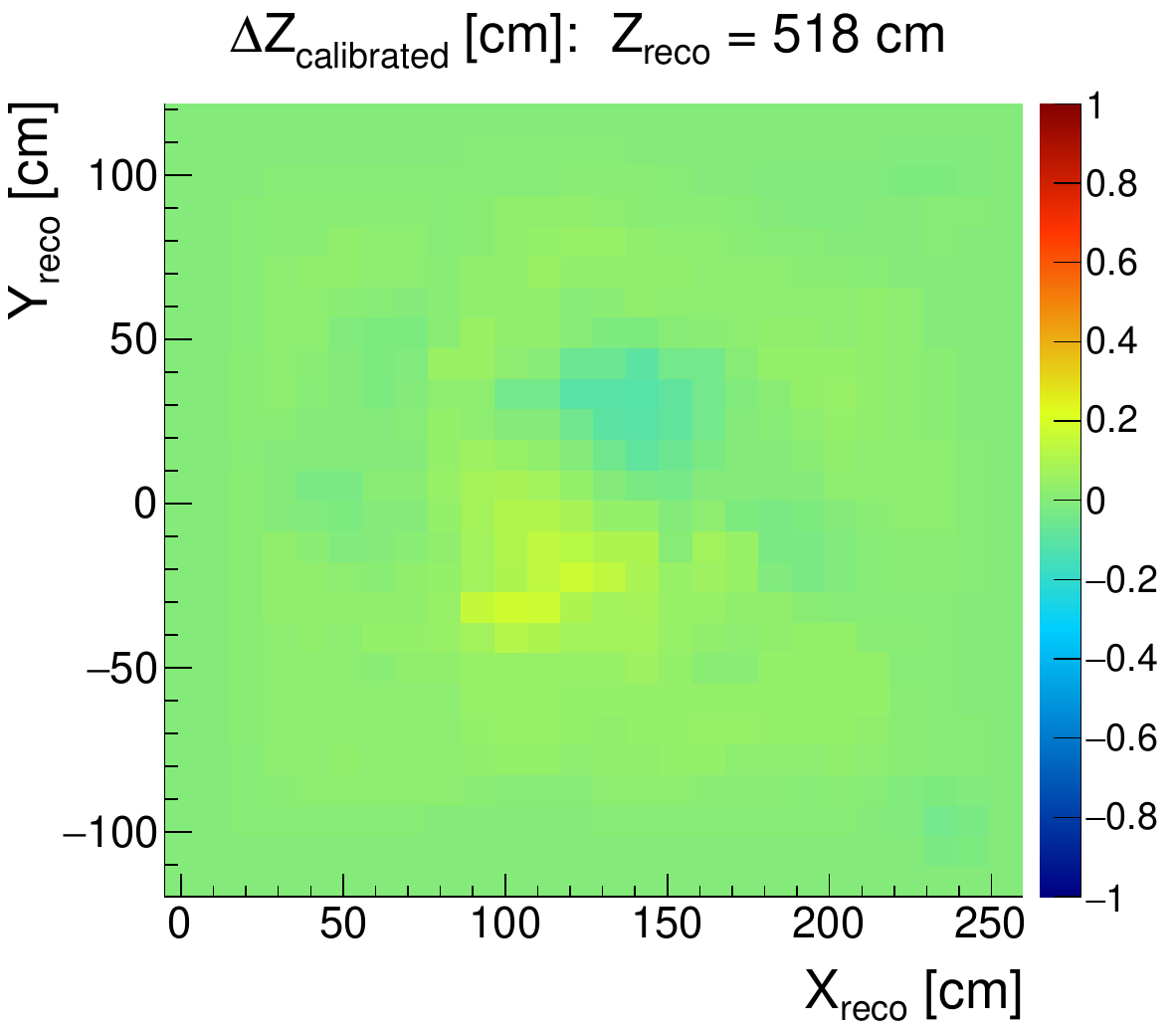}
    \caption{}
  \end{subfigure}
  \\
\Put(36,570){\fontfamily{phv}\selectfont \textbf{MicroBooNE}}
\Put(36,545){\fontfamily{phv}\selectfont \textbf{Simulation}}
\caption{Comparison of (a, c, e) spatial offsets predicted from the SCE simulation to (b, d, f) the results of the TPC bulk calibration on Monte Carlo simulation events for a central slice in $z$.  Results are shown for spatial offsets in (a, b) $x$, (c, d) $y$, and (e, f) $z$.  The distortions in reconstructed ionization electron cluster position are shown in units of cm and are plotted as a function of the reconstructed position in the TPC.} \label{fig:MC_Results_CentralZ}
\end{figure}

\begin{figure}[p]
\centering
  \begin{subfigure}{0.41\textwidth}
    \centering
    \includegraphics[width=.99\textwidth]{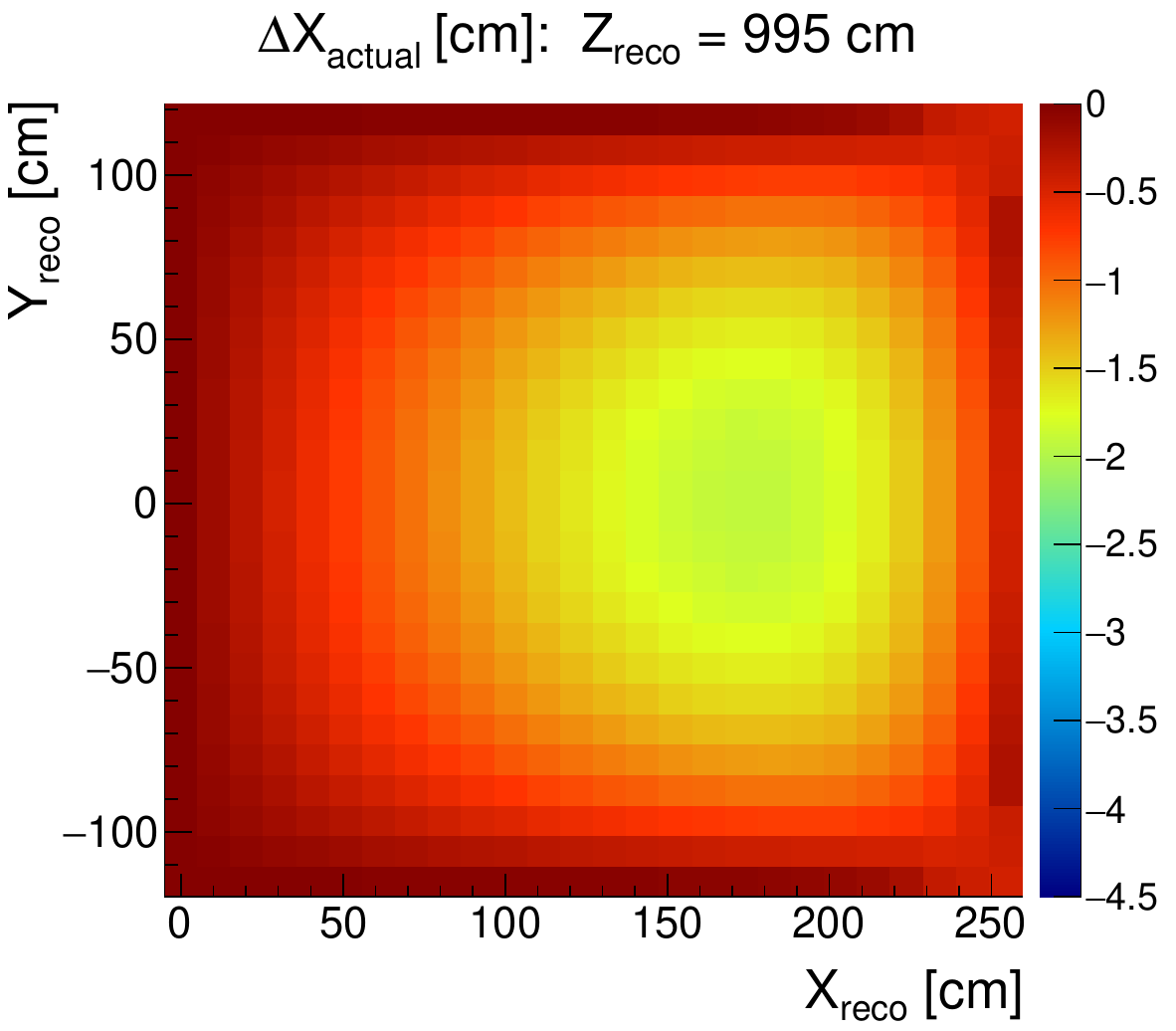}
    \caption{}
  \end{subfigure}
  \begin{subfigure}{0.41\textwidth}
    \centering
    \includegraphics[width=.99\textwidth]{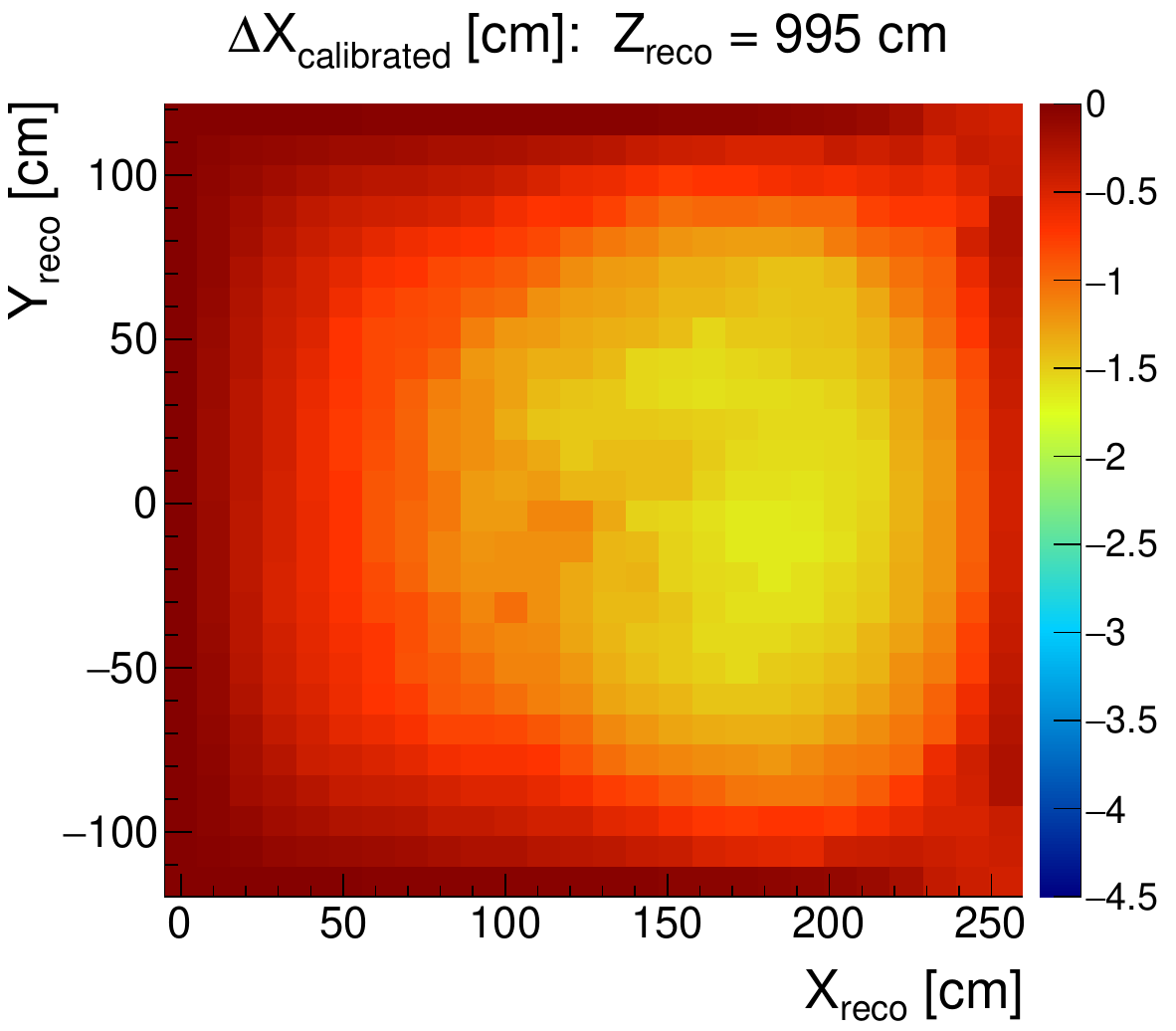}
    \caption{}
  \end{subfigure}
  \\
  \vspace{3mm}
  \begin{subfigure}{0.41\textwidth}
    \centering
    \includegraphics[width=.99\textwidth]{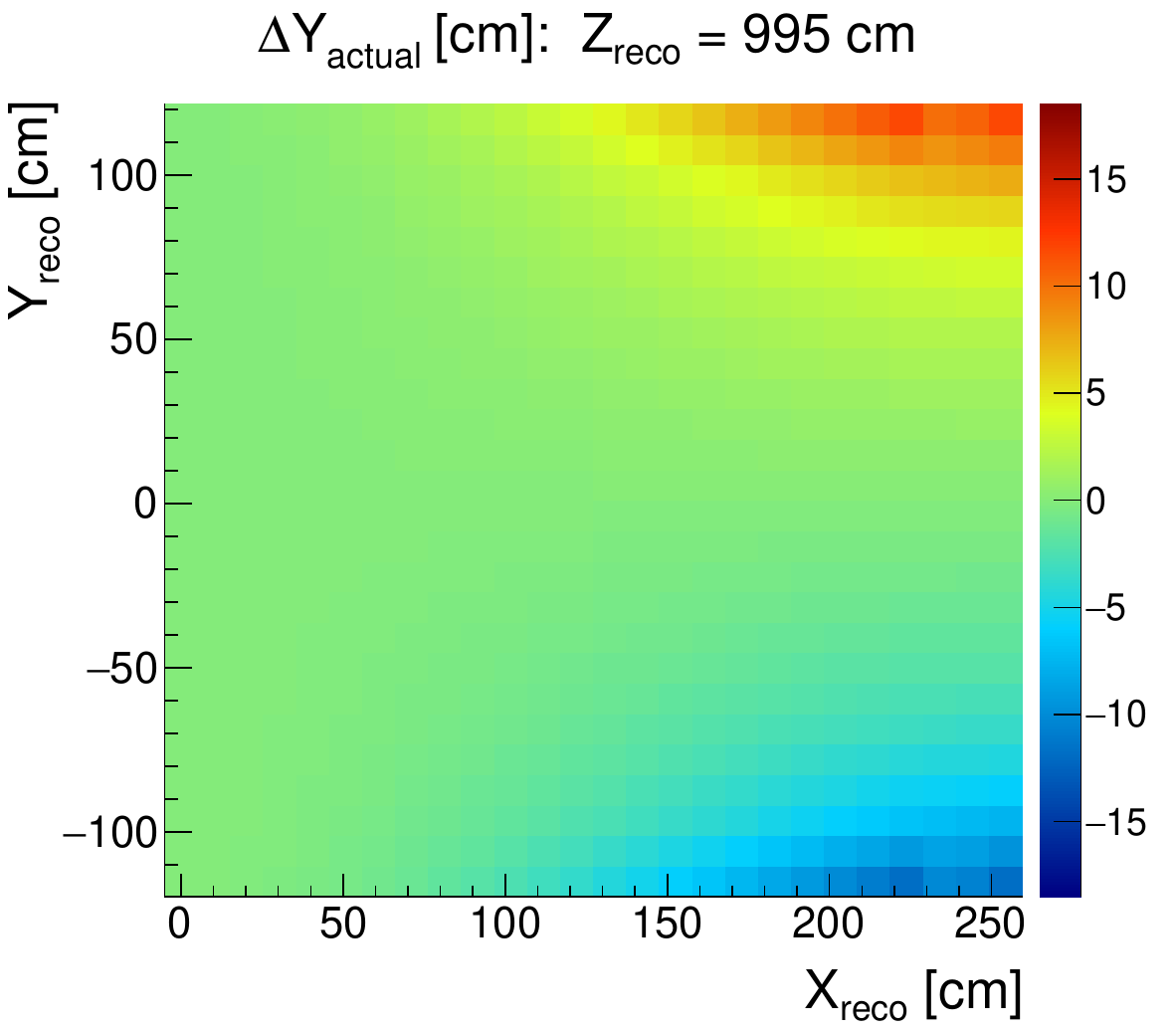}
    \caption{}
  \end{subfigure}
  \begin{subfigure}{0.41\textwidth}
    \centering
    \includegraphics[width=.99\textwidth]{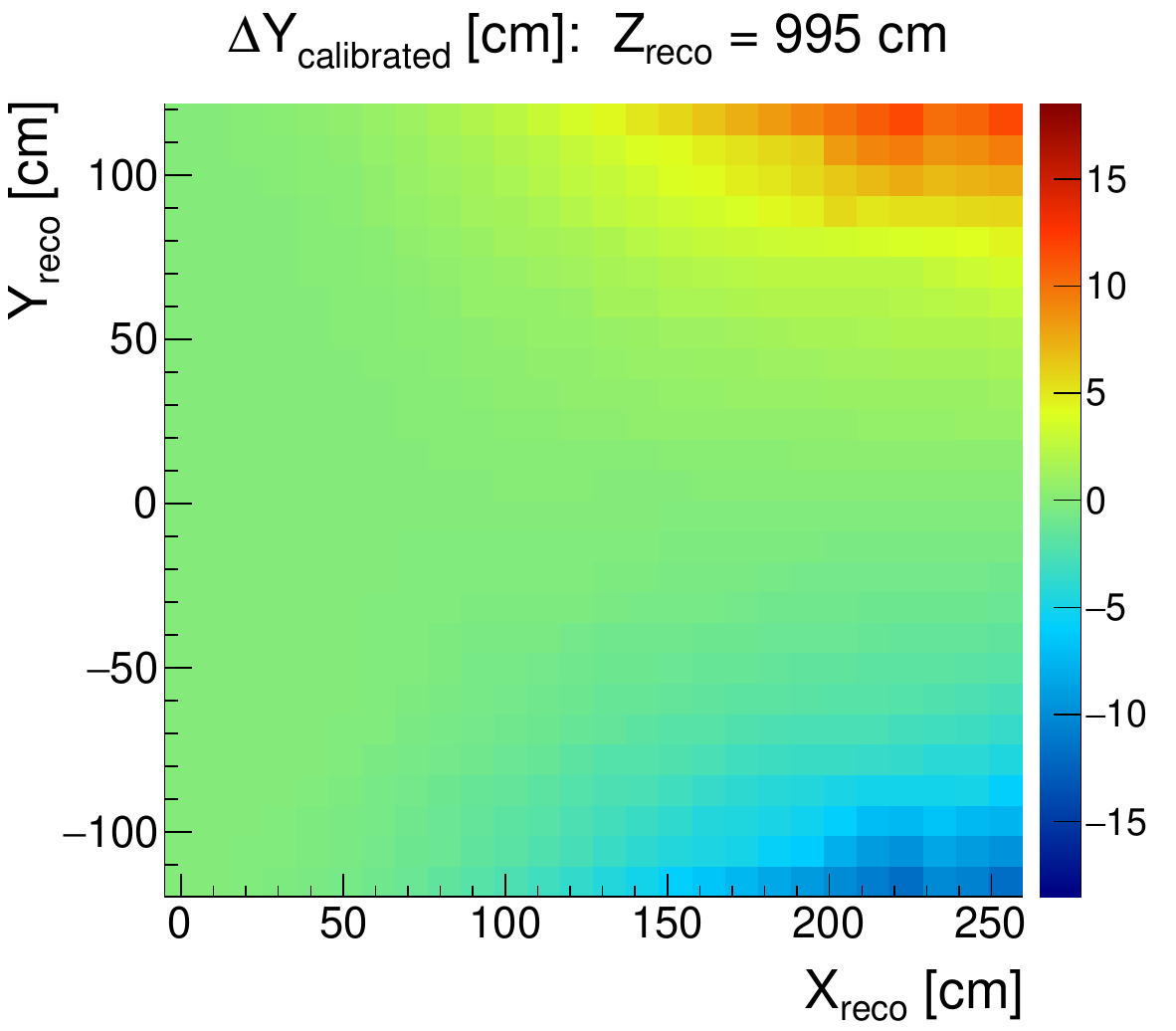}
    \caption{}
  \end{subfigure}
  \\
  \vspace{3mm}
  \begin{subfigure}{0.41\textwidth}
    \centering
    \includegraphics[width=.99\textwidth]{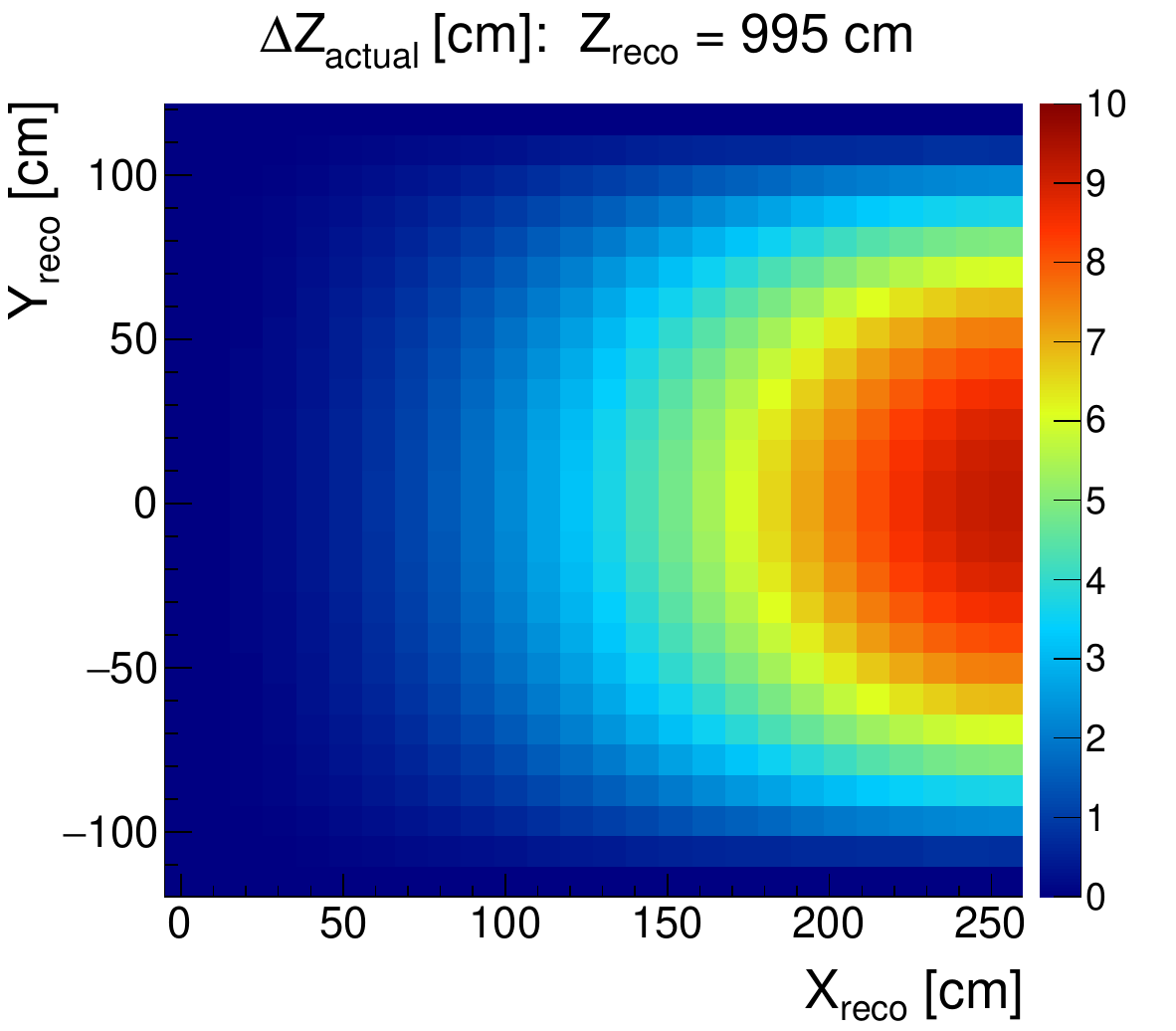}
    \caption{}
  \end{subfigure}
  \begin{subfigure}{0.41\textwidth}
    \centering
    \includegraphics[width=.99\textwidth]{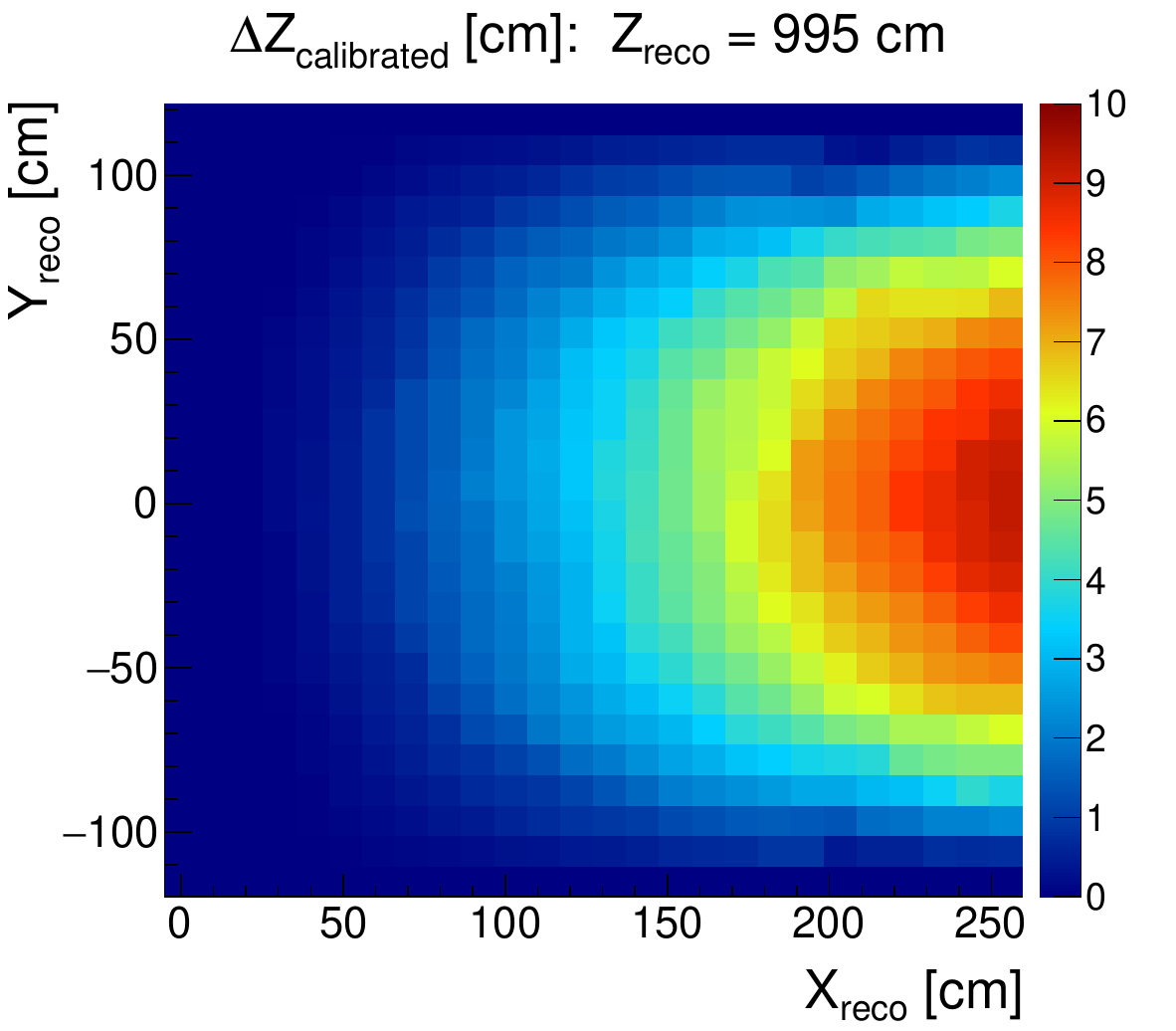}
    \caption{}
  \end{subfigure}
  \\
\Put(36,570){\fontfamily{phv}\selectfont \textbf{MicroBooNE}}
\Put(36,545){\fontfamily{phv}\selectfont \textbf{Simulation}}
\caption{Comparison of (a, c, e) spatial offsets predicted from the SCE simulation to (b, d, f) the results of the TPC bulk calibration on Monte Carlo simulation events for a slice in $z$ closer to the downstream end of the TPC.  Results are shown for spatial offsets in (a, b) $x$, (c, d) $y$, and (e, f) $z$.  The distortions in reconstructed ionization electron cluster position are shown in units of cm and are plotted as a function of the reconstructed position in the TPC.} \label{fig:MC_Results_EndZ}
\end{figure}

The results of the TPC bulk calibration for Monte Carlo simulation events are shown in figure~\ref{fig:MC_Results_CentralZ} and figure~\ref{fig:MC_Results_EndZ}, comparing the predictions from the simulation (``actual'' spatial offsets) to the results of the data-driven calibration (``calibrated'' spatial offsets) near the center of the detector and the downstream end of the detector, respectively.  The data-driven calibration largely reproduces the spatial distortion map predicted by simulation.

\begin{figure}[p]
\centering
  \begin{subfigure}{0.41\textwidth}
    \centering
    \includegraphics[width=.99\textwidth]{./figures/Dx_CentralZ_RecoMC.pdf}
    \caption{}
  \end{subfigure}
  \begin{subfigure}{0.41\textwidth}
    \centering
    \includegraphics[width=.99\textwidth]{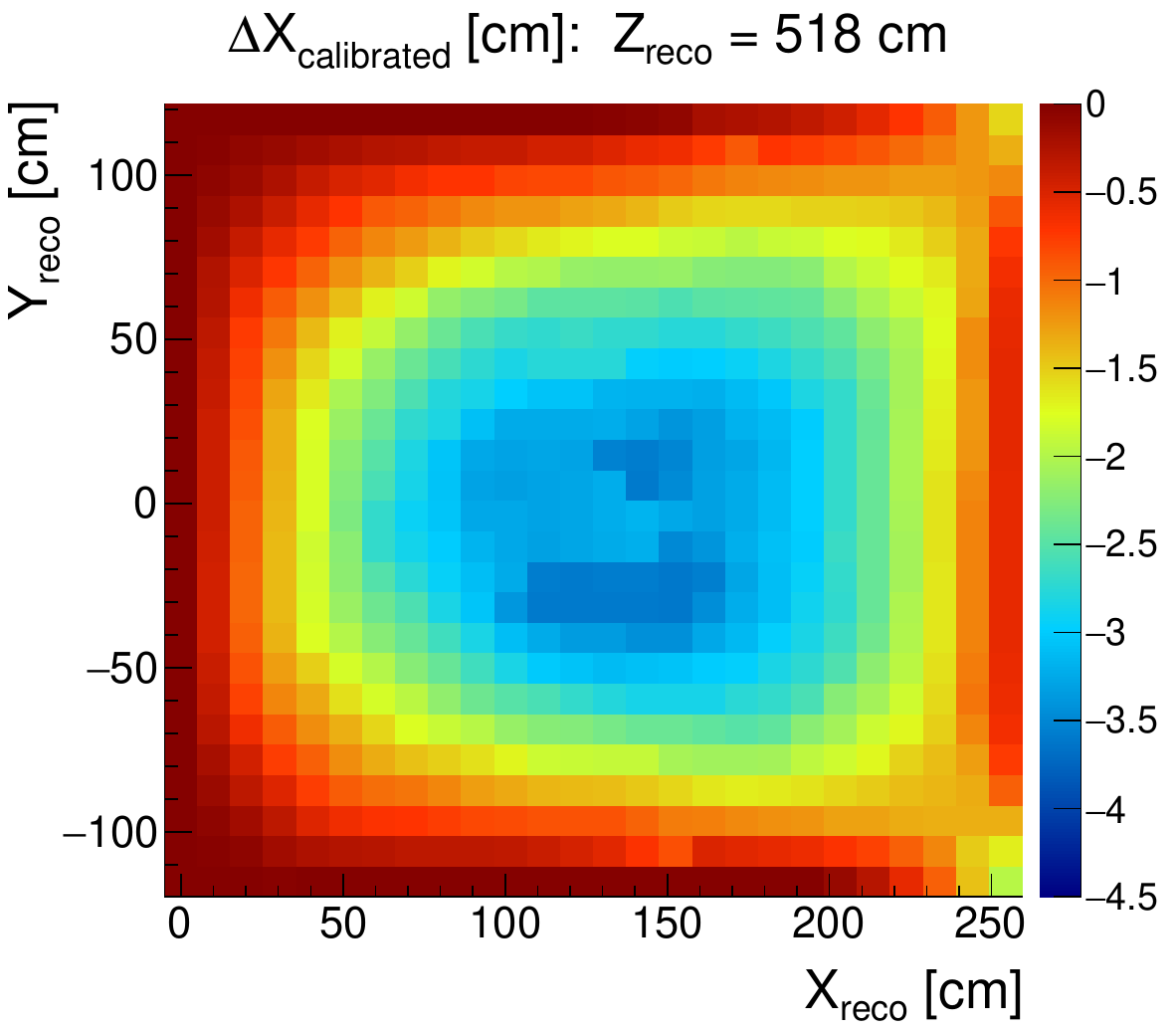}
    \caption{}
  \end{subfigure}
  \\
  \vspace{3mm}
  \begin{subfigure}{0.41\textwidth}
    \centering
    \includegraphics[width=.99\textwidth]{./figures/Dy_CentralZ_RecoMC.pdf}
    \caption{}
  \end{subfigure}
  \begin{subfigure}{0.41\textwidth}
    \centering
    \includegraphics[width=.99\textwidth]{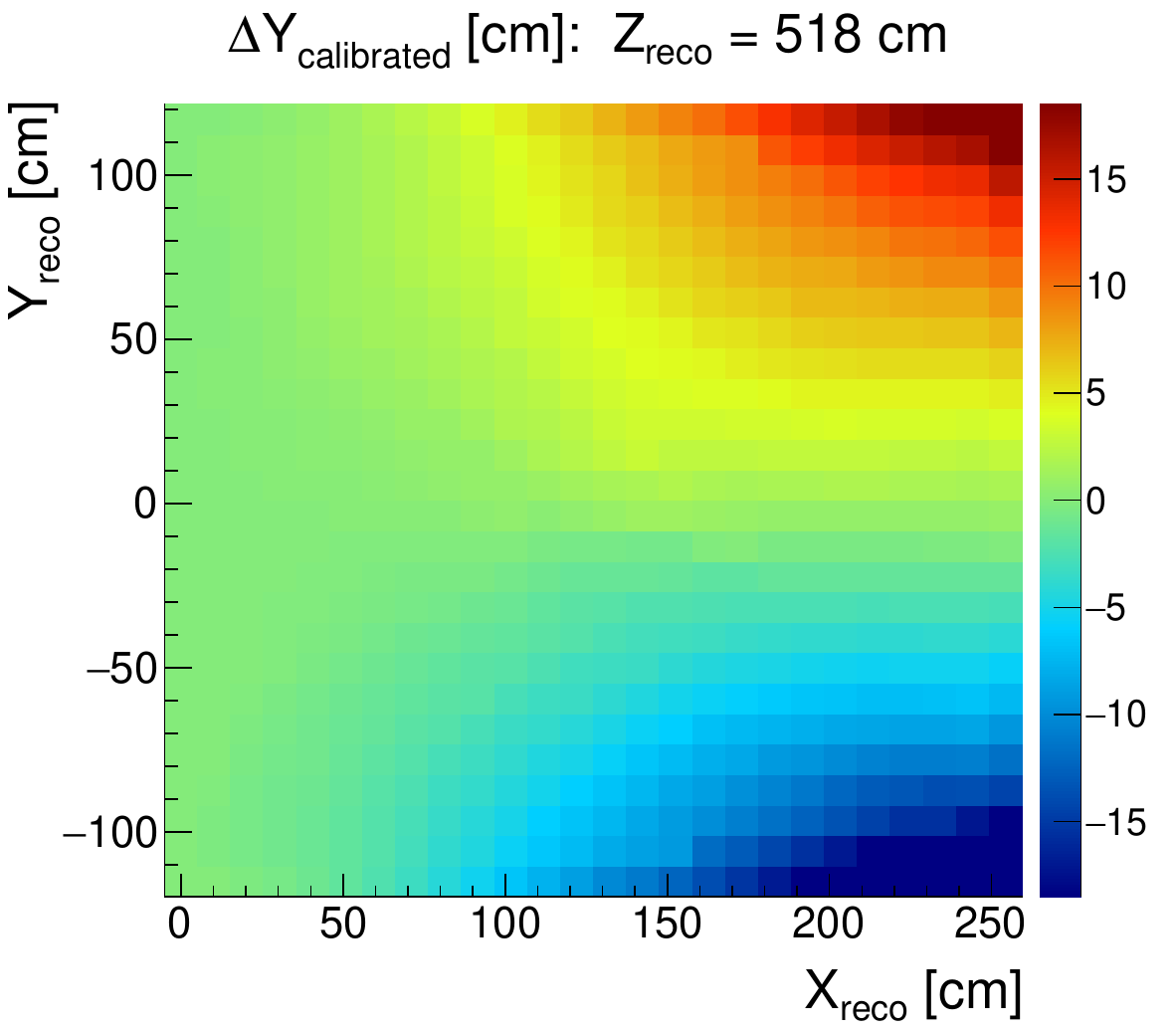}
    \caption{}
  \end{subfigure}
  \\
  \vspace{3mm}
  \begin{subfigure}{0.41\textwidth}
    \centering
    \includegraphics[width=.99\textwidth]{./figures/Dz_CentralZ_RecoMC.pdf}
    \caption{}
  \end{subfigure}
  \begin{subfigure}{0.41\textwidth}
    \centering
    \includegraphics[width=.99\textwidth]{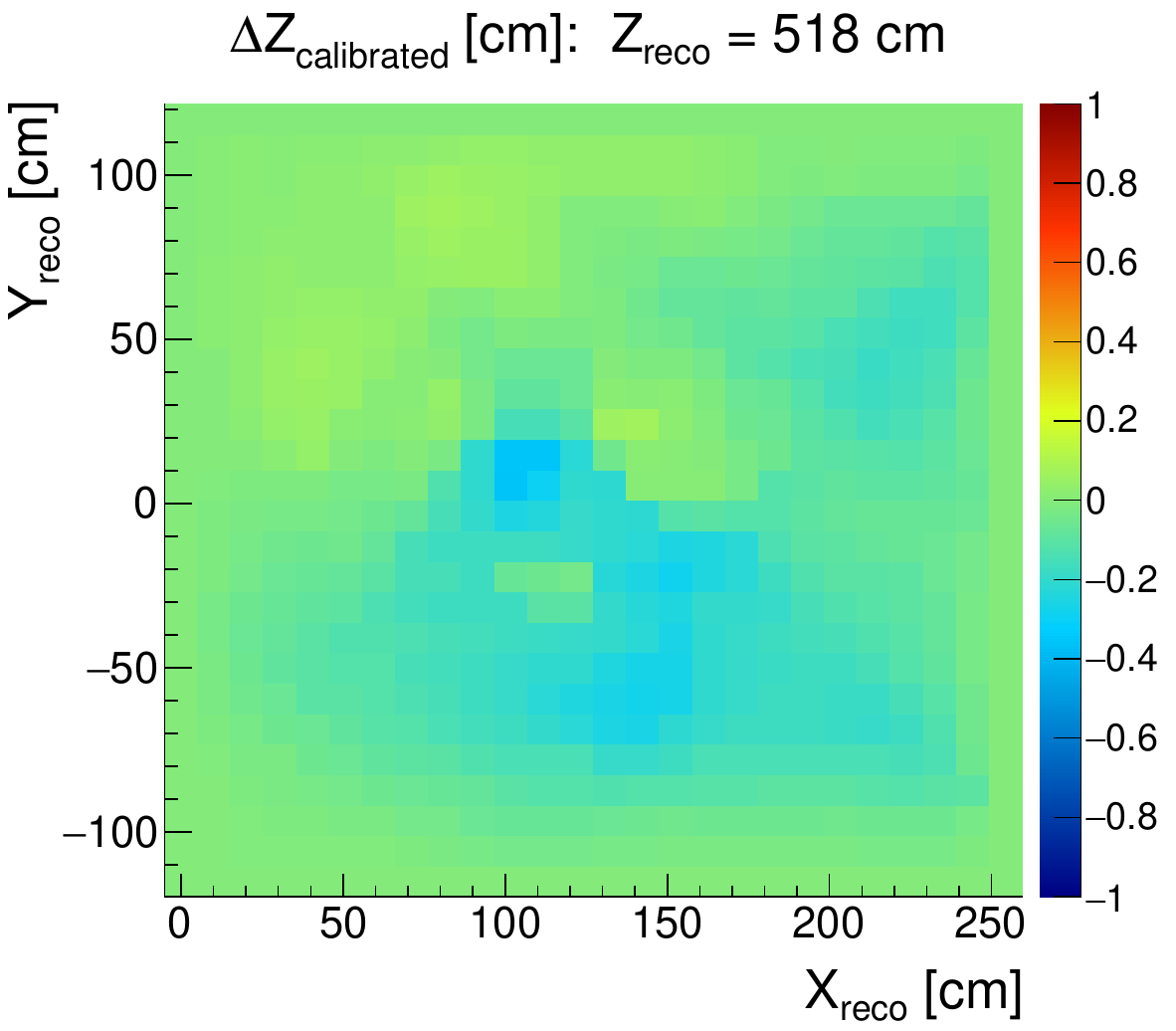}
    \caption{}
  \end{subfigure}
  \\
\Put(-142,570){\fontfamily{phv}\selectfont \textbf{MicroBooNE}}
\Put(-142,545){\fontfamily{phv}\selectfont \textbf{Simulation}}
\Put(33,570){\fontfamily{phv}\selectfont \textbf{MicroBooNE}}
\caption{Comparison of (a, c, e) the results of the TPC bulk calibration on Monte Carlo simulation events to (b, d, f) the results of the TPC bulk calibration on MicroBooNE data events for a central slice in $z$.  Results are shown for spatial offsets in (a, b) $x$, (c, d) $y$, and (e, f) $z$.  The distortions in reconstructed ionization electron cluster position are shown in units of cm and are plotted as a function of the reconstructed position in the TPC.} \label{fig:Data_Results_CentralZ}
\end{figure}

\begin{figure}[p]
\centering
  \begin{subfigure}{0.41\textwidth}
    \centering
    \includegraphics[width=.99\textwidth]{./figures/Dx_EndZ_RecoMC.pdf}
    \caption{}
  \end{subfigure}
  \begin{subfigure}{0.41\textwidth}
    \centering
    \includegraphics[width=.99\textwidth]{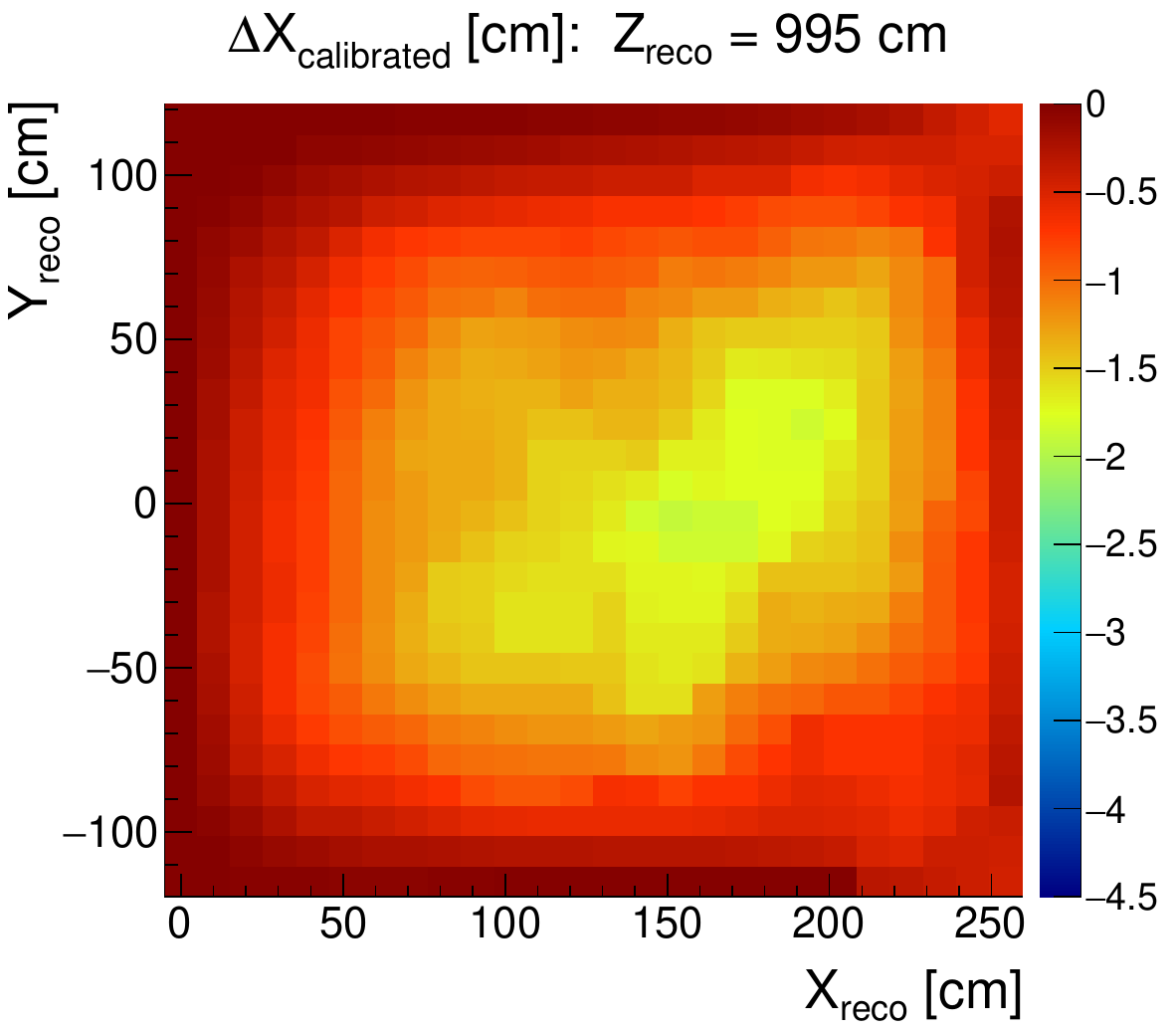}
    \caption{}
  \end{subfigure}
  \\
  \vspace{3mm}
  \begin{subfigure}{0.41\textwidth}
    \centering
    \includegraphics[width=.99\textwidth]{./figures/Dy_EndZ_RecoMC.pdf}
    \caption{}
  \end{subfigure}
  \begin{subfigure}{0.41\textwidth}
    \centering
    \includegraphics[width=.99\textwidth]{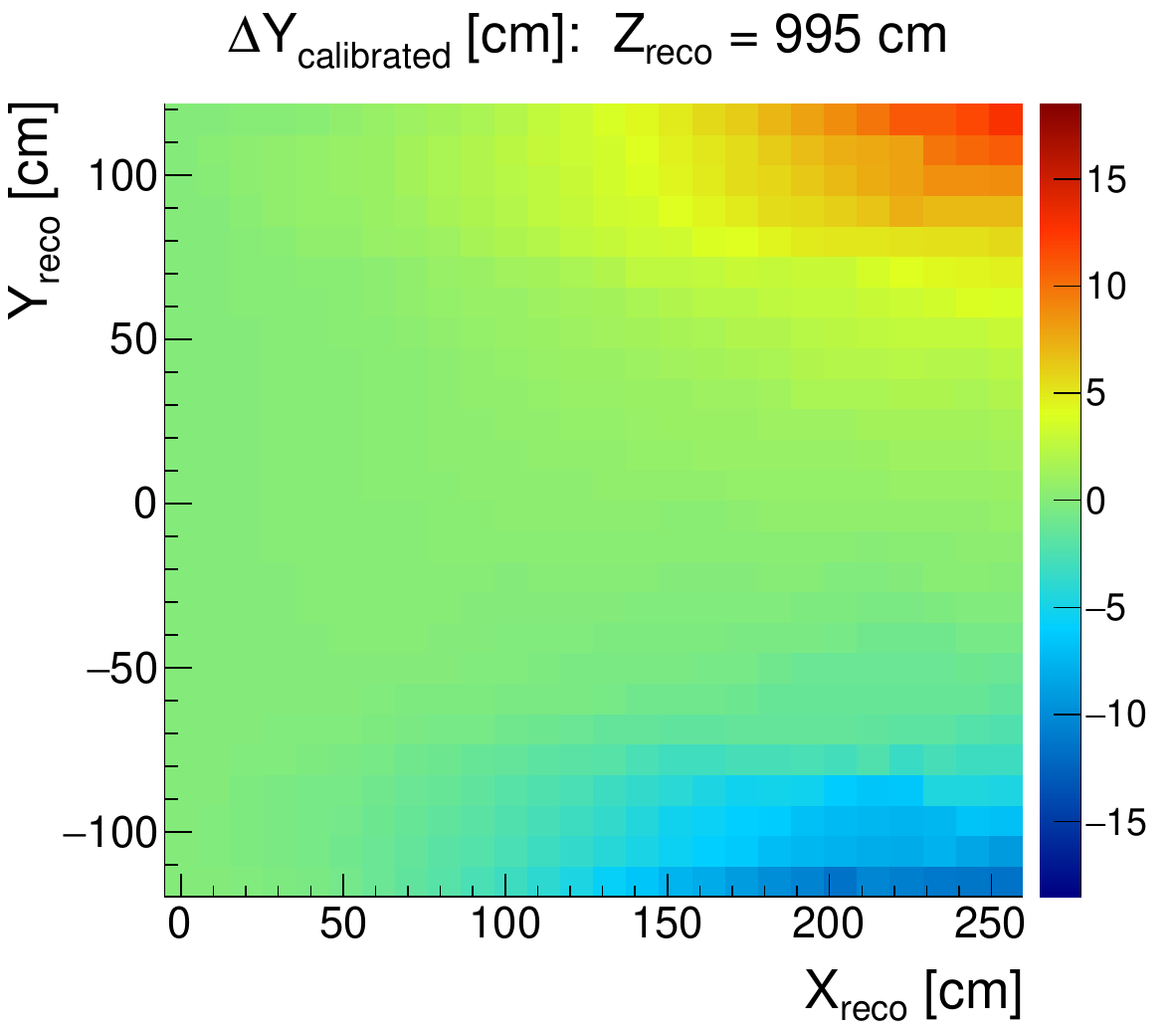}
    \caption{}
  \end{subfigure}
  \\
  \vspace{3mm}
  \begin{subfigure}{0.41\textwidth}
    \centering
    \includegraphics[width=.99\textwidth]{./figures/Dz_EndZ_RecoMC.pdf}
    \caption{}
  \end{subfigure}
  \begin{subfigure}{0.41\textwidth}
    \centering
    \includegraphics[width=.99\textwidth]{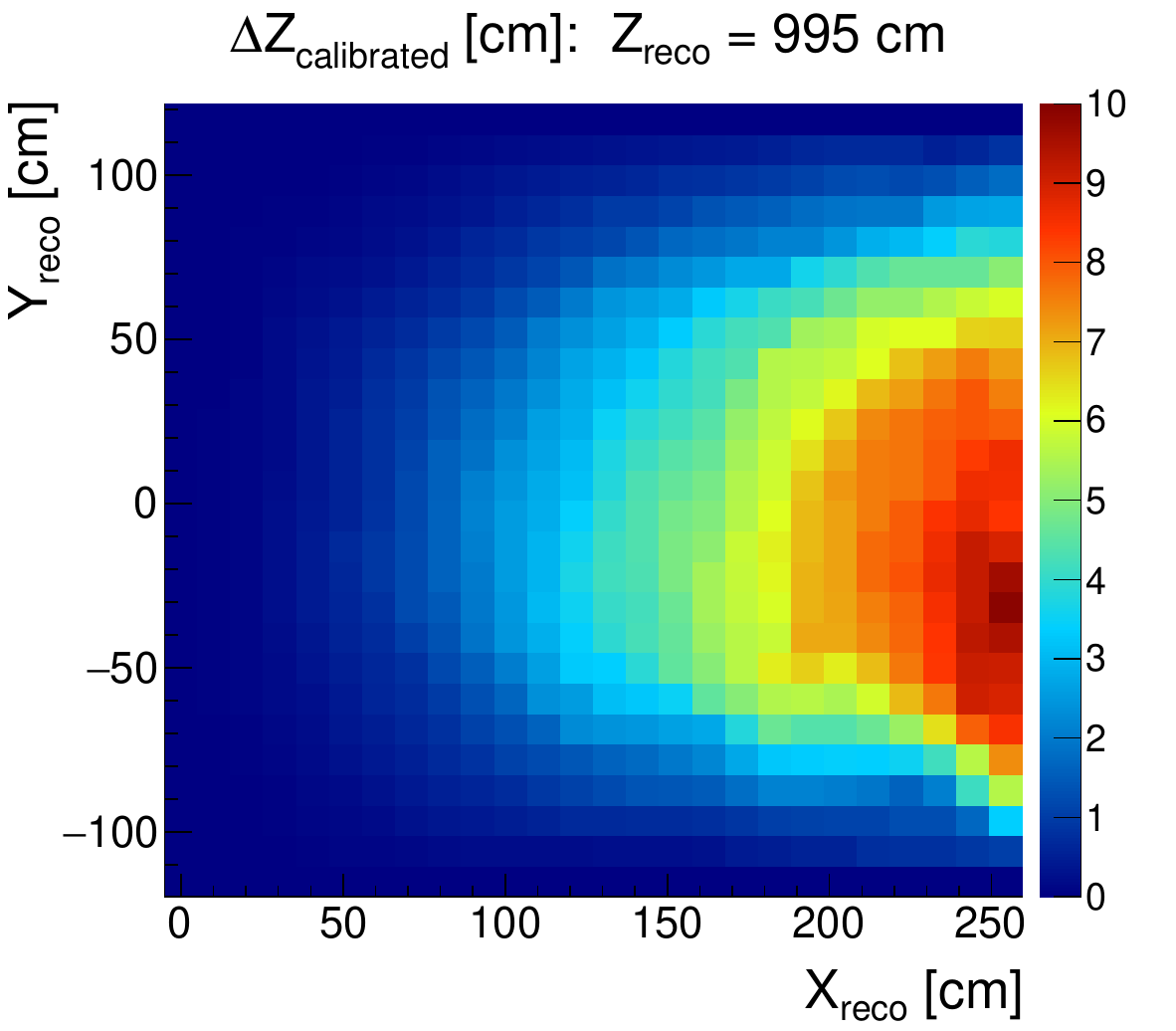}
    \caption{}
  \end{subfigure}
  \\
\Put(-142,570){\fontfamily{phv}\selectfont \textbf{MicroBooNE}}
\Put(-142,545){\fontfamily{phv}\selectfont \textbf{Simulation}}
\Put(33,570){\fontfamily{phv}\selectfont \textbf{MicroBooNE}}
\caption{Comparison of (a, c, e) the results of the TPC bulk calibration on Monte Carlo simulation events to (b, d, f) the results of the TPC bulk calibration on MicroBooNE data events for a slice in $z$ closer to the downstream end of the TPC.  Results are shown for spatial offsets in (a, b) $x$, (c, d) $y$, and (e, f) $z$.  The distortions in reconstructed ionization electron cluster position are shown in units of cm and are plotted as a function of the reconstructed position in the TPC.} \label{fig:Data_Results_EndZ}
\end{figure}

The spatial distortion maps associated with MicroBooNE data events are illustrated in figure~\ref{fig:Data_Results_CentralZ} and figure~\ref{fig:Data_Results_EndZ}.  These maps are compared with the corresponding distributions obtained from Monte Carlo simulation events that are also shown in figure~\ref{fig:MC_Results_CentralZ} and figure~\ref{fig:MC_Results_EndZ}.  The spatial distortion maps obtained using data are qualitatively very similar to those obtained using Monte Carlo simulation events.  As in the case of the TPC face calibration results shown in section~\ref{sec:results_faces}, there are some differences between data and simulation caused by the underlying space charge configuration not being exactly reproduced in the simulation.

\begin{figure}[tb]
\centering
\includegraphics[width=.55\textwidth]{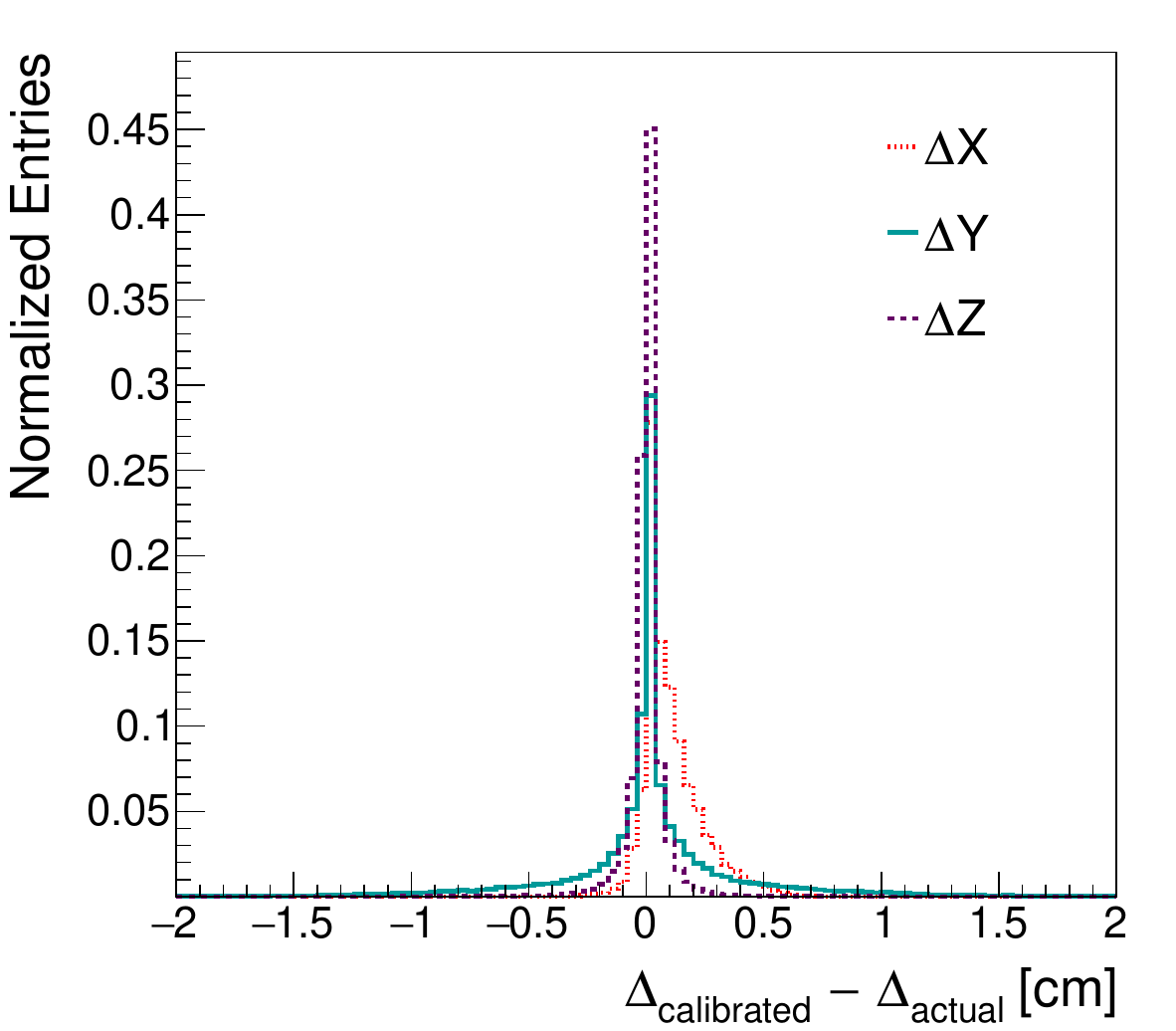}
\Put(-190,340){\fontfamily{phv}\selectfont \textbf{MicroBooNE}}
\Put(-190,315){\fontfamily{phv}\selectfont \textbf{Simulation}}
\caption{The distribution of differences between Monte Carlo simulation TPC bulk calibration results and simulated spatial offsets, $\Delta_{\mathrm{calibrated}} - \Delta_{\mathrm{actual}}$, across the entire TPC volume for spatial distortions in $x$ (red), $y$ (blue), and $z$ (green) in units of cm.} \label{fig:MC_perf}
\end{figure}

\begin{figure}[tb]
\centering
  \begin{subfigure}{0.49\textwidth}
    \centering
    \includegraphics[width=.99\textwidth]{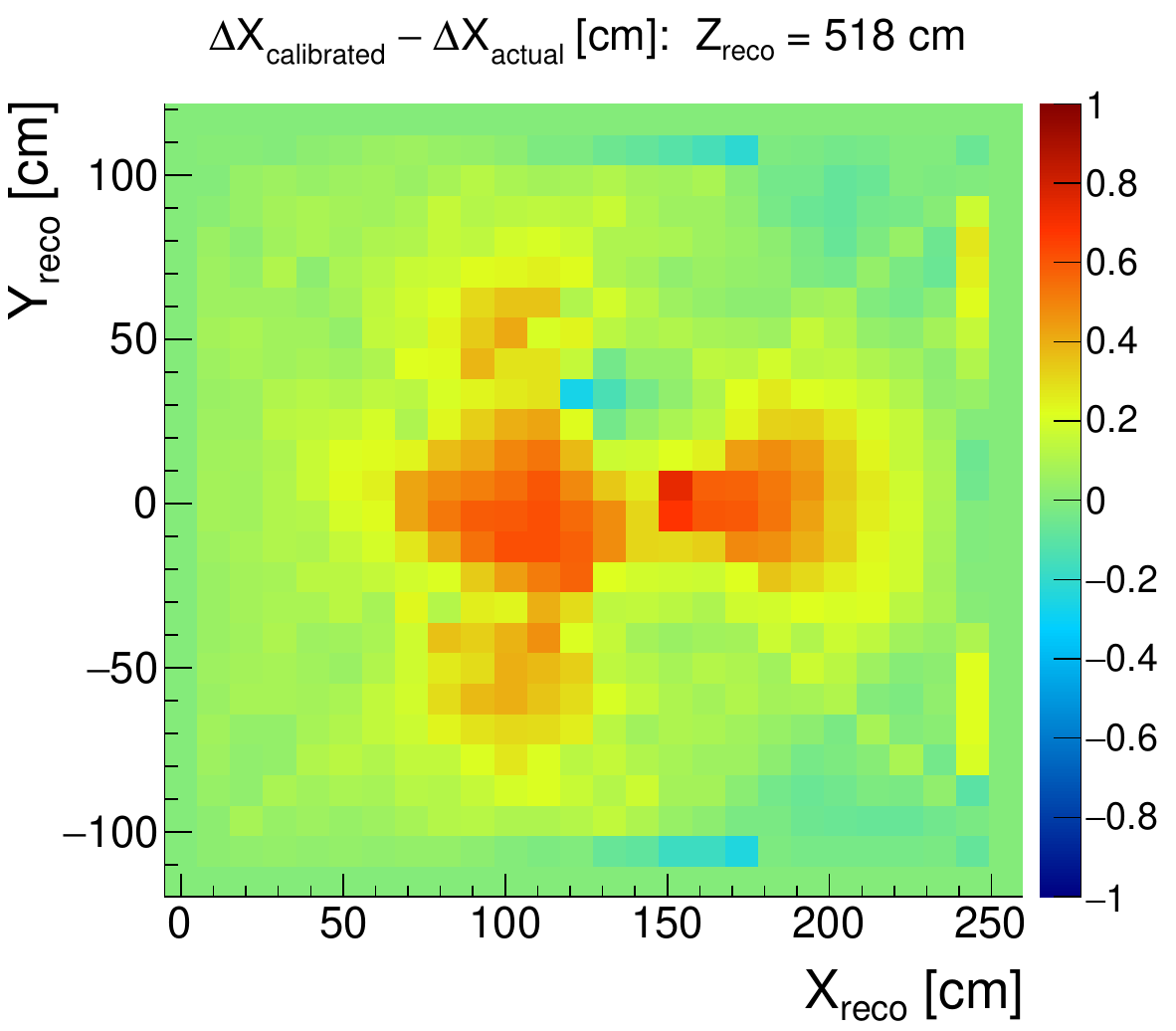}
    \caption{}
  \end{subfigure}
  \begin{subfigure}{0.49\textwidth}
    \centering
    \includegraphics[width=.99\textwidth]{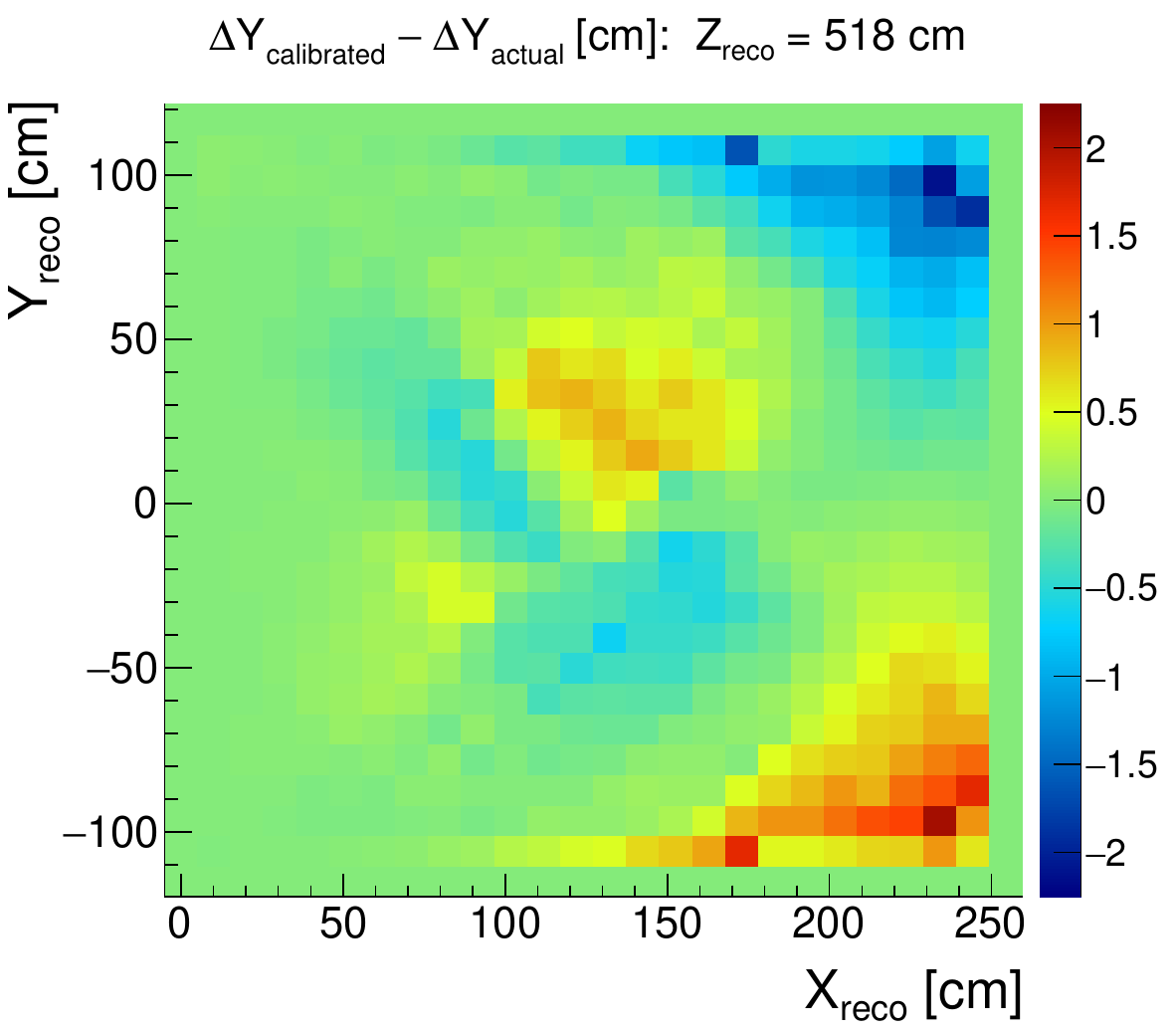}
    \caption{}
  \end{subfigure}
\Put(-390,145){\fontfamily{phv}\selectfont \textbf{MicroBooNE}}
\Put(-390,120){\fontfamily{phv}\selectfont \textbf{Simulation}}
\Put(37,345){\fontfamily{phv}\selectfont \textbf{MicroBooNE}}
\Put(37,320){\fontfamily{phv}\selectfont \textbf{Simulation}}
\caption{Difference between calibrated spatial offsets in Monte Carlo simulation events and actual spatial offsets obtained from a dedicated SCE simulation, showing the (a) ${\Delta}x$ bias and (b) ${\Delta}y$ bias in the $x-y$ plane for the central $z$ slice of the detector.}
\label{fig:MC_perf_2D}
\end{figure}

The performance of the data-driven calibration is illustrated in figure~\ref{fig:MC_perf} for all three dimensions and throughout the entire TPC, comparing the calibrated spatial offsets in Monte Carlo simulation events to the actual spatial offsets obtained from a dedicated SCE simulation.  A single entry in each histogram corresponds to a single voxel (with a spatial extent of approximately \SI{10}{cm} in each dimension) in the detector.  This comparison is also shown in the $x-y$ plane for ${\Delta}x$ and ${\Delta}y$, averaged over the $z$ dimension of the detector, in figure~\ref{fig:MC_perf_2D}.  The measured resolution and bias for each dimension, calculated using the standard deviation and mean of these distributions, respectively, are shown in table~\ref{tab:MC_perf}.  The resolution is <~\SI{4}{mm} and the bias is <~\SI{1}{mm} in all dimensions.  From figure~\ref{fig:MC_perf_2D} it is apparent that the tails of the $\Delta{y}$ distribution in figure~\ref{fig:MC_perf} are associated with the edges of the TPC near the cathode, where the spatial offsets are largest in the detector.

\begin{table}[tb]
  \centering
  \begin{tabu}{c|[2pt]c|c}
    Measured Offset      & Resolution [cm] & Bias [cm] \\ \tabucline[2pt]{-}
    ${\Delta}x$          & 0.13            & 0.09 \\ \hline
    ${\Delta}y$          & 0.38            & 0.01  \\ \hline
    ${\Delta}z$          & 0.10            & 0.00  
  \end{tabu}
  \caption{Estimation of the bias and resolution in measured spatial offsets as found using the TPC bulk calibration procedure, described in section~\ref{sec:method_step3}, on Monte Carlo simulation events.} \label{tab:MC_perf}
\end{table}

The metric shown in figure~\ref{fig:MC_perf} requires the use of Monte Carlo simulation events.  In order to evaluate the performance of the calibration procedure on data, laser tracks from the MicroBooNE UV laser system are used.  As the true trajectories of the laser tracks are known~\cite{ubLaser}, the reconstructed laser tracks~\cite{laser_calib} can be corrected with the results of the space charge effect calibration and subsequently compared to the true trajectories.  This comparison is made by calculating ``laser track residuals'' that are defined at each point along the reconstructed laser track as the projected distance to the true laser trajectory, either before or after spatial space charge corrections are applied; this calculation is performed on a point-by-point basis along the entire extent of each reconstructed laser track, and for each laser track associated with a full scan throughout the entire TPC volume.  If the space charge effect calibration is performing well, it should produce straighter reconstructed laser tracks and smaller laser track residuals.

\begin{figure}[tb]
\centering
  \begin{subfigure}{0.49\textwidth}
    \centering
    \includegraphics[width=.99\textwidth]{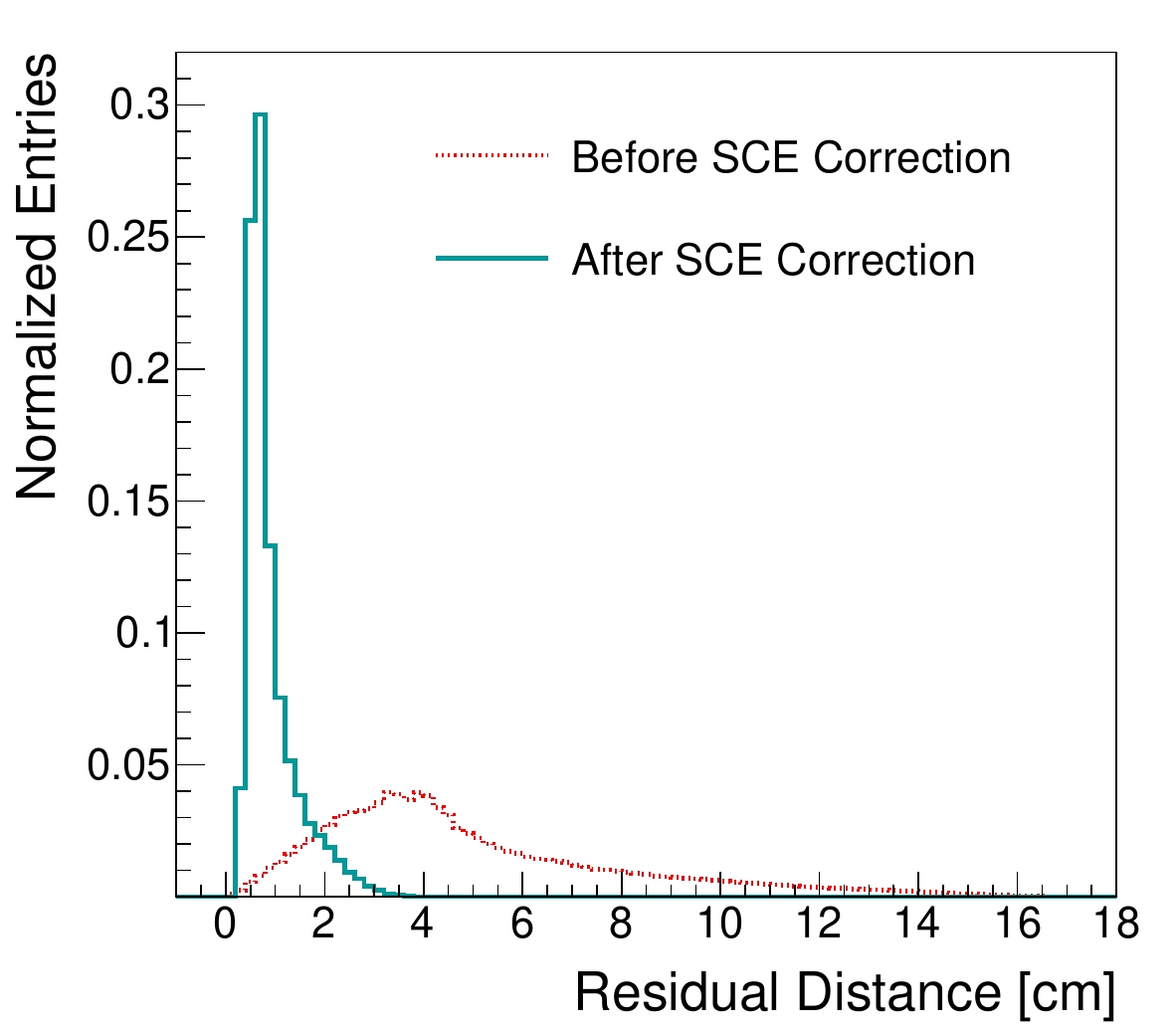}
    \caption{}
  \end{subfigure}
  \begin{subfigure}{0.49\textwidth}
    \centering
    \includegraphics[width=.99\textwidth]{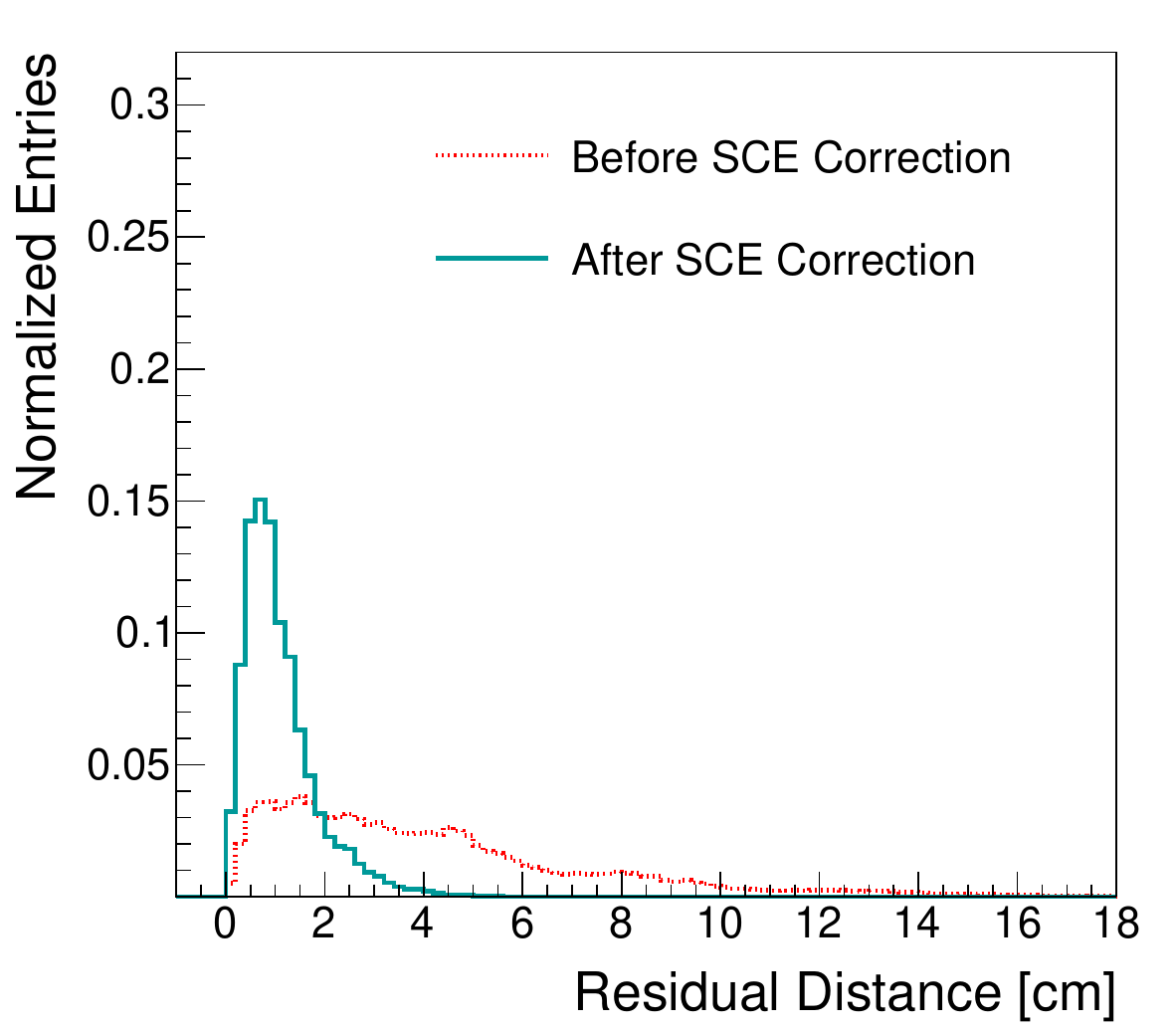}
    \caption{}
  \end{subfigure}
\Put(-310,-27){\fontfamily{phv}\selectfont \textbf{MicroBooNE}}
\Put(-310,-52){\fontfamily{phv}\selectfont \textbf{Simulation}}
\Put(117,180){\fontfamily{phv}\selectfont \textbf{MicroBooNE}}
\caption{Laser track residuals for (a) Monte Carlo simulation and (b) data events, shown both before and after applying spatial SCE corrections.}
\label{fig:LaserResiduals}
\end{figure}

The calculated laser track residuals are shown in figure~\ref{fig:LaserResiduals} for both simulated and MicroBooNE data events, showing the impact of the space charge effect calibration on this metric.  The laser track residuals become significantly smaller and closer to zero after the application of the space charge effect calibration.  The performance of the calibration on data events is worse than for the same calibration applied to Monte Carlo simulation events.  The additional degradation of spatial resolution associated with the data calibration is found to be roughly \SI{4}{mm}, determined by applying a Gaussian smearing to the simulation result until it is in agreement with the MicroBooNE data result.

Additional discussion of systematic uncertainties associated with the measurement of spatial distortions due to space charge effects is presented in section~\ref{sec:systbias}.

\subsection{Electric Field Distortion Results} \label{sec:results_efield}

\begin{figure}
\centering
  \begin{subfigure}{0.49\textwidth}
    \centering
    \includegraphics[width=.99\textwidth]{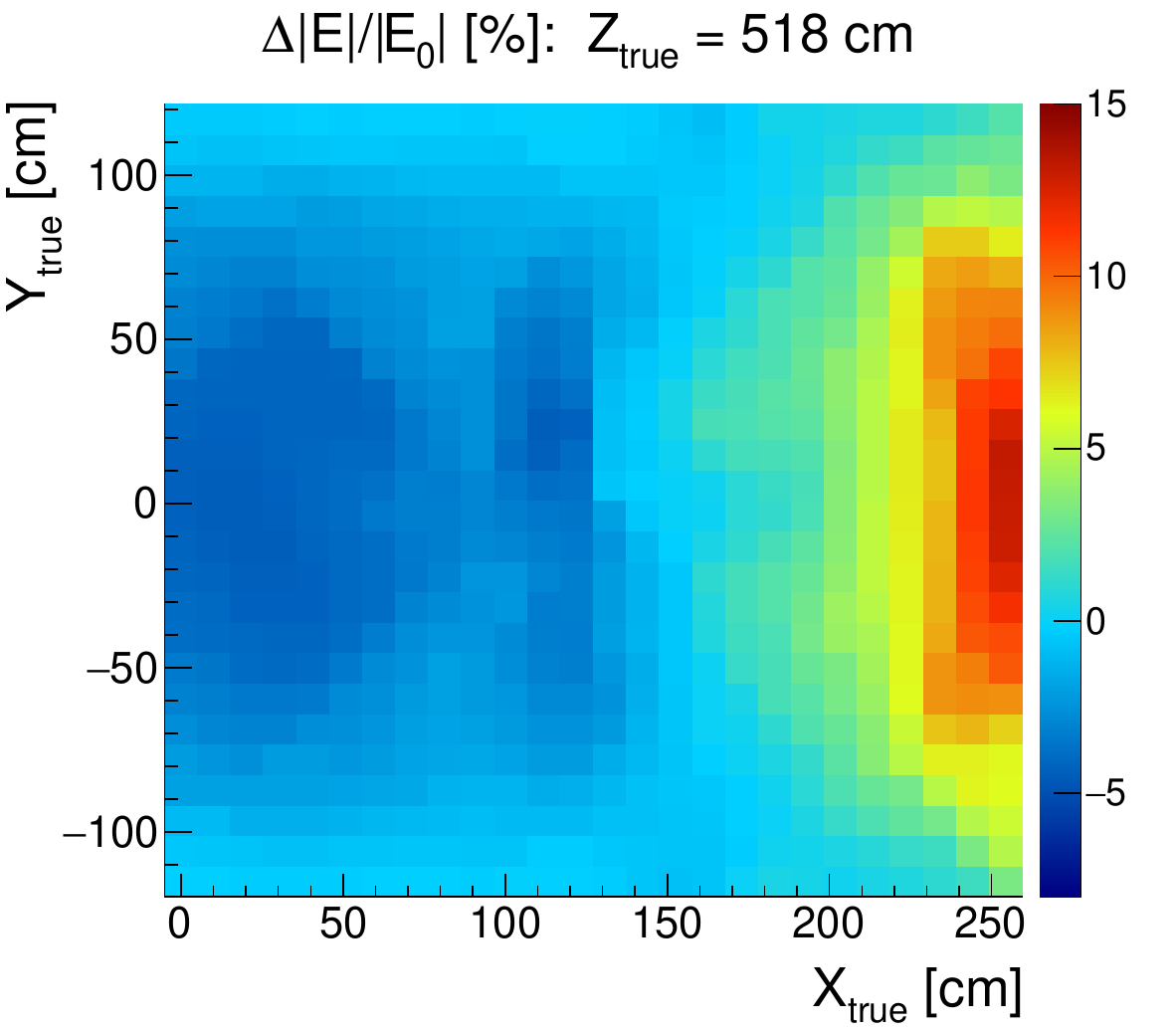}
    \caption{}
  \end{subfigure}
  \begin{subfigure}{0.49\textwidth}
    \centering
    \includegraphics[width=.99\textwidth]{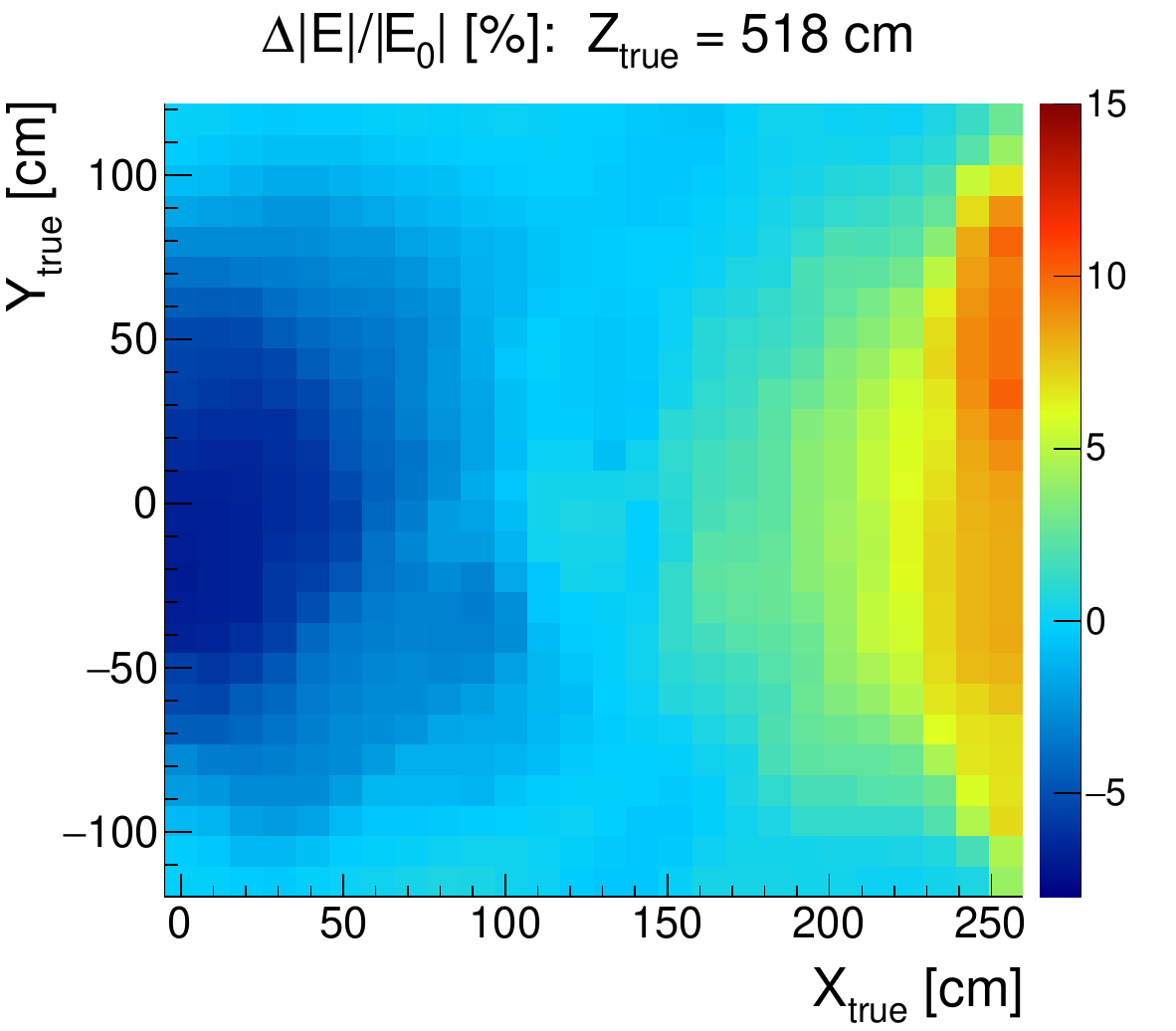}
    \caption{}
  \end{subfigure}
\Put(-390,147){\fontfamily{phv}\selectfont \textbf{MicroBooNE}}
\Put(-390,122){\fontfamily{phv}\selectfont \textbf{Simulation}}
\Put(37,345){\fontfamily{phv}\selectfont \textbf{MicroBooNE}}
\caption{Results of the calculation of electric field distortion magnitude for a central slice in $z$, comparing (a) Monte Carlo simulation to (b) data and shown as the percentage change with respect to the nominal MicroBooNE electric field magnitude, $|E_{0}| = \SI{273.9}{V/cm}$.}
\label{fig:Efield_Results_CentralZ}
\end{figure}

As discussed in section~\ref{sec:method_efield}, with the spatial distortion map determined throughout the TPC volume, the electric field distortions associated with space charge effects can be computed.  The results of this calculation are shown in figure~\ref{fig:Efield_Results_CentralZ} for simulation and MicroBooNE data, looking in a central slice of the detector in $z$.  These results are presented as the percentage change with respect to the nominal MicroBooNE electric field magnitude of \SI{273.9}{V/cm}.  It is observed that, while the general features of the electric field distortion map are similar when comparing the simulation result to that of data, there is a slight downward shift in the electric field magnitude in data across the entire TPC.  The electric field magnitude varies by no more than 10\% across the entire MicroBooNE TPC, with maximal change in the electric field magnitude near the cathode, as expected.

Additional discussion of systematic uncertainties associated with the measurement of electric field distortions due to space charge effects is presented in section~\ref{sec:systbias}.

\section{Systematic Bias Studies} \label{sec:systbias}

In section~\ref{sec:results}, results of the space charge effect calibration using cosmic muons are presented.  These results contain biases associated with both the calibration methodology and reliance on the simulation that may qualitatively deviate from the true space charge distribution in MicroBooNE data.  We utilize the UV laser system to establish a data-driven systematic uncertainty on the SCE calibration result.  The relevant methodology is discussed in section~\ref{sec:systbias_lasertracks}, and the associated impact on stopping muon $dE/dx$ measurements (as a case study) is presented in section~\ref{sec:systbias_stopmu}.

\subsection{Estimating Systematic Bias with UV Laser System} \label{sec:systbias_lasertracks}

In order to set a systematic uncertainty on the spatial offset calibration, the laser track residuals discussed in section~\ref{sec:results_bulk} are used, as the UV laser system provides a secondary handle on the impact of space charge effects in MicroBooNE data events.  The laser track residuals throughout the TPC are used as an estimator of the bias associated with local spatial offset measurements, which in turn serves as a systematic uncertainty on measurements making use of cosmic rays to estimate space charge effects in the detector.  This systematic uncertainty is determined in three dimensions across the entire TPC as follows:
\begin{itemize}
\item a pass through all reconstructed laser track points is made, calculating at each point the laser track residual from the three-dimensional projection of the reconstructed laser track point to the associated true laser trajectory (this projection also defines a three-dimensional vector that will be referred to as the ``projection vector'');
\item at each reconstructed laser track point, a three-dimensional vector is formed using the spatial offset measurements presented in section~\ref{sec:results_bulk}, which is then rescaled in magnitude such that the component of this vector aligned with the projection vector described in the above step is equal to the previously calculated laser track residual in terms of magnitude;
\item the corrected three-dimensional vector from the previous step is compared to the original spatial offset vector, with the component-wise differences divided by the original spatial offset vector magnitude yielding a relative systematic uncertainty for spatial offsets in each of the three dimensions ($x$, $y$, and $z$) at the point in question;
\item after all reconstructed laser track points are looped over, a separate pass through all voxels in the TPC is made in order to determine a systematic uncertainty for spatial offsets in all three dimensions throughout the entire TPC;
\item for each voxel in the TPC, a secondary pass through all reconstructed laser track points is made in order to compute a weighted average of relative systematic uncertainties (in each of the three dimensions) for the voxel in question, using weight factors of $r^{-2}$ where $r$ is the distance between the voxel and a given reconstructed laser track point;
\item the distance-weighted relative systematic uncertainty computed in the previous step for each dimension is multiplied by the corresponding spatial offset at that voxel to determine the absolute systematic uncertainty in the spatial offset measurement; and
\item finally, once all voxels are looped over, an absolute systematic uncertainty on the spatial offset measurement is available at all points in the detector for each of the three dimensions.
\end{itemize}
As the UV laser system only has partial coverage throughout the TPC~\cite{ubLaser}, gaps in the coverage are accounted in the above procedure by extrapolating the calculated relative systematic uncertainties from neighboring regions of the TPC where there is coverage by the UV laser system, weighted by relative proximity.

\begin{figure}[tb]
\centering
\includegraphics[width=.55\textwidth]{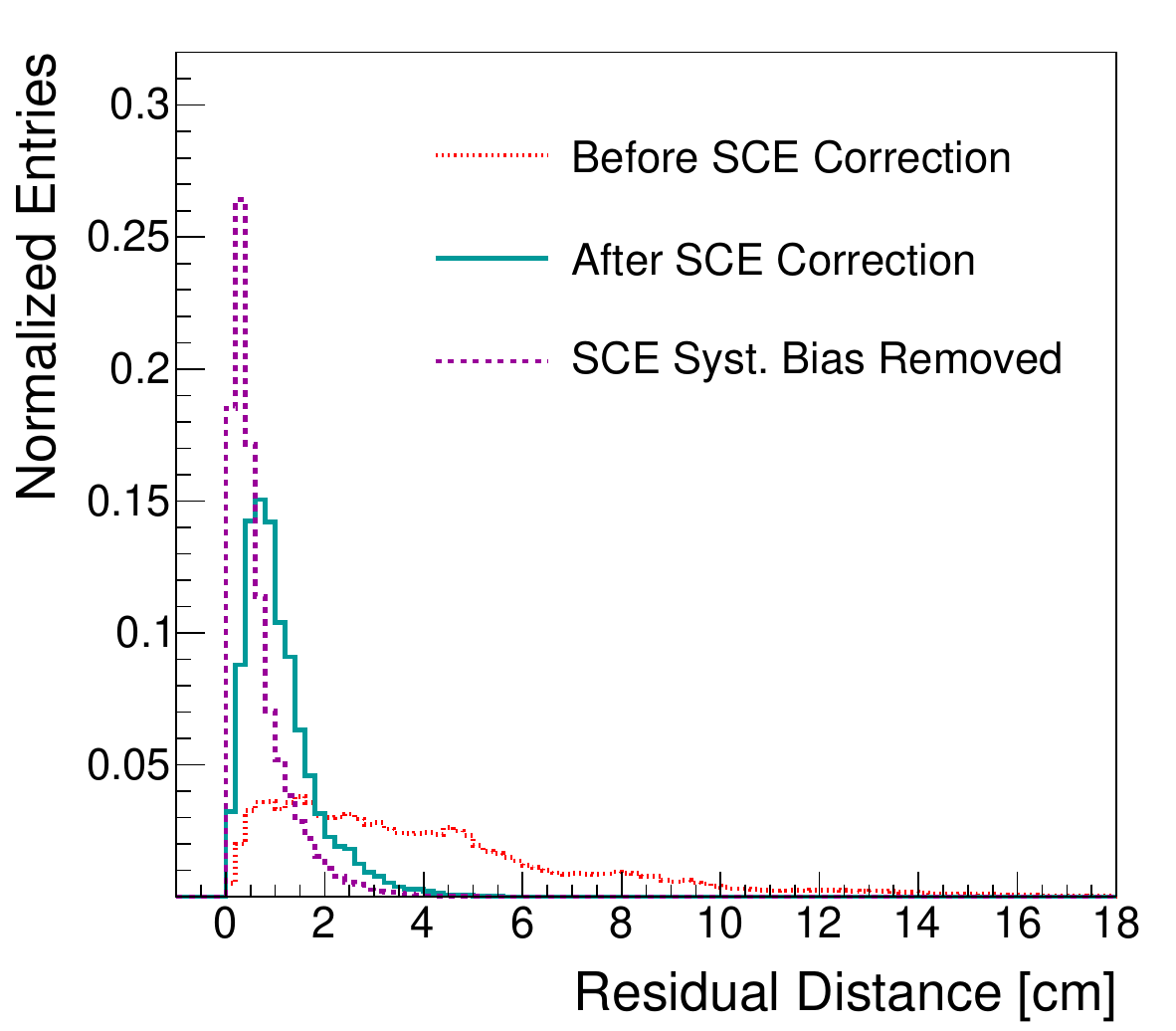}
\Put(-95,130){\fontfamily{phv}\selectfont \textbf{MicroBooNE}}
\caption{Laser track residuals for data events before and after applying spatial SCE corrections, now also including the result after removing systematic bias in the spatial SCE correction that is described in the text.}
\label{fig:LaserResidualsSyst}
\end{figure}

\begin{figure}[tb]
\centering
  \begin{subfigure}{0.49\textwidth}
    \centering
    \includegraphics[width=.99\textwidth]{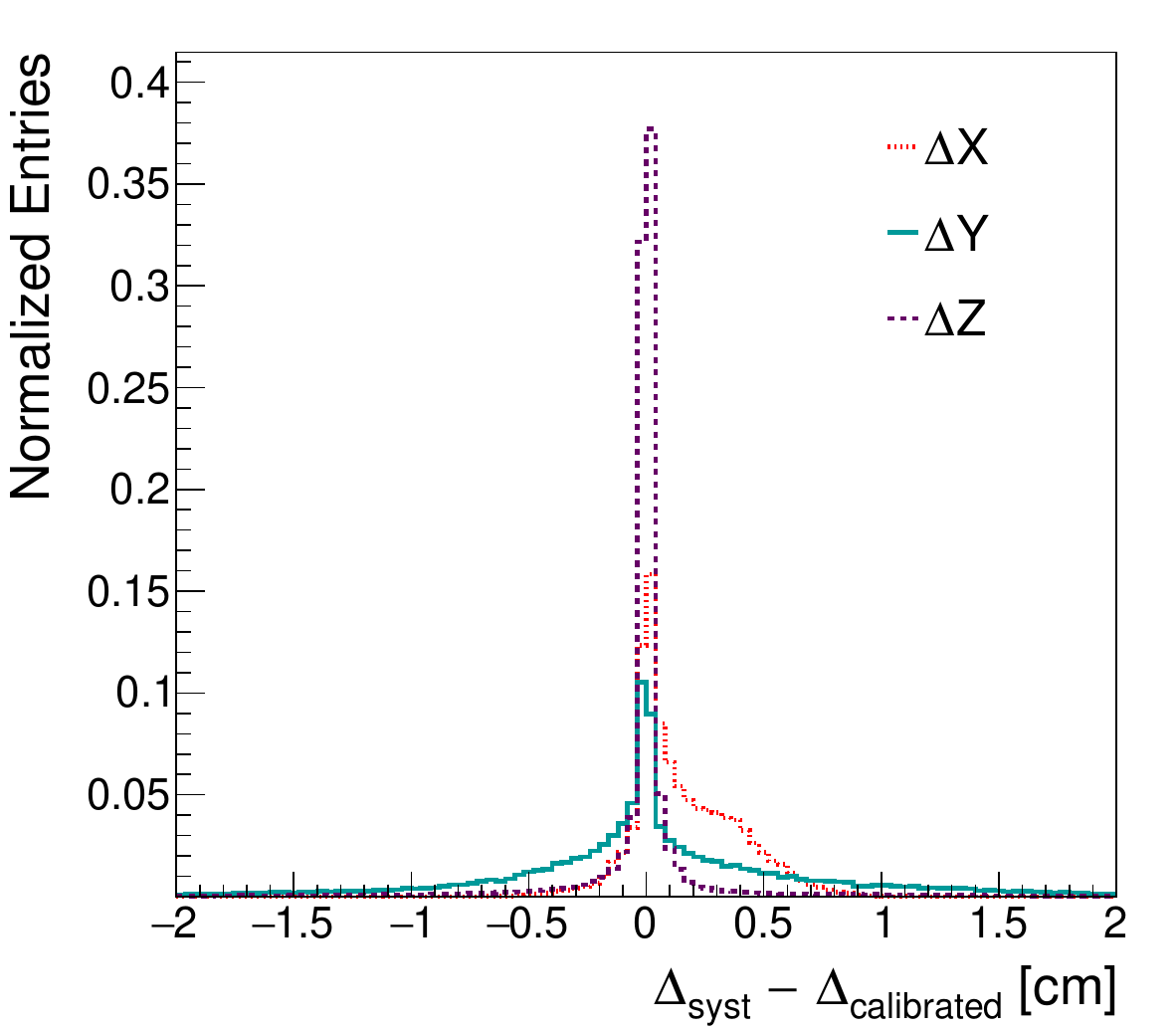}
    \caption{}
  \end{subfigure}
  \begin{subfigure}{0.49\textwidth}
    \centering
    \includegraphics[width=.99\textwidth]{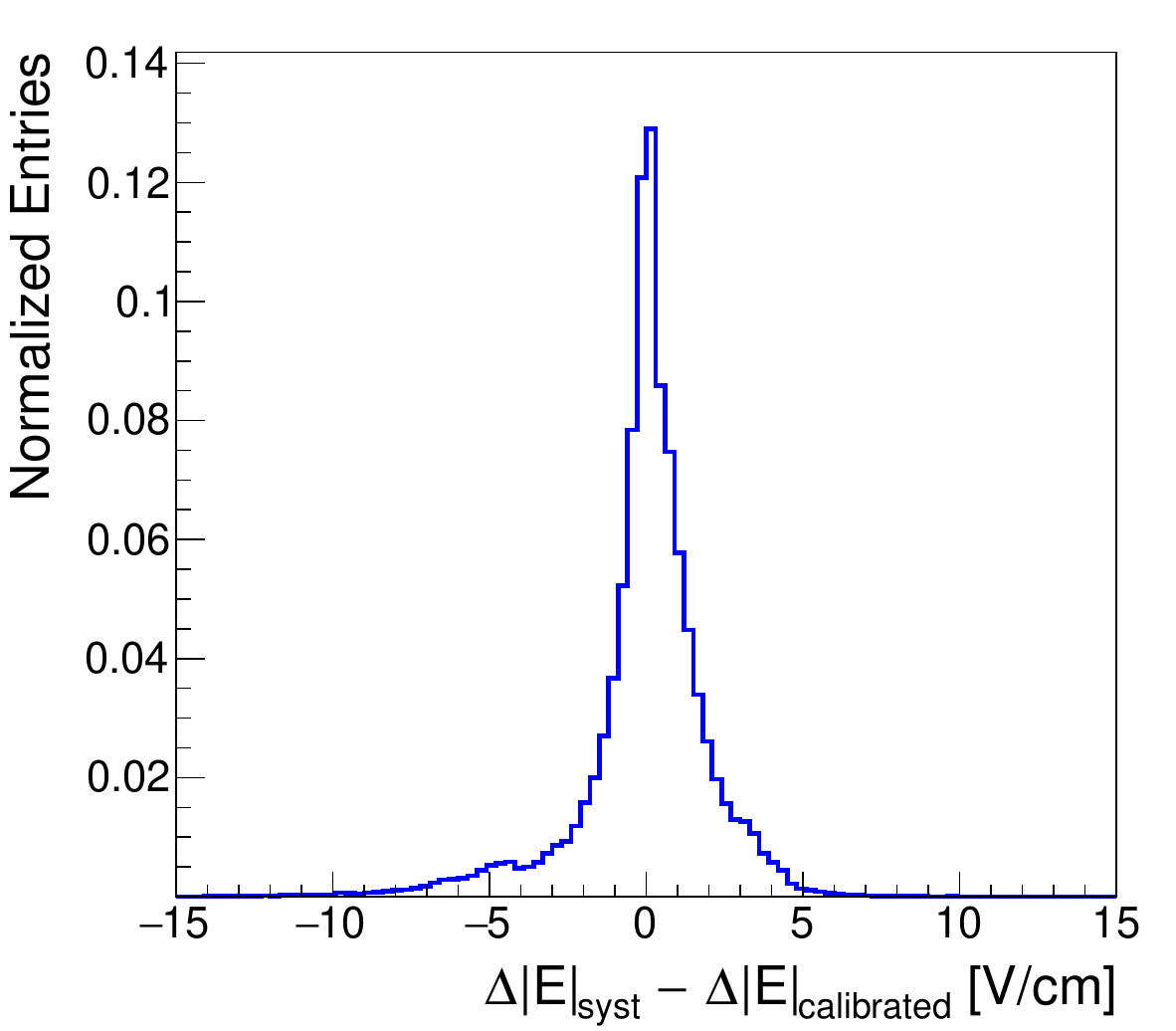}
    \caption{}
  \end{subfigure}
\Put(-387,140){\fontfamily{phv}\selectfont \textbf{MicroBooNE}}
\Put(-176,140){\fontfamily{phv}\selectfont \textbf{MicroBooNE}}
\caption{(a) Distributions of the systematic biases in the spatial offset measurement in data, using the method described in the text; (b) distribution of systematic bias in the electric field magnitude offset measurement in data using the same method, in units of \SI{}{V/cm}.}
\label{fig:SCECompSyst}
\end{figure}

\begin{figure}[tb]
\centering
  \begin{subfigure}{0.49\textwidth}
    \centering
    \includegraphics[width=.99\textwidth]{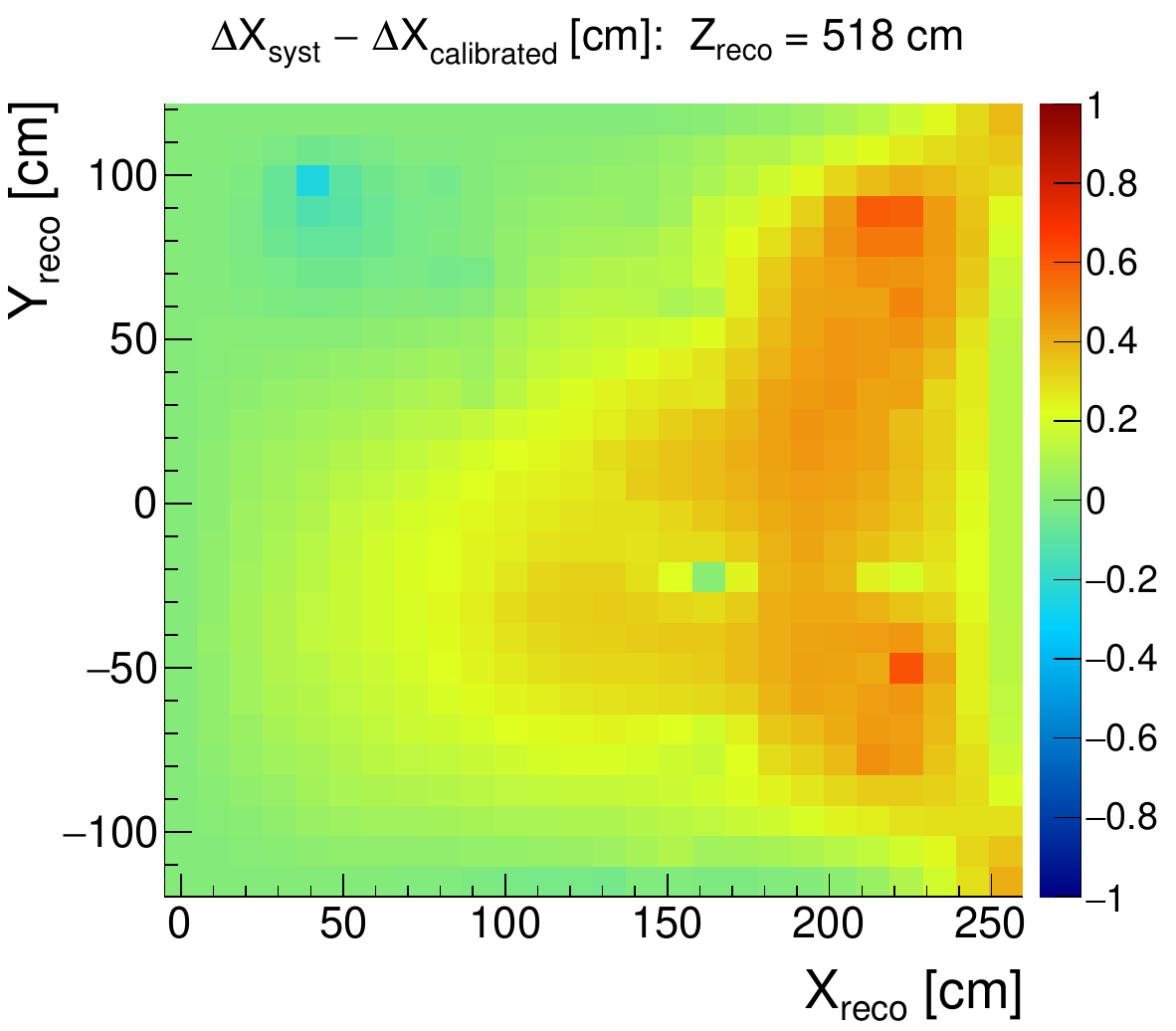}
    \caption{}
  \end{subfigure}
  \begin{subfigure}{0.49\textwidth}
    \centering
    \includegraphics[width=.99\textwidth]{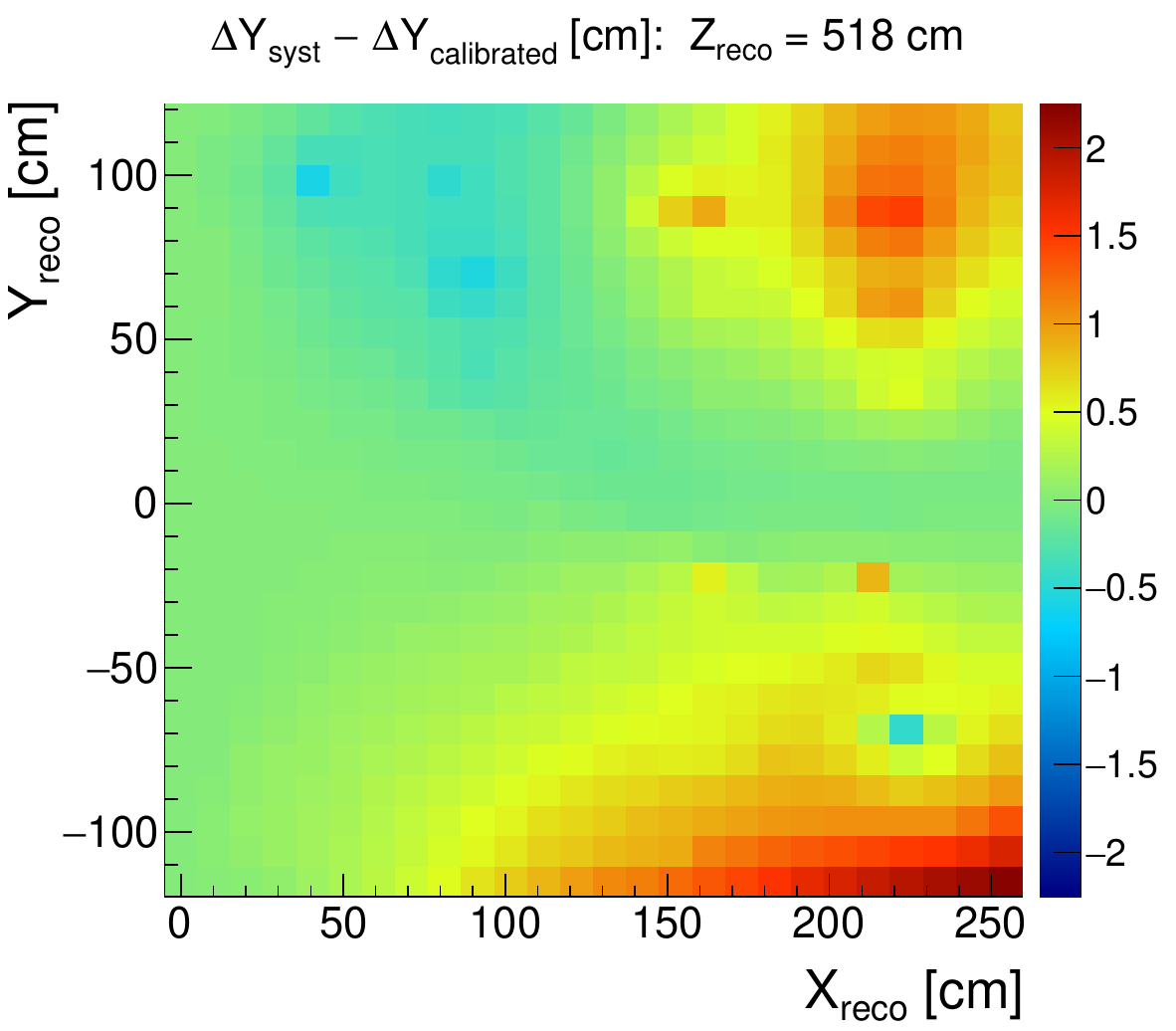}
    \caption{}
  \end{subfigure}
  \\
  \vspace{3mm}
  \begin{subfigure}{0.49\textwidth}
    \centering
    \includegraphics[width=.99\textwidth]{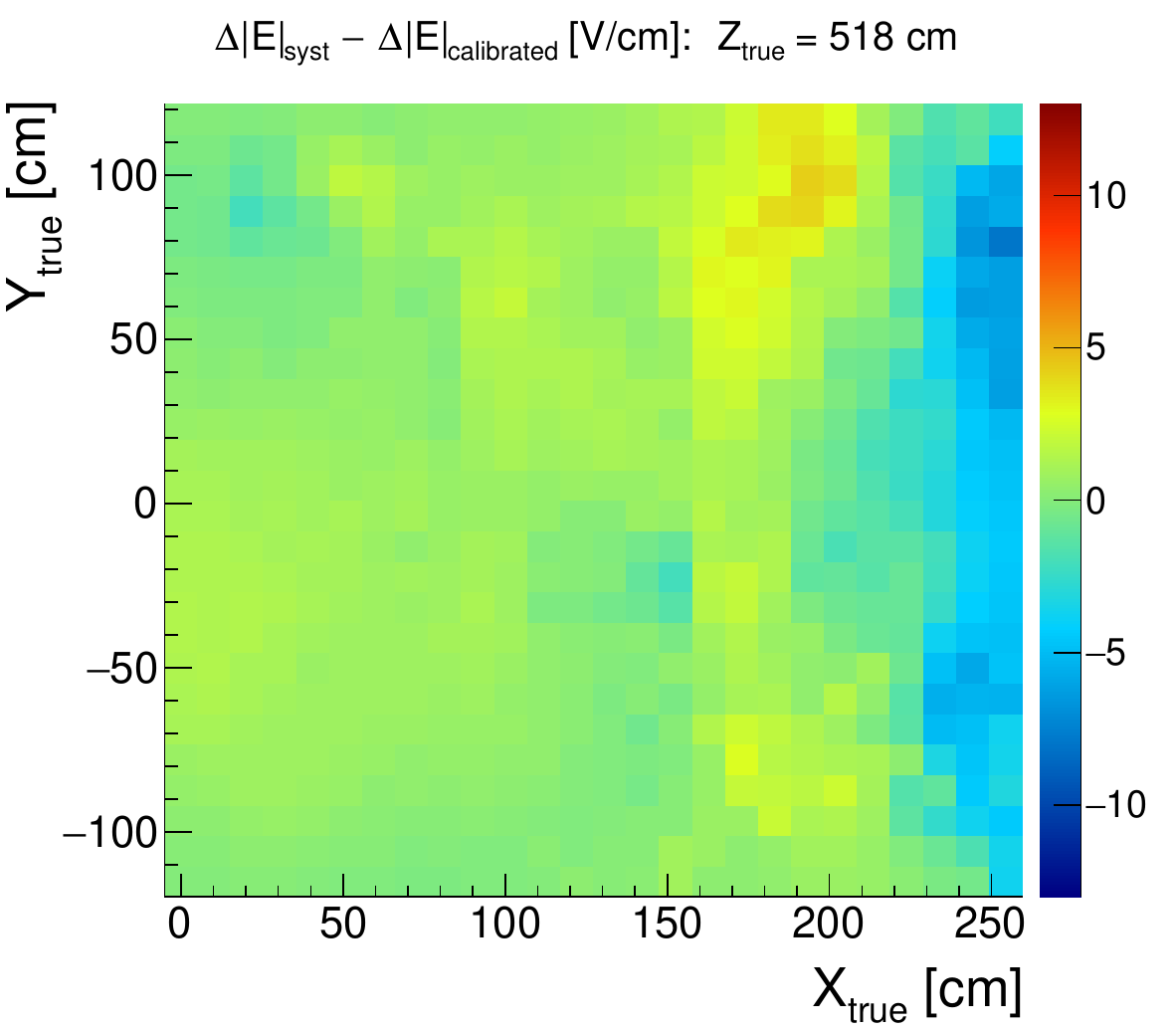}
    \caption{}
  \end{subfigure}
\Put(-280,568){\fontfamily{phv}\selectfont \textbf{MicroBooNE}}
\Put(-69,568){\fontfamily{phv}\selectfont \textbf{MicroBooNE}}
\Put(-182,145){\fontfamily{phv}\selectfont \textbf{MicroBooNE}}
\caption{Systematic bias in the spatial offset measurement in data using the method described in the text, showing the (a) ${\Delta}x$ bias and (b) ${\Delta}y$ bias in the $x-y$ plane for the central $z$ slice of the detector; (c) systematic bias in electric field magnitude offset measurement in data using the same method and looking at the same slice of the detector, in units of \SI{}{V/cm}.}
\label{fig:SCECompSyst2D}
\end{figure}

This systematic uncertainty, calculated everywhere throughout the TPC for all three dimensions of the spatial offsets, is an estimated bias which can be corrected on a point-by-point basis.  The results of correcting this bias are shown in figure~\ref{fig:LaserResidualsSyst}, leading to laser track residuals in data that are much closer to the Monte Carlo simulation distribution shown in figure~\ref{fig:LaserResiduals}.  This study serves as a check to ensure that the method is being carried out correctly, as it is expected that the overall magnitude of the laser track residuals decreases given that this information is being used in the estimation of the systematic bias.  This systematic uncertainty associated with the spatial offset measurement can be propagated to the electric field calculation as well, which is done by taking the bias-corrected spatial offset map and applying the methodology discussed in section~\ref{sec:method_efield}.  The impact of the systematic bias with respect to the nominal measurements of spatial offsets and electric field distortions for MicroBooNE data are shown in figure~\ref{fig:SCECompSyst} and figure~\ref{fig:SCECompSyst2D} throughout the TPC and in the central $z$ slice of the detector, respectively.  These comparisons quantify the performance of the data-driven calibration technique described in section~\ref{sec:method} on MicroBooNE data events.  Furthermore, the bias-corrected spatial and electric field distortion maps serve as an improved calibration of space charge effects at MicroBooNE.

\subsection[Impact on Stopping Muon $dE/dx$ Measurements]{Impact on Stopping Muon \boldmath{$dE/dx$} Measurements} \label{sec:systbias_stopmu}

To further illustrate the impact of space charge effects on particle energy reconstruction at MicroBooNE, a high-purity selection of contained neutrino-induced stopping muons is utilized~\cite{Pandora}.  This selection makes use of the following purity-enhancing requirements for the reconstructed muon:
\begin{itemize}
\item the reconstructed muon track must be at least \SI{100}{cm} in length;
\item the length of each reconstructed muon track segment depositing ionization charge on a single TPC wire must be less than \SI{3}{cm} for all collection plane wires receiving signal from the muon;
\item the median measured $dE/dx$ value in the last \SI{10}{cm} of the reconstructed muon track must be greater than \SI{2.5}{MeV/cm}; and
\item the $dE/dx$ profile of the reconstructed muon track must satisfy $\chi^{2}_{\mu}<4$.
\end{itemize}
$\chi^{2}_{\mu}$ is calculated by comparing the measured $dE/dx$ profile of the reconstructed muon track to the Bethe-Bloch expectation for a muon in liquid argon using a $\chi^2$ test, normalizing the $\chi^2$ to the number of hits in the reconstructed muon track~\cite{calibration}.

\begin{figure}[tb]
\centering
  \begin{subfigure}{0.49\textwidth}
    \centering
    \includegraphics[width=.99\textwidth]{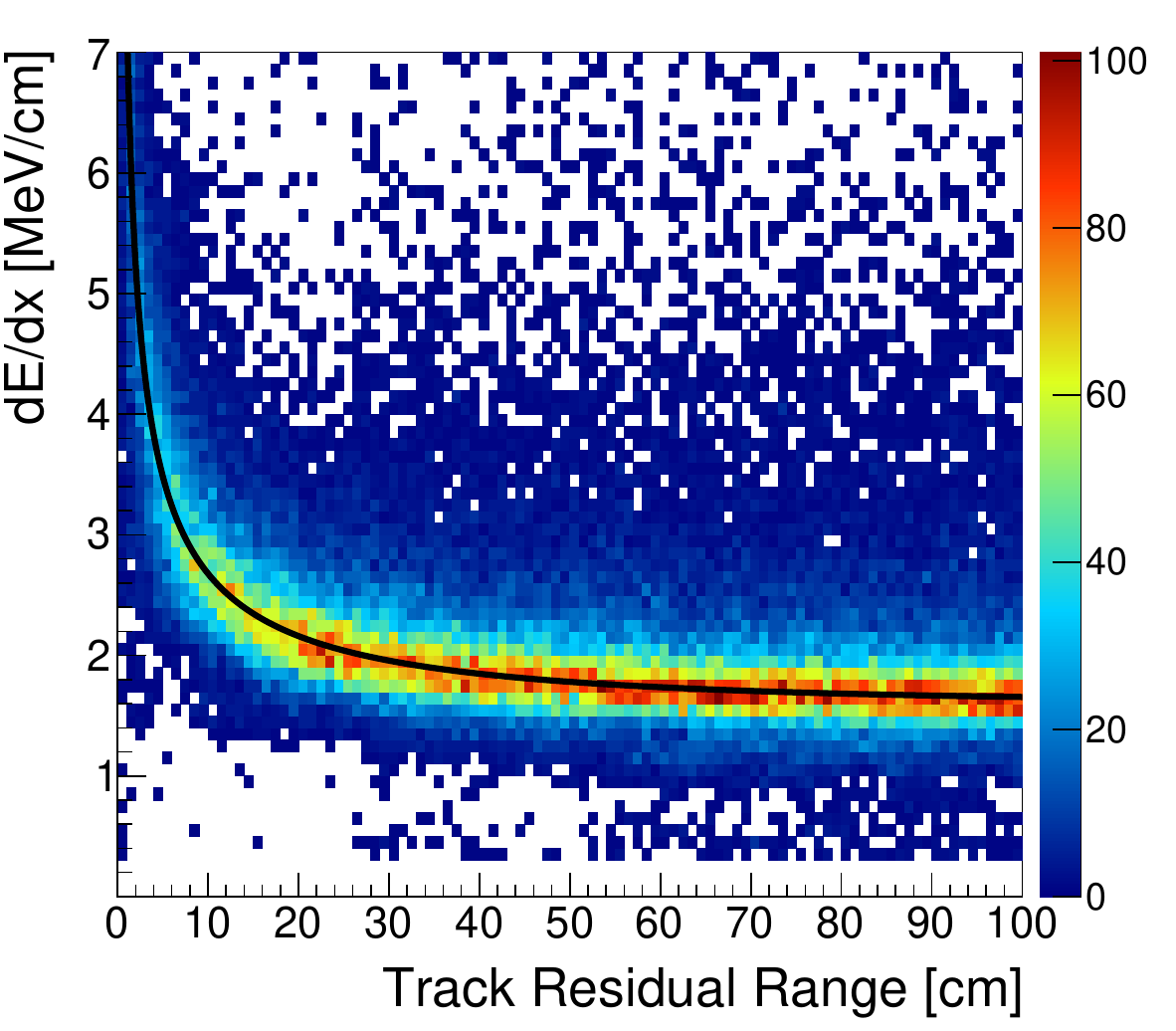}
    \caption{}
  \end{subfigure}
  \begin{subfigure}{0.49\textwidth}
    \centering
    \includegraphics[width=.99\textwidth]{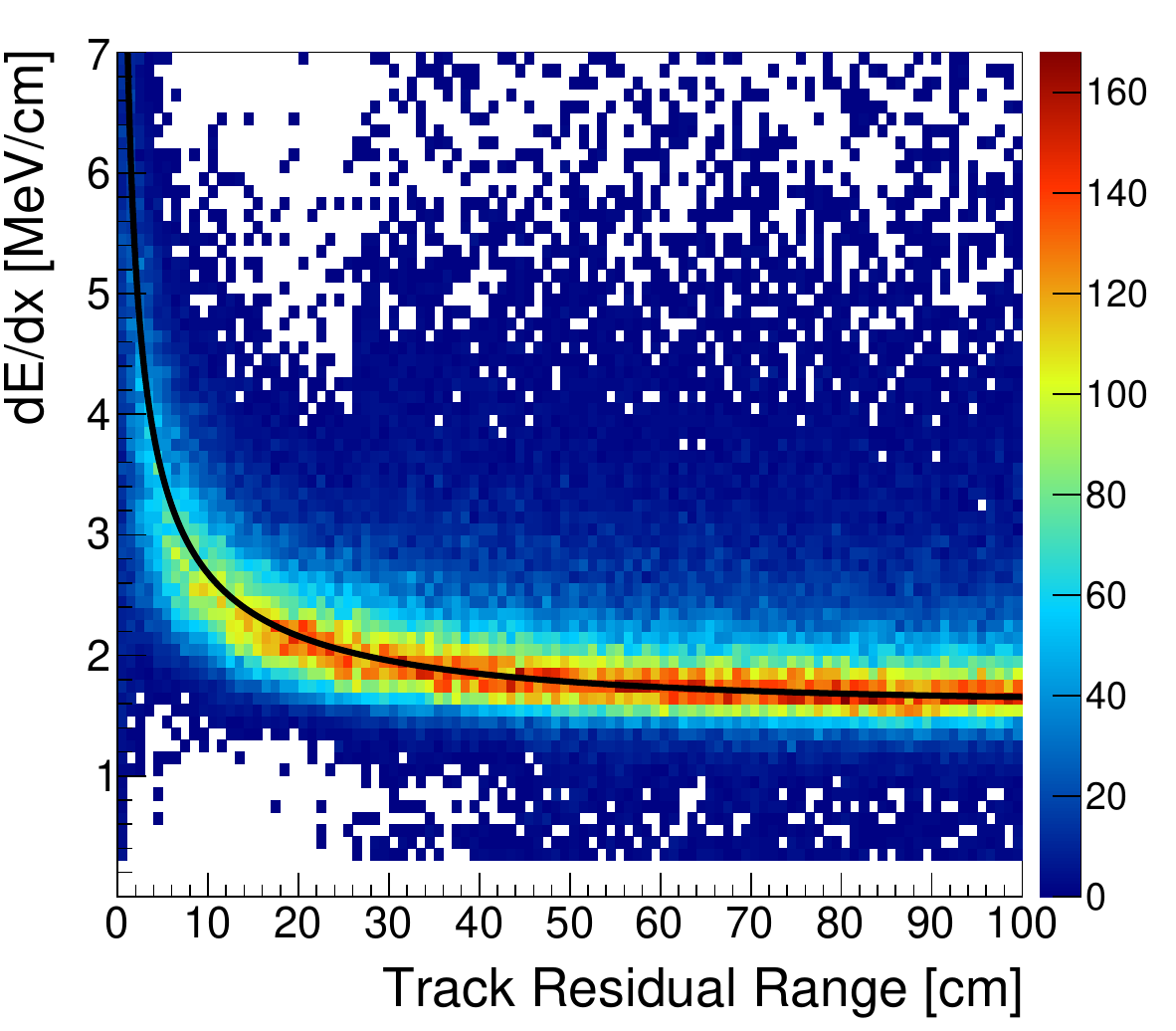}
    \caption{}
  \end{subfigure}
\Put(-405,203){\fontfamily{phv}\selectfont \textbf{On-beam}}
\Put(-312,203){\fontfamily{phv}\selectfont \textbf{MicroBooNE}}
\Put(22,402){\fontfamily{phv}\selectfont \textbf{Off-beam}}
\Put(115,402){\fontfamily{phv}\selectfont \textbf{MicroBooNE}}
\caption{$dE/dx$ as a function of residual range for a pure selection of stopping muons obtained using both (a) on-beam and (b) off-beam events.  The $dE/dx$ measurement, shown after application of the SCE $dE/dx$ calibration described in the text, is made using ionization signals from the collection plane of the MicroBooNE TPC.  The black curves show the Landau-Vavilov most probable energy loss per unit length associated with a detector thickness of \SI{4.5}{mm}, the median reconstructed muon track segment length associated with charge deposition on a single TPC wire.}
\label{fig:StoppingMuonRR}
\end{figure}

The stopping muon $dE/dx$ in the last \SI{100}{cm} of the reconstructed track, calculated using ionization charge signals from the collection plane, is shown in figure~\ref{fig:StoppingMuonRR} for two data samples: one when the neutrino beam is entering the TPC (``on-beam'' sample), and another making use of collected events when the neutrino beam was not running (``off-beam'' sample).  In the case of the off-beam sample, the selected muons are cosmic muons, though the selection is performed identically to the case of the on-beam sample.  As a result, the off-beam sample is characteristic of cosmogenic background events one would find in studies of charged-current muon neutrino events at MicroBooNE.  The distributions shown in figure~\ref{fig:StoppingMuonRR} demonstrate the high purity of the stopping muon selection, as the energy deposition profile of the selected tracks are largely consistent with that of stopping muons.

\begin{figure}[tb]
\centering
  \begin{subfigure}{0.49\textwidth}
    \centering
    \includegraphics[width=.99\textwidth]{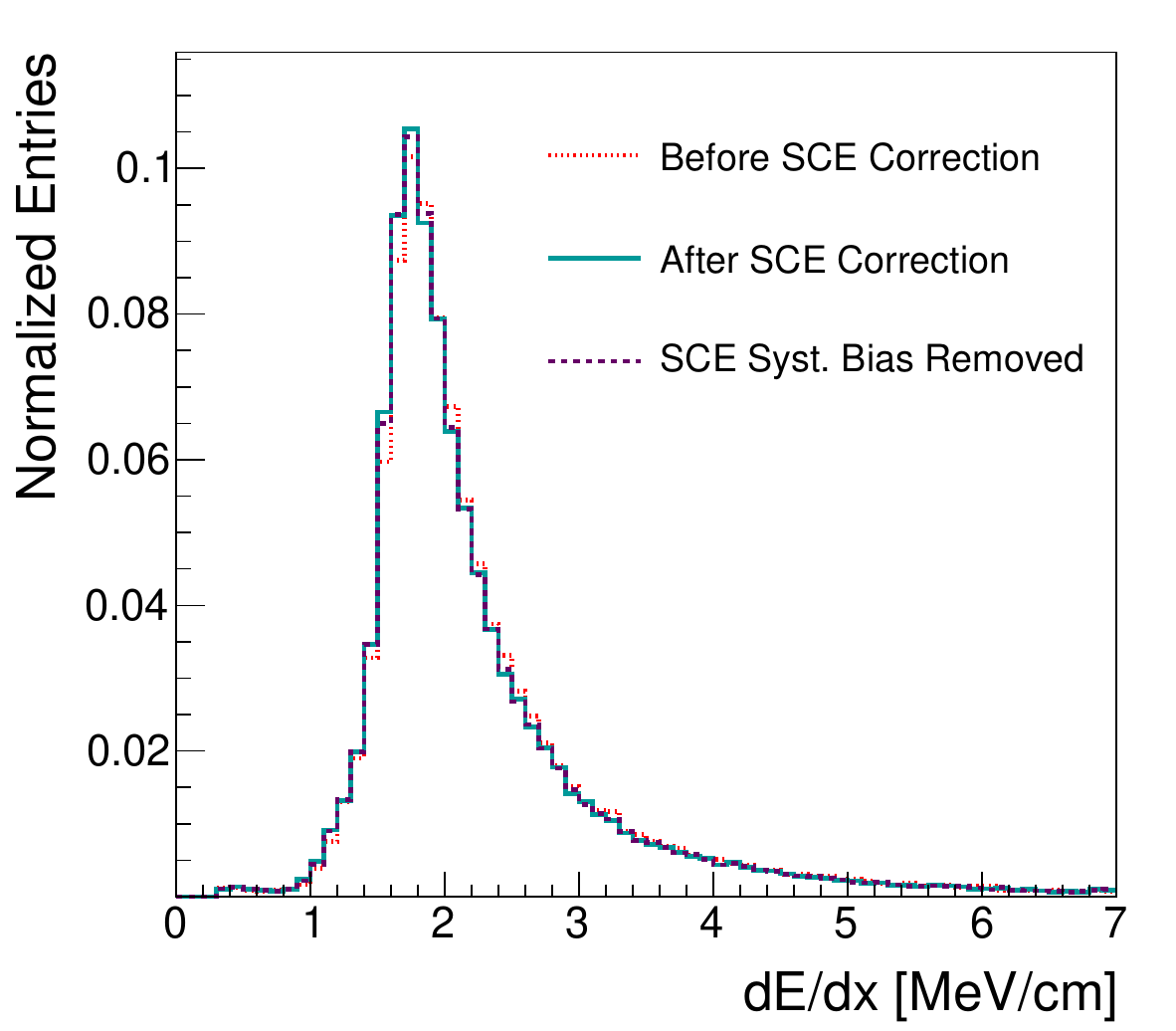}
    \caption{}
  \end{subfigure}
  \begin{subfigure}{0.49\textwidth}
    \centering
    \includegraphics[width=.99\textwidth]{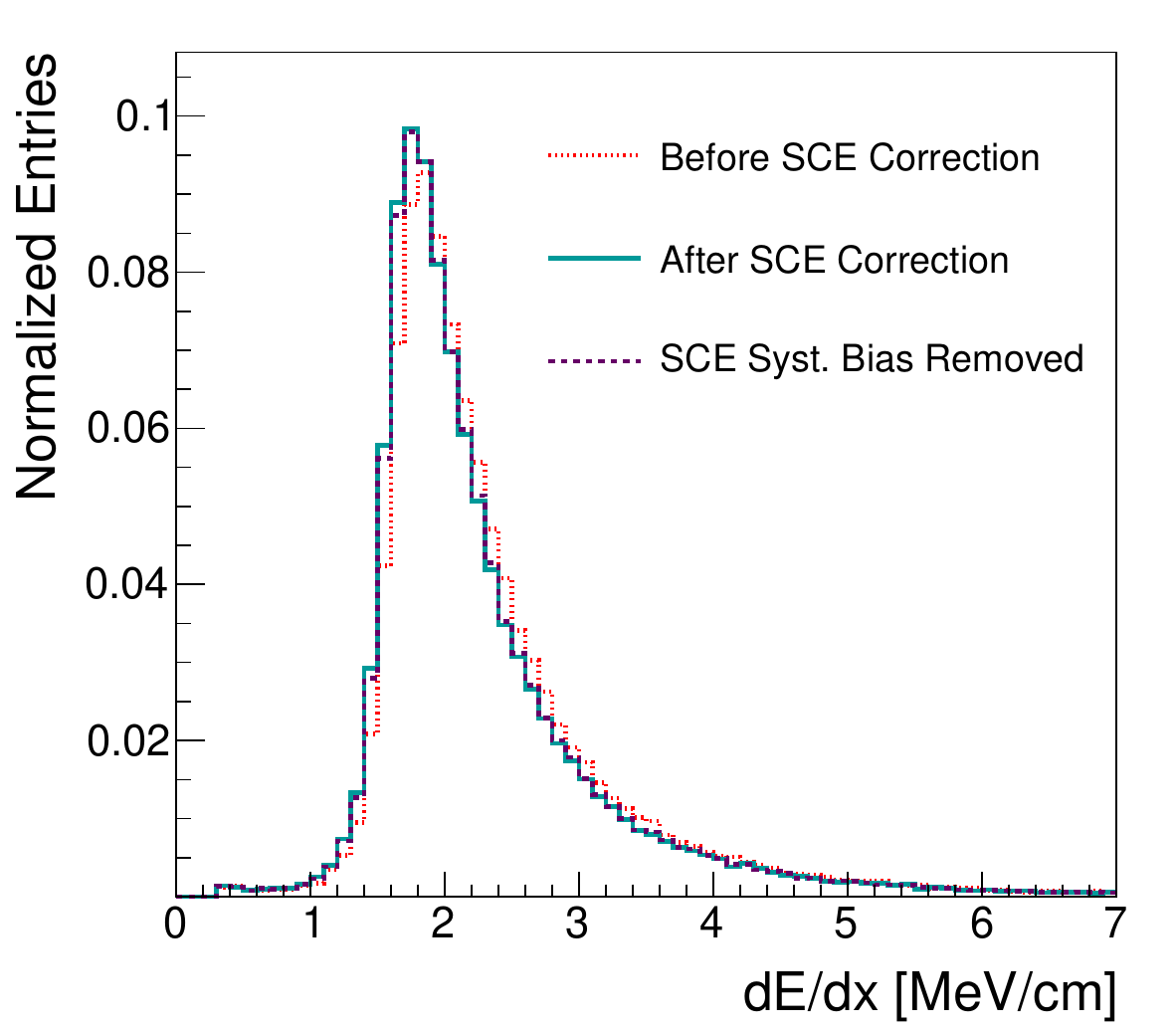}
    \caption{}
  \end{subfigure}
\Put(-293,20){\fontfamily{phv}\selectfont \textbf{On-beam}}
\Put(-303,-50){\fontfamily{phv}\selectfont \textbf{MicroBooNE}}
\Put(134,218){\fontfamily{phv}\selectfont \textbf{Off-beam}}
\Put(124,156){\fontfamily{phv}\selectfont \textbf{MicroBooNE}}
\caption{$dE/dx$ distribution for stopping muons in both (a) on-beam and (b) off-beam events, looking both before and after the SCE $dE/dx$ calibration described in the text is applied; these distributions are also shown after accounting for the estimated systematic bias in the measurement.}
\label{fig:StoppingMuonDEDX}
\end{figure}

Figure~\ref{fig:StoppingMuonDEDX} illustrates the impact of a space charge effect calibration on stopping muon $dE/dx$ measurements made using the collection plane wire signals, including the impact from the systematic bias calculation discussed above.  Two distinct corrections are applied: the spatial offset correction discussed in section~\ref{sec:results_bulk}, which accounts for squeezing/stretching of the track and associated bias of the energy deposition per unit length of the reconstructed track, and a second correction accounting for variations in electron-ion recombination due to electric field distortions changing the local electric field magnitude.  The second correction adjusts the measured ionization charge at a given point along the reconstructed muon track by a factor that accounts for the electric-field dependence of electron-ion recombination.  The recombination correction makes use of the MicroBooNE effective recombination parameters~\cite{calibration} for the modified Box model~\cite{ANrecomb}.

Stopping muons in on-beam events should be less impacted by space charge effects because they are more oriented in the direction of the neutrino beam, and as a result experience much less spatial squeezing that shifts $dE/dx$ measurements to larger values (due to $dx$ being smaller on average).  Stopping muons in off-beam events are largely downward-going cosmic muons that experience more vertical squeezing in the associated reconstructed track, especially impactful as spatial offsets in the vertical ($y$) direction are on average largest throughout the MicroBooNE TPC.  This implies an angular dependence of the SCE corrections to $dE/dx$ measurements.  Figure~\ref{fig:StoppingMuonDEDX} shows that the nominal SCE $dE/dx$ calibration leads to a shift of 1\% (4\%) in the mean of the $dE/dx$ distribution for on-beam (off-beam) events.  As expected, neutrino-induced stopping muons are less impacted by space charge effects than cosmogenic stopping muons, primarily a result of different levels of spatial squeezing along the muon tracks as explained above.  The variation in electron-ion recombination throughout the MicroBooNE TPC, due to variation in the electric field magnitude associated with space charge effects, has a subleading role in biasing $dE/dx$ measurements.  As shown in figure~\ref{fig:StoppingMuonDEDX}, the mean $dE/dx$ shift associated with the systematic bias in the SCE $dE/dx$ calibration is much smaller: 0.1\% (0.4\%) for on-beam (off-beam) events, or one tenth of the overall systematic $dE/dx$ bias associated with space charge effects.

\section{Time Dependence Study} \label{sec:timedep}

We have carried out an additional study to determine if there is significant time dependence of space charge effects in MicroBooNE.  Unless a calibration is performed frequently with respect to the rate at which underlying changes in the space charge configuration occur, this time dependence will ultimately serve as a lower bound on the level of precision achievable with any calibration technique.

In order to study the long-term time dependence of space charge effects at MicroBooNE, spatial distortions at various faces of the TPC are probed using the entry and exit points of $t_{0}$-tagged cosmic muon tracks.  The spatial offsets orthogonal to these TPC faces are most suitable for studying week-by-week variations of space charge effects in the detector as only about one hundred cosmic tracks are required to estimate the magnitude of the effect within a relatively small region (roughly \SI{1}{m^{2}}) on a TPC face.  Four different spots of the MicroBooNE TPC are probed in this way: two at the top face of the TPC and two at the bottom face of the TPC, with two different locations in the $z$ direction in both cases, looking at average measured value of the ${\Delta}y$ spatial offset from the TPC face in all cases. This study is carried out using off-beam MicroBooNE data collected between February~2016 and September~2018.

\begin{figure}[tb]
\centering
  \begin{subfigure}{0.49\textwidth}
    \centering
    \includegraphics[width=.99\textwidth]{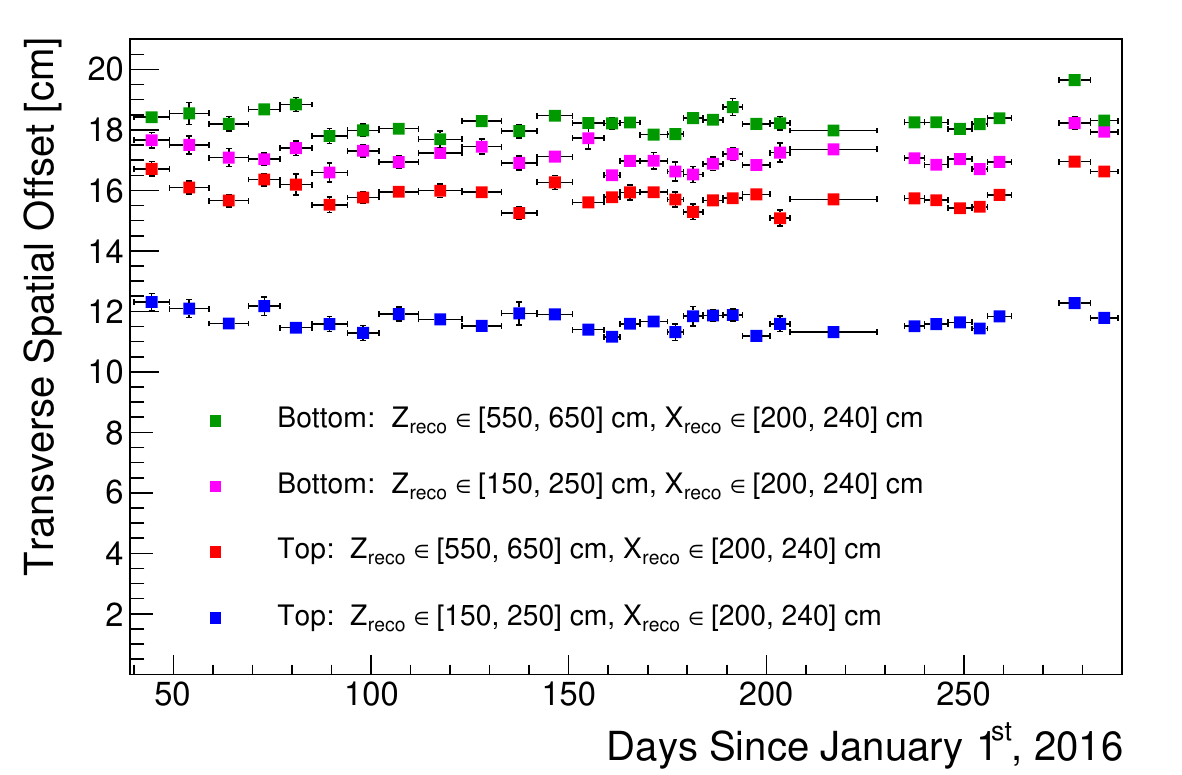}
    \caption{}
  \end{subfigure}
  \begin{subfigure}{0.49\textwidth}
    \centering
    \includegraphics[width=.99\textwidth]{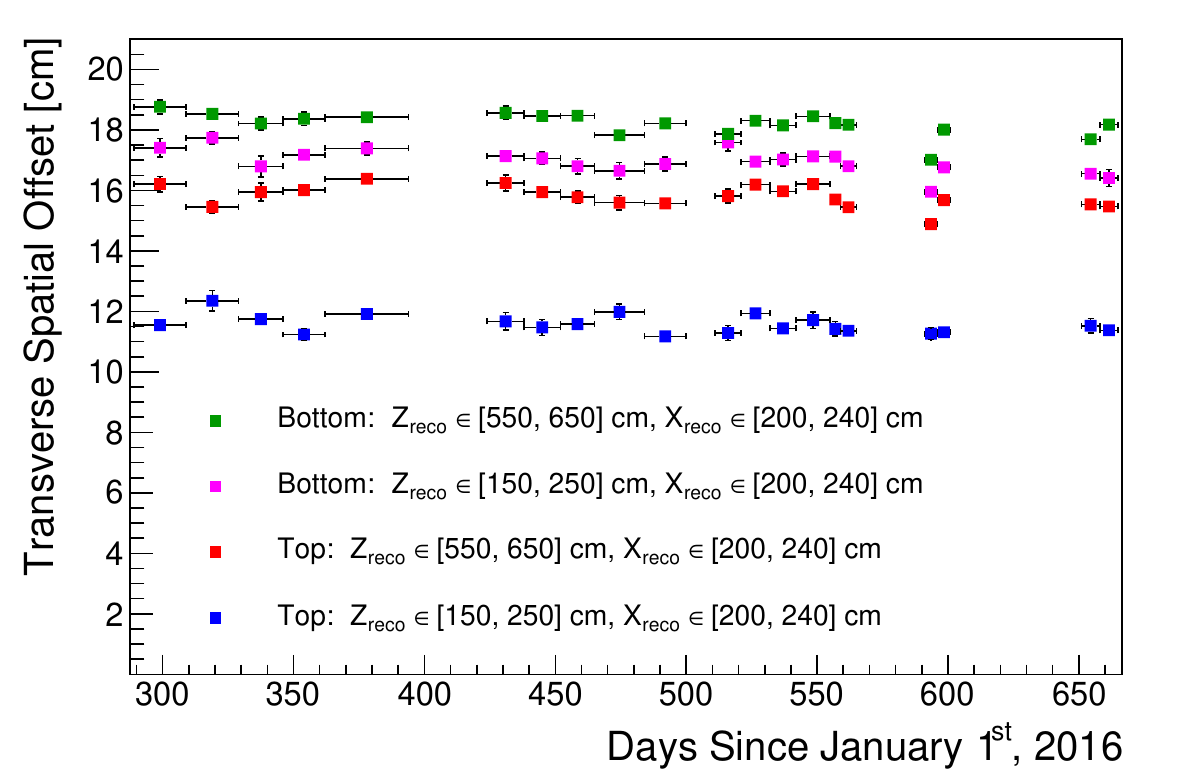}
    \caption{}
  \end{subfigure}
  \\
  \vspace{3mm}
  \begin{subfigure}{0.49\textwidth}
    \centering
    \includegraphics[width=.99\textwidth]{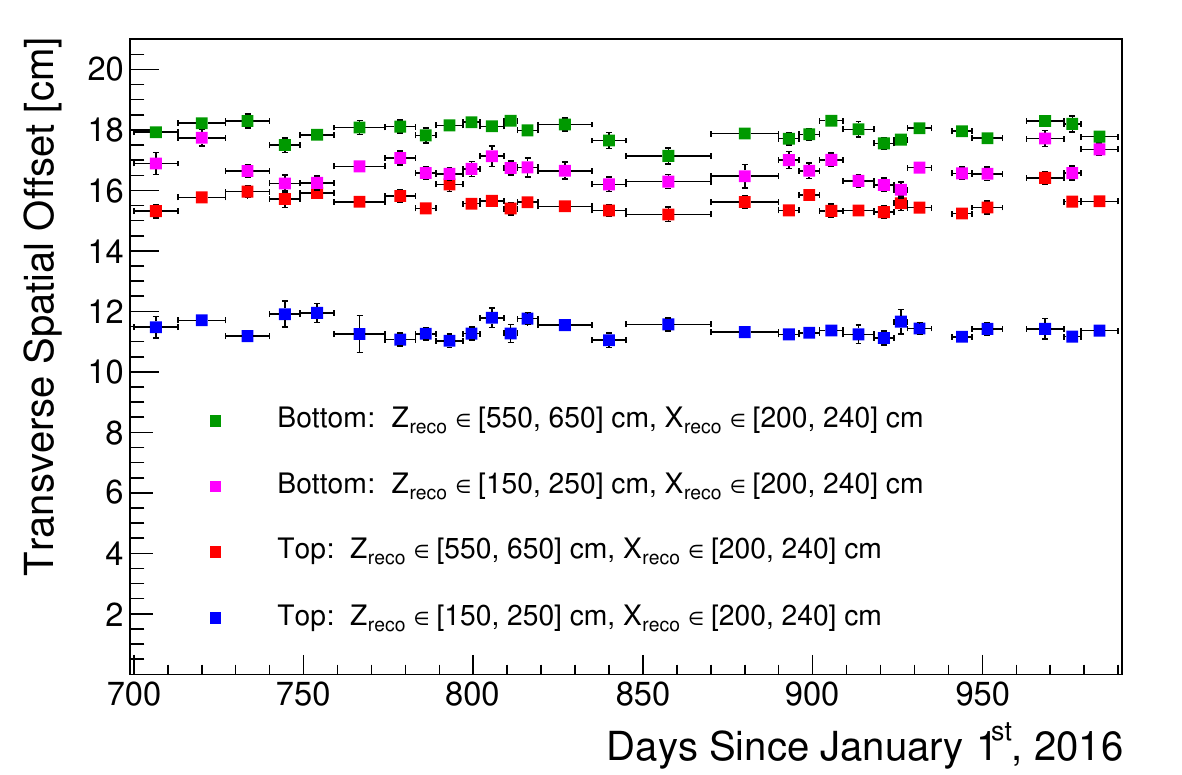}
    \caption{}
  \end{subfigure}
\Put(-287,493){\fontfamily{phv}\selectfont \textbf{Run 1}}
\Put(-183,493){\fontfamily{phv}\selectfont \textbf{MicroBooNE}}
\Put(-79,493){\fontfamily{phv}\selectfont \textbf{Run 2}}
\Put(25,493){\fontfamily{phv}\selectfont \textbf{MicroBooNE}}
\Put(-200,161){\fontfamily{phv}\selectfont \textbf{Run 3}}
\Put(-95,161){\fontfamily{phv}\selectfont \textbf{MicroBooNE}}
\caption{Time dependence of space charge effects in the MicroBooNE detector; shown are transverse spatial offsets (${\Delta}y$ in this case) at the top and bottom of the detector near two different values of $z$: one near the upstream part of the detector and another closer to the center of the detector in the beam direction.  Distributions are shown for three different time periods: (a) Run~1, (b) Run~2, and (c) Run~3, measured in days since January 1$\mathrm{^{st}}$, 2016.  Gaps in time are due to detector maintenance.  The overall level of variation in spatial offsets is less than 4\% across the entire data-taking period.}
\label{fig:TimeDep}
\end{figure}

The results of the time dependence study are shown in figure~\ref{fig:TimeDep}; these measurements take into account the \SI{4.5}{cm} average offset between the field cage and instrumented TPC volume, described in section~\ref{sec:method}.  It is observed that the magnitude of spatial offsets at each of the four probed regions in the TPC varies by no more than 4\% over time relative to the total magnitude of the effect.  An overall decrease in the magnitude of the spatial offsets is also observed over time, included as part of this 4\% variation.  It is expected that seasonal variations in the cosmic muon rate (which is proportional to the amount of space charge deposited in the detector) should account for no more than 1\% of this variation over time, bounded by measurements of seasonal cosmic muon rate variations at the MINOS near detector~\cite{MINOSCosmicRatePaper} and noting that the variation should be smaller at the surface, where MicroBooNE operates, due to the lower average cosmic muon energy.  One possible explanation for the larger time dependence of space charge effects in the detector than expected from seasonal variations in the cosmic muon rate is potential impact from shifting liquid argon flow patterns in the cryostat.  The velocity of convective argon flow in the detector is of similar scale to the argon ion drift velocity in the electric field, several millimeters per second.  The interference of these two effects may lead to changes in the space charge configuration in the detector, resulting in temporal variations in observed space charge effects.  However, this has not been definitively concluded as the cause for the long-term time dependence, and as such remains a speculative explanation.

Given the observed long-term time dependence of space charge effects, it is reasonable to target a level of precision of roughly 5\% or better with respect to measurement of spatial offsets, which is demonstrated to be the case for the methodology discussed in section~\ref{sec:method} within the majority of the MicroBooNE TPC, as shown in section~\ref{sec:systbias}.  In principle it is possible to improve the accuracy of the space charge effect calibration by performing the full calibration at different points in time, applying different calibrations to different periods of the total collected dataset.  No study of short-term variations (less than a day) of SCE over time is presented in this work, as it is difficult to do with any reasonable precision using cosmic muons given the relatively small cosmic muon sample obtainable on such timescales.  For timescales much shorter than a day, the UV laser system at MicroBooNE is better equipped to study the time dependence of SCE~\cite{laser_calib}.  However, techniques using cosmic muons, such as those discussed in this work, allow for continuous long-term monitoring of space charge effects in LArTPC detectors without interrupting data-taking.

\section{Conclusions} \label{sec:conclusions}

Cosmic muon tracks reconstructed in the MicroBooNE TPC have been shown to be particularly useful in measuring spatial distortions due to underlying electric field non-uniformities throughout the detector.  By comparing the results of a data-driven calibration method making use of cosmic tracks with predictions from a dedicated simulation of SCE in the detector, we have shown that the spatial distortions observed in MicroBooNE data are similar in nature to those that would arise from the presence of space charge in the detector.  However, the features of the spatial distortion map differ in detail from the predictions of a simple simulation.  This underscores the necessity of using a data-driven calibration procedure, such as the one presented in this work, at large LArTPC detectors operating near the surface in order to correct for these effects, which can impact the reconstruction of particle trajectories and measured ionization charge per unit length.

Spatial offsets in reconstructed particle trajectories as large as \SI{15}{cm} have been observed in MicroBooNE data, associated with underlying electric field distortions as large as 10\% with respect to the nominal MicroBooNE electric field of \SI{273.9}{V/cm}.  The calibration methodology presented in this work is shown to significantly improve the estimation of particle trajectories, studied in MicroBooNE data using measured track residuals associated with laser tracks from the UV laser system before and after the SCE calibration is applied.  A data-driven determination of systematic uncertainty on the measurement is also derived by using these laser track residuals, consistent with a spatial smearing of \SI{4}{mm} after the calibration is applied.  The associated uncertainty on the calculated underlying electric field is less than 1\% of the nominal MicroBooNE electric field, or 10\% of the total electric field distortion arising due to space charge effects in the TPC.  The impact of SCE on reconstructed muon $dE/dx$ measurements in the detector varies from 1\% to 4\%, heavily dependent on the angle of the reconstructed muon tracks, with a relative residual uncertainty of one tenth of the systematic effect after the SCE $dE/dx$ calibration is carried out.  Long-term temporal variations in the underlying space charge profile on the order of 4\% have been observed, which cannot be explained by seasonal variation in the cosmic muon rate at MicroBooNE.  It is possible that convective flow of liquid argon in the detector is responsible for these variations over time.

These calibration methods, developed at MicroBooNE, are foreseen to be useful to other running and future large LArTPC detectors.  This includes the single-phase~\cite{PDSP} and dual-phase~\cite{PDDP} ProtoDUNE detectors that serve as prototypes for the Deep Underground Neutrino Experiment (DUNE) far detector~\cite{DUNE} as well as the other two detectors of the Short-Baseline Neutrino (SBN) program~\cite{SBN} at Fermilab, SBND and ICARUS.  While the single-phase far detector and LArTPC near detector of DUNE are expected to experience negligible space charge effects due to being located deep underground and having a TPC with a short ionization drift length (\SI{0.5}{m}), respectively, the dual-phase far detector of DUNE will have such a long ionization drift length (\SI{12}{m}) that it may still see significant space charge effects due to positive ion build-up from \Ar{39} beta decays.  In all of these cases, the work presented here should provide a useful starting point for additional study.

\acknowledgments

This document was prepared by the MicroBooNE collaboration using the
resources of the Fermi National Accelerator Laboratory (Fermilab), a
U.S. Department of Energy, Office of Science, HEP User Facility.
Fermilab is managed by Fermi Research Alliance, LLC (FRA), acting
under Contract No. DE-AC02-07CH11359.  MicroBooNE is supported by the
following: the U.S. Department of Energy, Office of Science, Offices
of High Energy Physics and Nuclear Physics; the U.S. National Science
Foundation; the Swiss National Science Foundation; the Science and
Technology Facilities Council of the United Kingdom; and The Royal
Society (United Kingdom).  Additional support for the laser
calibration system and cosmic ray tagger was provided by the Albert
Einstein Center for Fundamental Physics (Bern, Switzerland).

\end{document}